\documentclass[11pt,letterpaper,fleqn]{article}
\usepackage{graphicx}
\usepackage{tabulary}
\usepackage{lineno}
\usepackage{hyperref}
\usepackage{subfigure}
\usepackage{cite}
\usepackage{xcolor,colortbl}
\usepackage{wrapfig}
\usepackage{enumitem} 

\selectcolormodel{rgb}
\definecolor{LHCb dark}{rgb}{0.0000,0.3412,0.6549}
\definecolor{UC red}{rgb}{0.8196,0.1176,0.2314} 
\definecolor{brickred}{rgb}{0.8, 0.25, 0.33}
\definecolor{LHCb dark}{rgb}{0.0000,0.3412,0.6549}
\definecolor{Gray}{gray}{0.85}
\definecolor{asparagus}{rgb}{0.53, 0.66, 0.42}
\definecolor{brightcerulean}{rgb}{0.11, 0.67, 0.84}
\definecolor{brightgreen}{rgb}{0.4, 1.0, 0.0}
\definecolor{candyapplered}{rgb}{1.0, 0.03, 0.0}

\let\oldenumerate\itemize
\renewcommand{\itemize}{
  \oldenumerate
  \setlength{\itemsep}{1pt}
  \setlength{\parskip}{1pt}
  \setlength{\parsep}{1pt}
}

\newcommand{\textpack}[1]{\textsc{#1}}

\newcommand{\statusnoteoff}[1]{}

\newcommand{\katznoteoff}[1]{}

\newcommand{\langenoteoff}[1]{}

\newcommand{\kenbloomnoteoff}[1]{}

\newcommand{\paolonoteoff}[1]{}

\newcommand{\carlosnoteoff}[1]{}

\newcommand{\tempnewpage}[0]{{\newpage}}
\def\s2i2{$S^2 I^2$}

\title{{\bf Strategic Plan for a \\ 
       Scientific Software Innovation Institute (\s2i2) \\ 
       for High Energy Physics }}
\date{\today}
\author{Peter Elmer (Princeton University)\\ 
Mark Neubauer (University of Illinois at Urbana-Champaign)\\
Michael D. Sokoloff (University of Cincinnati)}

\begin{document}


\vbox{
    \centering
    \includegraphics[width=1.0\textwidth]{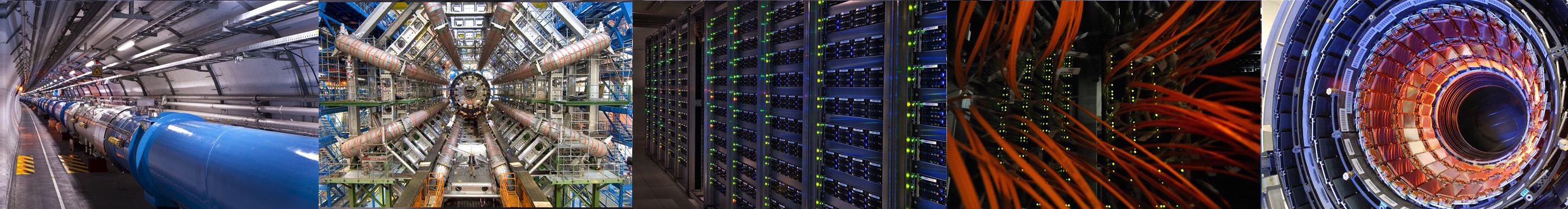}
    \maketitle 
\vskip 4.1in
\begin{tabulary}{1.0\textwidth}{lr}
\includegraphics[width=0.1\textwidth]{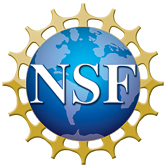} & \parbox{0.85\textwidth}{This report has been produced by the S2I2-HEP project (\url{http://s2i2-hep.org}) and supported by National Science Foundation grants ACI-1558216, ACI-1558219, and ACI-1558233. Any opinions, findings, conclusions or recommendations expressed in this material are those of the project participants and do not necessarily reflect the views of the National Science Foundation.}
\end{tabulary}
}

\thispagestyle{empty}
\newpage
\section*{Executive Summary}
\thispagestyle{empty}

The quest to understand the fundamental building blocks of nature
and their interactions is one of the oldest and most
ambitious of human scientific endeavors. Facilities such as CERN's Large 
Hadron Collider (LHC) represent a huge step forward in this quest.
The discovery of the Higgs boson, the observation of
exceedingly rare decays of $B$ mesons, and stringent constraints on
many viable theories of physics beyond the Standard Model (SM)
demonstrate the great scientific value of the LHC physics program. The
next phase of this global scientific project will be the
High-Luminosity LHC (HL-LHC) which will collect data starting
circa 2026 and continue into the 2030's.
The primary science goal is to search for physics beyond the
SM and, should it be discovered, to study its details and implications.
During the HL-LHC era, the ATLAS and CMS
experiments will record $\sim$10 times as much data from $ \sim 100 $
 times as
many collisions as in Run 1. 
The NSF and the DOE are planning large investments in detector
upgrades so the HL-LHC can operate in this high-rate environment.
A commensurate investment in R\&D for
the software for acquiring, managing, processing and analyzing HL-LHC
data will be critical
to maximize the return-on-investment in
the upgraded accelerator and detectors.

The strategic plan presented in this report is the result of
a conceptualization process carried out to
explore how a potential 
Scientific Software Innovation Institute (\s2i2) for High Energy Physics (HEP)
can play a key role in meeting HL-LHC
challenges. 
In parallel, a Community White Paper (CWP) describing the bigger picture
was
prepared under the auspices of the HEP Software Foundation (HSF).
Approximately 260
scientists and engineers participated in more than a dozen workshops
during 2016--2017, most jointly sponsored by both HSF and the \s2i2-HEP project. 

The conceptualization process concluded that the mission of an
Institute should be two-fold: it should serve  as an active center for software
R\&D {\em and}  as an intellectual hub for the 
larger software R\&D effort required to
ensure the success of the HL-LHC scientific program. 
Four high-impact R\&D areas were identified as highest priority
for the U.S.\ university community:
(1)~development of advanced algorithms for data reconstruction and
triggering; (2)~development of highly performant analysis systems
that reduce `time-to-insight' and maximize the HL-LHC physics
potential; (3)~development of data organization, management and access
systems for the Exabyte era; (4)~leveraging the recent
advances in Machine Learning and Data Science.
In addition,
sustaining the investments in the
fabric for distributed high-throughput computing was identified
as essential to current and future operations activities. 
A plan for managing and evolving an \s2i2-HEP 
identifies a set of activities and services that will
enable and sustain the Institute's mission.

As an intellectual hub, the Institute should
lead efforts in 
(1) developing
partnerships between HEP and the cyberinfrastructure communities
(including Computer Science, Software Engineering, 
Network Engineering, and Data
Science) for novel
approaches to meeting HL-LHC challenges, 
(2) bringing in new
effort from U.S.\ Universities emphasizing professional
development and training, and 
(3) sustaining HEP software and
underlying knowledge related to the algorithms
and their implementations  over the two decades required.
HEP is a global, complex,
scientific endeavor.
These activities will help ensure that the software
developed and deployed by a globally distributed
community will extend the science reach of the HL-LHC
and will be sustained over its
lifetime.

The strategic plan for an \s2i2 targeting HL-LHC physics presented in
this report reflects a community vision.
Developing, deploying, and maintaining 
sustainable software for the HL-LHC experiments 
has tremendous technical and social challenges. 
The campaign of R\&D, testing, and deployment 
should start as soon as possible to ensure
readiness for doing physics when the upgraded accelerator and
detectors turn on.
An NSF-funded, U.S.\ university-based \s2i2 to lead a
``software upgrade"
will complement the hardware investments being made.
In addition to enabling the best possible HL-LHC science,
an \s2i2-HEP will 
bring together the larger cyberinfrastucture and HEP 
communities to  study problems
and build
algorithms and software implementations to address issues
of general import for
Exabyte scale problems in big science.






\newpage

\section*{Contributors}
\thispagestyle{empty}

The PIs of the \s2i2-HEP conceptualization project (Peter Elmer, Mark 
Neubauer, Mike Sokoloff) 
served as overall editors for this document. Many of the general ideas 
regarding the roadmap came out during the Community White Paper (CWP) process 
from the many
participants (see the full list in Appendix~\ref{workshoplist}). The specifics
regarding an \s2i2-HEP came from the U.S.\ subset of the participants. The 
focus area text
in this document includes elements taken and adapted from (in particular) 
the CWP working group documents on ``Data Analysis and Interpretation'',
``Data Organization, Management and Access'',  ``Software Trigger and 
Event Reconstruction'' and ``Machine Learning''. Many people provided
feedback and comments during the drafting and editing process, but we
would like to call out explicitly Ken Bloom, Rob Gardner, Dan Katz, David 
Lange and Gordon Watts who helped with the editing or provided systematic 
comments on the whole document. Frank Wuerthwein organized elements regarding
OSG as well as other contributions in the document.
Miron Livny provided valuable feedback and ideas on the potential Institute 
role throughout the process. The title page images are courtesy of CERN.

\vskip 0.6in

\section*{Endorsers}
As of 4 April, 2018, the strategic plan described in this report has 
been explicitly endorsed by
Fernando Barreiro Megino (University of Texas at Arlington),
Lothar Bauerdick (Fermilab),
Matthew Bellis (Siena College),
Doug Benjamin (Duke University),
Riccardo Maria Bianchi (University of Pittsburgh),
Kenneth Bloom (University of Nebraska-Lincoln),
Brian Bockelman (University of Nebraska-Lincoln),
Joseph F. Boudreau (University of Pittsburgh),
Paolo Calafiura (Berkeley Lab),
Jeffrey Carver (University of Alabama),
J. Taylor Childers (Argonne National Laboratory),
Giuseppe Cerati (Fermilab),
Kyle Cranmer (New York University),
Kaushik De (University of Texas at Arlington),
Dave Dykstra (Fermilab),
Peter Elmer (Princeton University),
Amir Farbin (University of Texas at Arlington),
Robert Gardner (University of Chicago),
Sergei V. Gleyzer (University of Florida),
Dick Greenwood (Louisiana Tech University),
Oliver Gutsche (Fermilab),
Andrew Hanushevsky (SLAC),
Michael Hildreth (University of Notre Dame),
Benjamin Hooberman (University of Illinois at Urbana-Champaign),
Daniel S. Katz (University of Illinois at Urbana-Champaign),
Alexei Klimentov (BNL),
Markus Klute (MIT),
Slava Krutelyov (University of California, San Diego),
Valentin Kuznetsov (Cornell University),
David Lange (Princeton University),
Eric Lancon (BNL),
Steven Lantz (Cornell University),
Matthieu Lefebvre (Princeton University),
Miron Livny (University of Wisconsin - Madison),
Sudhir Malik (University of Puerto Rico Mayaguez),
Tanu Malik (DePaul University),
Usha Mallik (University of Iowa),
Mario Masciovecchio (University of California, San Diego),
Ruslan Mashinistov (University of Texas at Arlington),
Kevin McDermott (Cornell University),
Shawn McKee (University of Michigan),
Mark S Neubauer (University of Illinois at Urbana-Champaign),
Harvey Newman (Caltech),
Jason Nielsen (University of California, Santa Cruz),
Peter Onyisi (University of Texas at Austin),
Gabriel Perdue (Fermilab),
Jim Pivarski (Princeton University),
Fernanda Psihas (University of Texas at Arlington),
Dan Riley (Cornell University),
Eduardo Rodrigues (University of Cincinnati),
Terry Schalk (University of California, Santa Cruz),
Henry F Schreiner (University of Cincinnati),
Horst Severini (University of Oklahoma),
Elizabeth Sexton-Kennedy (Fermilab),
Michael D Sokoloff (University of Cincinnati),
Matevz Tadel (University of California, San Diego),
Douglas Thain (University of Notre Dame),
Ilija Vukotic (University of Chicago),
Gordon Watts (University of Washington),
Torre Wenaus (Brookhaven National Laboratory),
Mike Williams (MIT),
Peter Wittich (Cornell University),
Frank Wuerthwein (University of California, San Diego),
Avi Yagil (University of California, San Diego),
Wei Yang (SLAC)
and
Saul Youssef (Boston University).

\newpage

\thispagestyle{empty}
\newpage
\tableofcontents
\thispagestyle{empty}
\newpage
\setcounter{page}{1}
\section*{Document revisions}
\thispagestyle{empty}
This document has been published as arXiv 1712.06592.
The revision history is:
\begin{description}
\item[18 Dec 2017] Initial version submitted to NSF (arXiv v1) 
\item[14 Jan 2018] Add additional endorsers, add references to final CWP, minor fixes (arXiv v2)
\end{description}
\newpage


\section{Introduction}
\label{sec:intro}

The High-Luminosity Large Hadron Collider (HL-LHC) is scheduled
to start producing data in 2026 and extend the LHC physics program
through the 2030s.  Its primary science goal is to search for
Beyond the Standard Model (BSM) physics, or study its details if there
is an intervening discovery.  Although the basic constituents of
ordinary matter and their interactions are extraordinarily well
described by the Standard Model (SM) of particle physics, a quantum field 
theory built
on top of simple but powerful symmetry principles, it is incomplete.
For example, most of the gravitationally interacting matter in the
universe does not interact via electromagnetic or strong nuclear
interactions.  As it produces no directly visible signals, it is
called dark matter.  Its existence and its quantum nature lie outside
the SM.  Equally as important, the SM does not address fundamental
questions related to the detailed properties of its {\em own}
constituent particles or the specific symmetries governing their
interactions.  To achieve this scientific program, the HL-LHC will record 
data from 100 times as many proton-proton collisions as did Run 1 of the LHC. 

Realizing the full potential of the HL-LHC requires
large investments in upgraded hardware. The R\&D preparations for these 
hardware upgrades are underway and the full project funding for the construction
phase is expected to begin to flow in the next few years. 
The two general purpose detectors at the LHC, ATLAS and CMS, are
operated by collaborations of more than 3000 scientists each.  
U.S.\ personnel constitute about 30\% of the collaborators on these
experiments.  Within the U.S., funding for the construction and
operation of ATLAS and CMS is jointly provided by the Department
of Energy (DOE) and the National Science Foundation (NSF). 
Funding for U.S.\ participation in the LHCb 
experiment is provided only by the NSF. 
The NSF is also planning a major role in the hardware upgrade of the
ATLAS and CMS detectors for the HL-LHC.  This would use the Major Research
Equipment and Facilities Construction (MREFC) mechanism with a
possible start in 2020.  

Similarly, the HL-LHC will require
commensurate investment in the research and development necessary
to develop and deploy the software to acquire, manage, process, and analyze 
the data. 
Current estimates
of HL-LHC computing needs significantly exceed what will be possible 
assuming Moore's Law and more or less constant operational budgets.~\cite{Campana2016}
The underlying nature of computing hardware (processors, storage, networks)
is also evolving, the quantity of data to be processed is increasing
dramatically, its complexity is increasing, and more sophisticated
analyses will be required to maximize the HL-LHC physics yield.
The magnitude of the HL-LHC computing problems to be solved will require
different approaches. 
In planning for the
HL-LHC, it is critical that all parties agree on the
software goals and priorities, and that the efforts tend to complement
each other.  
In this spirit, the HEP Software Foundation (HSF) began
a planning exercise in late 2016 to prepare a Community White Paper
(CWP).  Its goal is to provide a roadmap for software R\&D in
preparation for the HL-LHC era which would identify and prioritize
the software research and development investments required:
\begin{enumerate}
\itemsep0em
  \item 
    to enable new approaches to computing and software that 
    can  radically extend the physics reach of the detectors; and
  \item 
    to achieve improvements in software efficiency, scalability, and 
    performance, and to make use of the advances in CPU, storage, and
    and network technologies;
  \item 
    to ensure the long term sustainability of the software through 
    the lifetime of the HL-LHC.
\end{enumerate}
In parallel to the global CWP exercise and with funding from the NSF, the U.S.\ community executed a conceptualization process to produce a Strategic Plan
for how a Scientific Software Innovation Institute (\s2i2) for high-energy physics (HEP) could
help meet the HL-LHC challenges. Specifically, the \s2i2-HEP conceptualization 
process~\cite{S2I2HEP} had three additional goals:
\begin{enumerate}
\itemsep0em
 \item
  to identify specific focus areas for R\&D efforts that could be part of an \s2i2 in the U.S.\ university community; 
 \item
  to build a consensus within the U.S.\ HEP software community for a 
  common effort; and
 \item
  to engage with experts from the related fields of scientific computing and
  software engineering to identify topics of mutual interest and
  build teams for collaborative work to advance the scientific
  interests of all the communities. 
\end{enumerate}
This document, the {\it ``Strategic Plan for a Scientific Software Innovation 
Institute (\s2i2) for High Energy Physics''}, is the result of the 
\s2i2-HEP process.

The existing computing system of the LHC experiments is the result of almost
20 years of effort and experience. In addition to addressing the 
significant future challenges, sustaining the fundamental aspects of what 
has been built to date is also critical.
Fortunately, the collider nature of this physics program implies that 
essentially all computational challenges are pleasantly parallel.
The large LHC collaborations each produce tens of billions of
events per year through a mix of simulation and data triggers
recorded by their experiments, and all events are mutually independent
of each other. This intrinsic simplification from the science itself permits 
aggregation of distributed
computing resources and is well-matched to the use of {\it high throughput 
computing} to meet LHC and HL-LHC computing needs.
In addition, the LHC today requires more computing resources than will be
provided by funding agencies in any single location (such as CERN).
Thus {\it distributed high-throughput computing} (DHTC) will continue to 
be a fundamental characteristic of the HL-LHC. Continued support for DHTC is 
essential for the HEP community.


Developing, maintaining and deploying sustainable software for the HL-LHC
experiments, given these constraints, is both a technical and a
social challenge.  An NSF-funded, U.S.\ university-based \s2i2 can play a primary leadership role in
the international HEP community to prepare the ``software upgrade'' which
should run in parallel to the hardware upgrades planned for the HL-LHC.
The Institute will exist within a larger context of international
and national projects.
It will  help build a more cooperative, community process
for developing, prototyping, and deploying software.
It will drive research and development in a specific
set of focus  areas (see Section~\ref{sec:focusareas})
using its own resources directly,
and also leveraging them through collaborative efforts.
In addition, the Institute will
  serve as an intellectual hub for the
  larger community effort
  in HEP software and computing --
it will
serve as a center for
  disseminating knowledge related to the current software and computing
  landscape, emerging technologies, and tools (see Section~\ref{sec:role}).
It will work closely with  its partners to evolve a common vision for
future work (see Section~\ref{sec:partnerships}).
To achieve its specific goals,
the Executive Director and core personnel will support backbone
activities;
Area Managers will organize the day to day activities 
of distributed efforts within each focus
area.
Goals and resources allocated to all projects will be reviewed
on an annual basis, and updated with advice from stakeholders via the 
Institute's Steering Board (see Section~\ref{sec:orggov}).
Altogether, the  Institute
should serve as both an
active software research and development center
and as an intellectual hub for the larger software R\&D effort required
to ensure that the HL-LHC is able to address its Science Driver questions
(see Section~\ref{science_drivers}).



\tempnewpage

\section{Science Drivers} 
\label{science_drivers}



An \s2i2 focused on software required for an upgraded HL-LHC is
primarily intended to enable
the discovery of BSM physics,
or study its details, if there
is a discovery before the upgraded accelerator
and detectors turn on.
To understand why discovering and elucidating
BSM physics will be transformative, we need to start with
the key concepts of the SM, what they explain, what they do not,
and how the HL-LHC will address the latter.

In the past 200 years, physicists have discovered the basic
constituents of ordinary matter and they have developed a very
successful theory to describe the interactions (forces) among
them.  All atoms, and the molecules from which they are built, can
be described in terms of these constituents.  The nuclei of atoms
are bound together by strong nuclear interactions.  Their decays
result from strong and weak nuclear interactions.  Electromagnetic
forces bind atoms together, and bind atoms into molecules.  The
electromagnetic, weak nuclear, and strong nuclear forces are described
in terms of quantum field theories.  The predictions of these
theories are extremely precise, generally speaking, and they have been validated
with equally precise experimental measurements.  The electromagnetic
and weak nuclear interactions are intimately related to each other,
but with a fundamental difference: the particle responsible for the
exchange of energy and momentum in electromagnetic interactions
(the photon) is massless while the corresponding particles responsible
for the exchange of energy and momentum in weak  interactions (the
$ W $ and $ Z $ bosons) are about 100 times more massive than the
proton.  A critical element of the SM is the
prediction (made more than 50 years ago) that a qualitatively new
type of particle, called the Higgs boson, would give mass to the
$ W $ and $ Z $ bosons.  Its discovery at the LHC by the ATLAS and CMS Collaborations in 2012~\cite{HIGG-2012-27,Chatrchyan:2012xdj}
confirmed experimentally
the last critical element of the SM.

The SM describes essentially all known physics very well, but its
mathematical structure and some important empirical evidence tell
us that it is incomplete.  These observations motivate a large
number of SM extensions, generally  using the formalism of quantum
field theory, to describe  BSM physics.
For example, ``ordinary" matter accounts for only 5\% of the
mass-energy budget of the universe, while dark matter, which interacts
with ordinary matter gravitationally, accounts for 27\%.  While we
know something about dark matter at macroscopic scales, we know
nothing about its microscopic, quantum nature, {\em except} that
its particles are not found in the SM and they lack electromagnetic
and SM nuclear interactions.  BSM physics also addresses a key
feature of the observed universe: the apparent dominance of matter
over anti-matter.  The fundamental processes of leptogenesis and
baryongenesis (how electrons and protons, and their heavier cousins,
were created in the early universe) are not explained by the SM,
nor is the required level of CP violation (the asymmetry between
matter and anti-matter under charge and parity conjugation).
Constraints on BSM physics come from ``conventional" HEP experiments
plus others searching for dark matter particles either directly or
indirectly.

The LHC was designed to search for the Higgs boson and for BSM
physics -- goals in the realm of discovery science.  The ATLAS and
CMS detectors are optimized to observe and measure the direct
production and decay of massive particles.  They have now begun to
measure the properties of the Higgs boson more precisely to test
how well they accord with SM predictions.

Where ATLAS and CMS were primarily designed to study high mass particles
directly, LHCb  was designed to study heavy flavor physics where
quantum influences of very high mass particles, too massive to be directly detected at LHC, are manifest in lower
energy phenomena.  Its primary goal is to look for BSM physics in
CP violation (CPV, defined as asymmetries in the decays of particles and their
corresponding antiparticles) and rare decays of beauty and charm
hadrons.  As an example of how one can relate flavor physics to
extensions of the SM, Isidori, Nir, and Perez \cite{Isidori:2010kg}
have considered model-independent BSM constraints from measurements
of mixing and CP violation.  They assume the new fields are heavier
than SM fields and construct an effective theory.  Then, they
``analyze all realistic extensions of the SM in terms of a limited
number of parameters (the coefficients of higher dimensional
operators)." They determine bounds on an effective coupling strength
couplings 
of their results is that kaon, $ B_d $, $ B_s $, and $ D^0 $ mixing
and CPV measurements provide powerful constraints that are complementary
to each other and often constrain BSM physics more powerfully than
direct searches for high mass particles.

The Particle Physics Project Prioritization Panel (P5) issued their
{\em Strategic Plan for U.S.\ Particle Physics}~\cite{p5Final} in
May 2014.  It was very quickly endorsed by the High Energy
Physics Advisory Panel and submitted to the DOE and the NSF.  The
report says,\katznoteoff{use quotes instead of italics?}
{\em we have identified five compelling lines of inquiry that show great
promise for discovery over the next 10 to 20 years.
These are the Science Drivers:
\begin{itemize}
 \item
  Use the Higgs boson as a new tool for discovery
 \item
  Pursue the physics associated with neutrino mass
 \item
  Identify the new physics of dark matter
 \item
  Understand cosmic acceleration: dark matter and inflation
 \item
  Explore the unknown: new particles, interactions, and 
  physical principles.
\end{itemize}
}  
The HL-LHC will address the first, third, and fifth of these
using data acquired at
twice the energy of Run 1 and with 100 times the
luminosity.
As the P5 report says,

\vskip 0.1in
\begin{quotation}
{\em
The recently discovered Higgs boson is a form of matter never
before observed, and it is mysterious. What principles determine
its effects on other particles? How does it interact with neutrinos
or with dark matter? Is there one Higgs particle or many? Is the
new particle really fundamental, or is it composed of others? The
Higgs boson offers a unique portal into the laws of nature, and
it connects several areas of particle physics. Any small deviation
in its expected properties would be a major breakthrough.

The full discovery potential of the Higgs will be unleashed by
percent-level precision studies of the Higgs properties. The
measurement of these properties is a top priority in the physics
program of high-energy colliders. The Large Hadron Collider
(LHC) will be the first laboratory to use the Higgs boson as a
tool for discovery, initially with substantial higher energy running
at 14 TeV, and then with ten times more data at the High-
Luminosity LHC (HL-LHC). The HL-LHC has a compelling and
comprehensive program that includes essential measurements
of the Higgs properties.
}  
\end{quotation}

In addition to HEP experiments,
the LHC hosts the one of world's foremost nuclear physics
experiments.
``The ALICE Collaboration has built a dedicated heavy-ion detector to 
exploit the unique physics potential of nucleus-nucleus interactions at 
LHC energies. 
[Their] aim is to study the physics of strongly interacting matter at 
extreme energy densities, where the formation of a new phase of matter, 
the quark-gluon plasma, is expected. 
The existence of such a phase and its properties are key issues in 
QCD for the understanding of confinement and of chiral-symmetry
restoration."~\cite{alice}
In particular, 
these collisions reproduce the temperatures and pressures
of hadronic matter in the very early universe,
and so provide a unique window into the physics of that era. 

\vskip 0.10in
\noindent
{\bf Summary of Physics Motivation:}\ \
The ATLAS and CMS collaborations published letters of intent to do
experiments at the LHC in October 1992, about 25 years ago.  At the
time, the top quark had not yet be discovered; no one knew if the
experiments would discover the Higgs boson, supersymmetry, technicolor,
or something completely different.  Looking forward, no one can say
what will be discovered in the HL-LHC era.  However, with data from
100 times the number of collisions recorded in Run 1 
the next 20 years are likely to
bring even more exciting discoveries.

\tempnewpage

\section{Computing Challenges}
\label{sec:computing_challenges}

During the HL-LHC era (Run 4, starting circa 2026/2027), 
the ATLAS and CMS experiments intend to record about
$ 10 \times $  as much data from $ 100 \times $ as many collisions as they did in
in Run 1, and at twice the energy:
the Run 1 integrated luminosity for each of these experiments was  $ {\cal L}_{\rm int} \sim 30 \, 
{\rm fb}{}^{-1} $ at 7 and 8 TeV; for Run 4 it is designed to be $ {\cal L}_{\rm int} \sim 3000 \,
{\rm fb}{}^{-1} $ at 14 TeV by 2035.
Mass storage costs will not improve sufficiently to record so much more data,
and the projection is that budgets will allow the experiments to collect only a
factor of 10 more.
{For the LHCb experiment, this $ 100 \times $ increase in data and processing 
over that of Run 1 will start in  Run 3 (beginning circa 2021).}
The software and computing budgets for these experiments are
projected to remain flat. Moore's Law (a doubling number of transistors on integrated circuits every two years), even if it continues to hold, 
will not provide
the required increase in computing power to enable fully processing
all the data.
Even assuming the experiments significantly reduce the
amount of data stored per event,
the total size of the datasets will be well into the exabyte
scale;
they will be constrained primarily by costs and funding levels,
not by scientific interest.
{\em The overarching goal of an \s2i2 for HEP 
will be to maximize the return-on-investment in the upgraded accelerator
and detectors
 to enable break-through
scientific discoveries}.

\begin{wraptable}{c}{0.52\textwidth}
\begin{center}
\vspace{-2.0em}
\caption{
Estimated mass storage to be used by the LHC experiments
in 2018, at the end of 
Run 2 data-taking.
Numbers extracted from
the CRSG report
to CERN's RRB in April 2016~\cite{CRSG2016} 
for ALICE, ATLAS, \& CMS and taken from
LHCb-PUB-2017-019~\cite{LHCb-PUB-2017-019}
for LHCb.
\label{tab:CSRG2016}
}
\resizebox{0.52\textwidth}{!}{
\begin{tabular}{| l |c|c| c |}\hline
Experiment & Disk Usage (PB) & Tape Usage (PB) & Total (PB) \\ \hline
 ALICE         &  98 &    86  &  184 \\
 ATLAS         & 164 &   324  &  488 \\
 CMS           & 141 &   247  &  388 \\
 LHCb          &  41 &    79  &  120 \\ \hline
 {\bf Total}   & 444 &   736 & {\bf 1180} \\ \hline
\end{tabular}
}
\end{center}
\end{wraptable}

Projections for the HL-LHC start with the operating experience of
the LHC to date, and account for the increased luminosity
to be provided by the accelerator and the increased sophistication
of the detectors.
Run 2 started in the summer of 2015, with the bulk of the luminosity
being delivered in 2016--2018.
The April 2016 Computing Resources Scrutiny Group (CRSG) report
to CERN's Resource Review Board (RRB) report~\cite{CRSG2016} estimated 
the ALICE, ATLAS, and CMS usage
for the full period 2016--2018.
A summary is shown in Table \ref{tab:CSRG2016},
along with corresponding numbers for LHCb taken from  their
2017 estimate~\cite{LHCb-PUB-2017-019}.
Altogether, the LHC experiments will be saving more than
an exabyte of data in mass storage by the end of Run 2.
In their April 2017 report~\cite{CRSG2017}, the CRSG says that 
``growth equivalent to 20\%/year [...] towards HL-LHC [...] should be
assumed".

\begin{figure}
\includegraphics[width=0.48\linewidth]{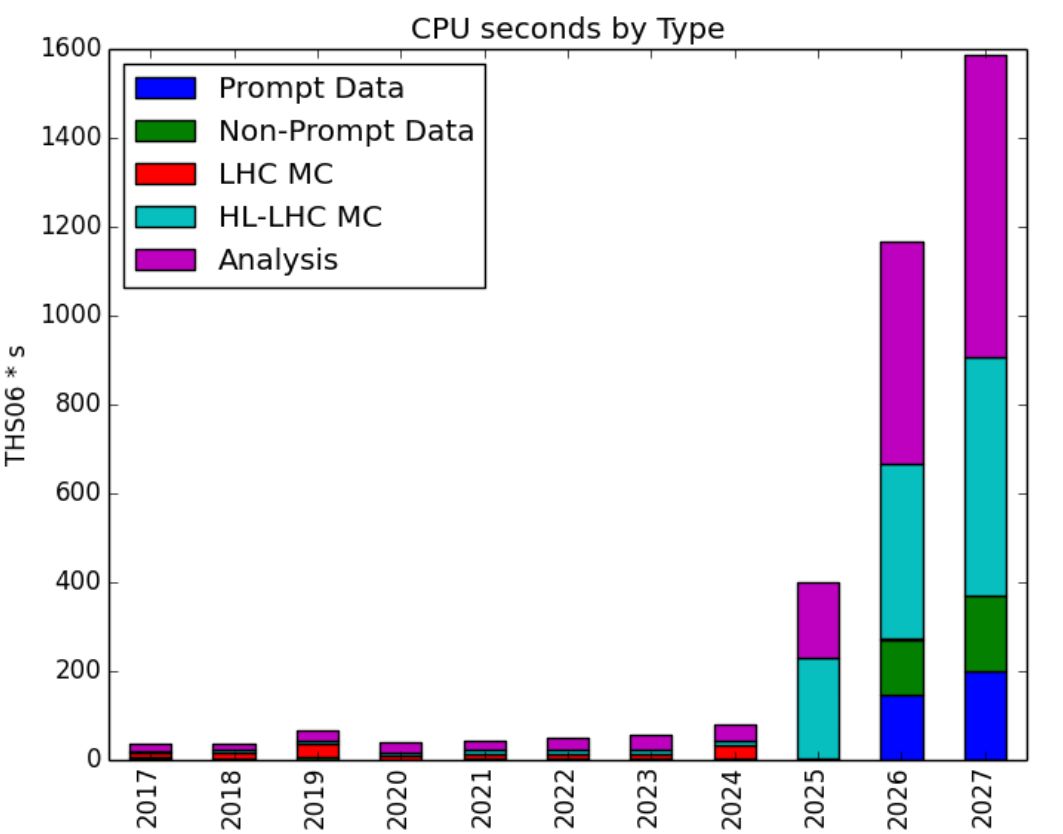}
\quad
\includegraphics[width=0.42\linewidth]{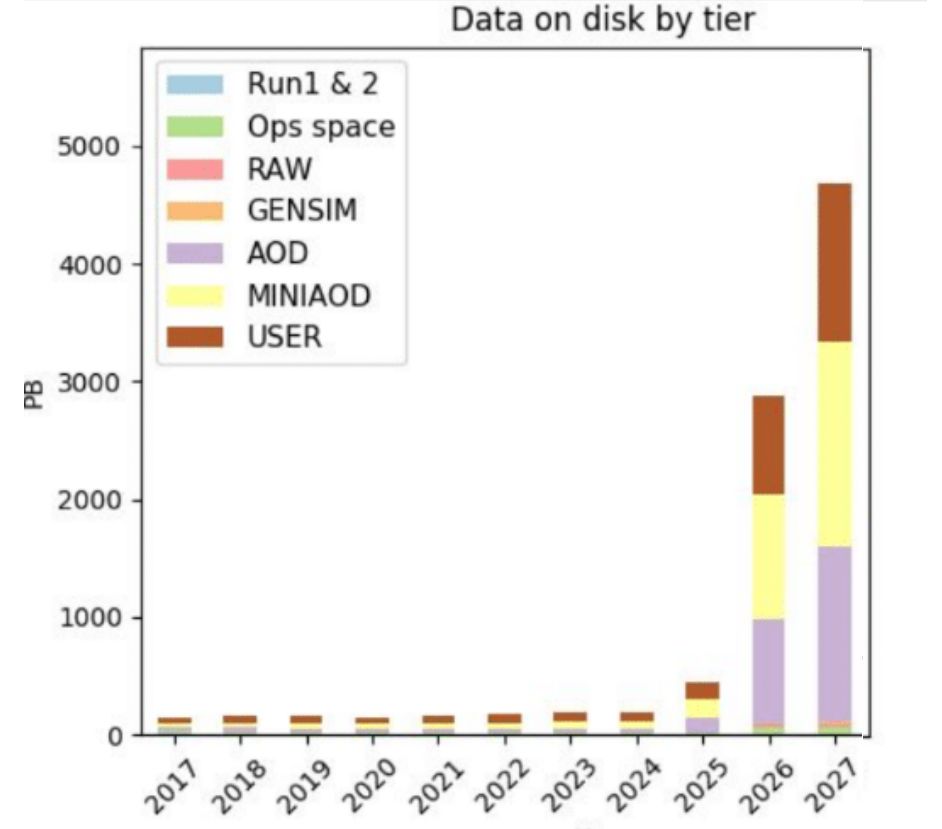}
\caption{
\label{fig:cms}
CMS CPU and disk requirement evolution into the first two years of HL-LHC~\cite{SEXTON2017}}
\end{figure}
\begin{figure}
\centering
\includegraphics[width=0.72\linewidth]{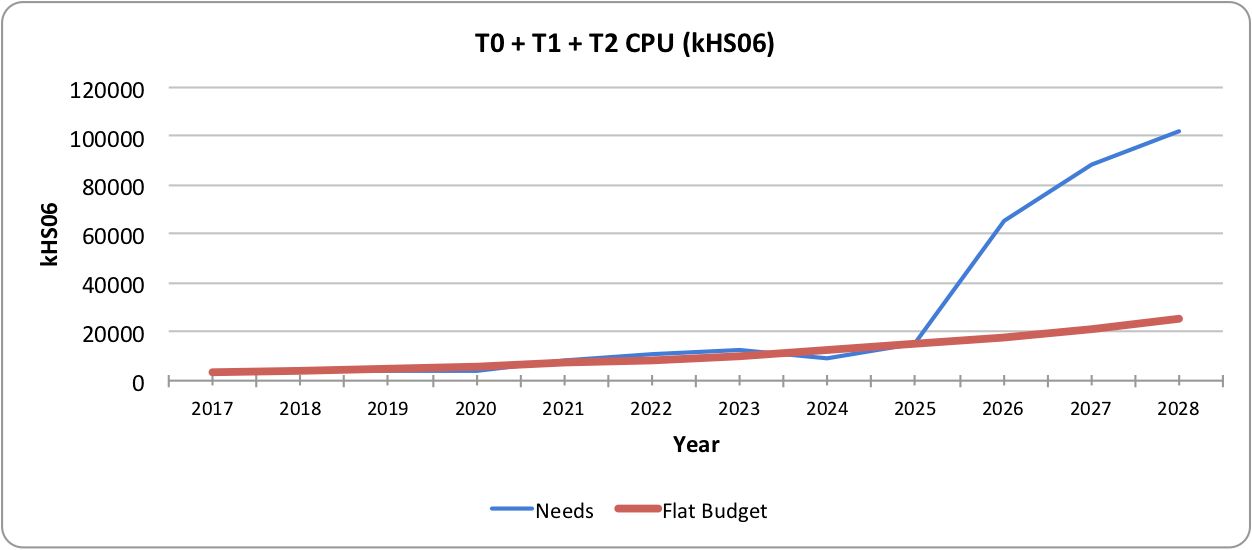}
\includegraphics[width=0.72\linewidth]{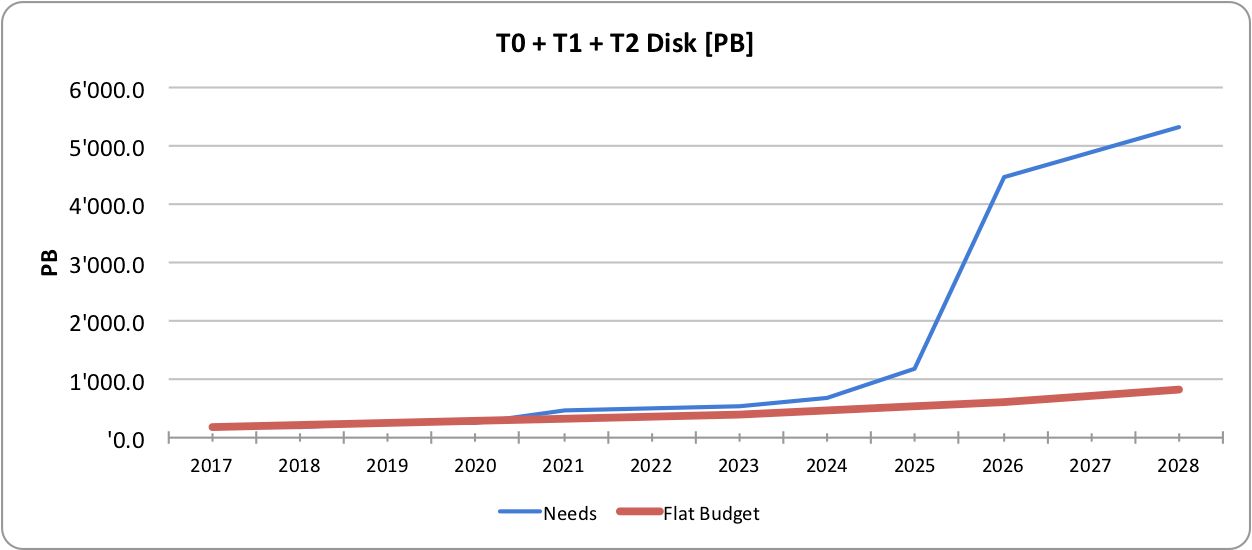}
\caption{
ATLAS CPU and disk requirement evolution into the first three years of HL-LHC,
compared to growth rate assuming flat funding~\cite{CAMPANA2017}.
\label{fig:atlas}
}
\end{figure}

While no one expects such projections to be accurate over 10 years,
simple exponentiation predicts a factor of six growth.
Naively extrapolating resource requirements
using today's software and computing models,
the experiments project significantly greater needs.
The magnitude of the discrepancy is illustrated in Figures.~\ref{fig:cms}
and \ref{fig:atlas} for CMS and ATLAS, respectively.
The CPU usages are specified in kHS06 years where a
``standard" modern core corresponds to about 10 HS06 units.
The disk usages are specified in PB. 
Very crudely, the experiments need five times greater resources
than will be available to achieve their full science reach.
An aggressive and coordinated software R\&D program, such as would be
possible with an \s2i2 for HEP, can help mitigate this problem.

The challenges for processor technologies are well known~\cite{GAMEOVER}.
While the number of transistors on integrated circuits doubles
every two years (Moore's Law), power density limitations
and aggregate power limitations 
lead to a situation where ``conventional'' sequential processors are
being replaced
by vectorized and even more 
highly parallel architectures.
To take of advantage of this increasing computing power
demands major
changes to the algorithms implemented in our software.
Understanding how emerging architectures (from low power
processors to parallel architectures like GPUs to more specialized
technologies like FPGAs) will allow HEP
computing to realize the dramatic growth in computing
power required to achieve our science goals
will be a central element of an \s2i2-HEP R\&D effort. 

Similar challenges
exist with storage and network at the scale of HL-LHC~\cite{SMSTORAGE},
with implications for the persistency of data and the 
computing models and the
software supporting them.
Limitations in affordable storage pose a major challenge, 
as does the I/O capacity of ever
larger hard disks. 
While wide area network capacity will probably continue to 
increase at the required
rate, the ability to use it efficiently will need a closer 
integration with applications. 
This will
require developments in software to support 
distributed computing (data and
workload management, software distribution and data access) 
and an increasing awareness of
the extremely hierarchical view of data, from long latency tape 
access and medium-latency
network access through to the CPU memory hierarchy.

The human and social challenges run in parallel with the technical
challenges.
All algorithms and software implementations are developed and maintained
by flesh and blood individuals, many with unique expertise.
What can the community do to
help these people contribute most effectively to the 
larger scientific enterprise?
\begin{itemize}
 \item
  How do we train
  large numbers of novice developers, and smaller numbers of
  more expert developers and architects,
  in appropriate software engineering and 
  software design principles and best practices?
 \item
   How do we foster effective collaboration within software
   development teams and across experiments? 
 \item
  How do we create a culture for designing, developing,
  and deploying sustainable  software?
\end{itemize}
\textcolor{black}{\em Learning how to work together as a coherent community, 
 and engage productively
with the larger scientific software community, will be critical to
the success of the R\&D enterprise preparing for the HL-LHC. An \s2i2
for HEP can play a central role in guaranteeing this success.}

\tempnewpage

\section{Summary of \s2i2-HEP Conceptualization Process}
\label{sec:Conceptualization}

The proposal ``Conceptualization of an \s2i2 Institute for High
Energy Physics (\s2i2-HEP)'' was submitted to the NSF in August 2015. Awards 
ACI-1558216, ACI-1558219, and ACI-1558233
were made in July 2016, 
and the \s2i2 conceptualization project
began in Fall 2016. Two major deliverables were foreseen
from the conceptualization process in the original \s2i2-HEP proposal:
\vskip 0.1in
\noindent (1) A {\bf Community White Paper (CWP)}~\cite{HSF2017} 
describing a global
vision for software and computing for the HL-LHC era;
this includes discussions of elements that are common to
the LHC community as a whole and those that  are specific
to the individual experiments.
It also discusses the relationship of the common elements
to the broader HEP and scientific computing communities.
Many of the topics discussed are relevant
for a HEP \s2i2.
The CWP document has been prepared and written as an initiative of 
the HEP 
Software Foundation (HSF).
As its purview is greater than an \s2i2 Strategic Plan,
it fully engaged the international HL-LHC community, including U.S.\
university and national labs personnel.
In addition, international and U.S.\ personnel associated
with other HEP experiments participated at all stages.
The  CWP  provides a roadmap for software R\&D in preparation for
the HL-LHC and for other HL-LHC era HEP experiments.
The charge from
the Worldwide LHC Computing Grid (WLCG) to the HSF and the LHC
experiments~\cite{WLCGCWPCHARGE}
says it should identify and 
prioritize the software research and development investments required:
\begin{itemize}
 \item
   to achieve improvements in software efficiency, scalability 
   and performance and to make
   use of the advances in CPU, storage and network technologies,
 \item
   to enable new approaches to computing and software that can 
   radically extend the
   physics reach of the detectors,
 \item
  to ensure the long term sustainability of the software through 
  the lifetime of the HL-LHC.
\end{itemize}
The Community White Paper was published with title ``A Roadmap for HEP 
Software and Computing R\&D for the 2020s''~\cite{CWPDOC} on 18 December, 2017.
\vskip 0.1in
\noindent (2) A separate {\bf Strategic Plan} identifying areas where the U.S.\ university
community can provide leadership and discussing
those issues required for an \s2i2
which are not (necessarily) relevant to the larger community. 
This is the
document you are currently reading. 
In large measure, it builds on the findings of the CWP.
In addition, it  addresses the following questions:
 \begin{itemize}
   \item
     where does the U.S.\ university community already have
     expertise and important leadership roles?
   \item
     which software elements and frameworks would provide
     the best educational and training opportunities
     for students and postdoctoral fellows?
   \item
     what types of programs (short courses, short-term
     fellowships, long-term fellowships, etc.)
     might enhance the educational
     reach of an \s2i2?
   \item
     possible organizational, personnel and management
     structures and operational processes; and
    \item
     how the investment in an \s2i2 can be judged
     and how the investment can be sustained
     to assure the scientific goals of the HL-LHC.
 \end{itemize}
The Strategic Plan has been prepared in collaboration with 
members of the  U.S.\ DOE Laboratory community as well as the
U.S.\ university community.  
Although it is not a project deliverable, an additional goal of the
conceptualization process has been to 
engage broadly with computer scientists and software engineers, 
as well as
high energy physicists, to build community interest in
submitting
an \s2i2 implementation proposal, should there be an
appropriate solicitation.

The process to produce these two documents has been built 
around a series of
dedicated workshops, meetings, and special outreach sessions 
in preexisting 
workshops. 
Many of these were organized under the umbrella of the HSF and
involved the full international community. 
A smaller, dedicated set of
workshops focused on \s2i2- or U.S.- specific topics, 
including interaction
with the Computer Science community. Engagement with the
computer science community has been an integral part of the \s2i2
process from the beginning, including two workshops dedicated to
fostering collaboration between HEP and computer scientists, the first
at the University of Illinois and National Center of
Supercomputing Applications in December 2016 (see the workshop report
at \cite{HEPCSWORKSHOPREPORT}) and the second at Princeton University
in May 2017. \s2i2-HEP project Participant Costs
funds were used to support the participation of relevant individuals
in all types of workshops. A complete list of the workshops held as
part of the CWP or to support the \s2i2-specific efforts is included
in Appendix~\ref{workshoplist}.


The community at large was engaged in the CWP and \s2i2 processes by building
on existing communication mechanisms. 
The involvement of the LHC experiments
(including in particular the software and computing coordinators) 
in the CWP
process allowed for communication using the pre-existing 
experiment channels.
To reach out more widely than just to the LHC experiments, 
specific contacts were made
with individuals with software and computing responsibilities in the 
FNAL 
muon and neutrino experiments, Belle-II, the Linear Collider community, 
as
well as various national computing organizations.
The HSF had, in fact, been building up mailing lists and contact people 
beyond LHC for about 2 years before the CWP process began. The CWP
process was able to build on that.

Early in the CWP process, a number of working groups were established on 
topics that were expected to be important parts of the HL-LHC roadmap:
{\it Careers, Staffing and Training; Computing Models, Facilities, and
Distributed Computing; Conditions Database; Data Organization,
Management and Access; Data Analysis and Interpretation; Data and
Software Preservation; Detector Simulation; Event Processing
Frameworks; Machine Learning; Physics Generators; Software Development,
Deployment and Validation/Verification; Software Trigger and Event
Reconstruction;} and {\it Visualization.}


In addition, a small set of working groups envisioned at the
beginning of the CWP process failed to gather significant community
interest or were integrated into the active working groups listed
above. These below-threshold working groups were: {\it Math Libraries; 
Data Acquisition Software; Various Aspects of Technical Evolution 
(Software Tools, Hardware, Networking); Monitoring; 
Security and Access Control;} and 
{\it Workflow and Resource Management.}

The CWP process began with a kick-off workshop at UCSD/SDSC in 
January 2017 
and concluded with a final workshop in June 2017 in Annecy, France. 
A large number of intermediate topical workshops and meetings were held
between these. 
The CWP process involved a total of $\sim260$ participants,
listed in 
Appendix~\ref{workshoplist}. 
The working groups
continued to meet virtually to produce their own white papers
with completion targeted for early fall 2017.
At the CWP kick-off workshop (in January 2017), each of the 
(active) working groups defined a charge for itself, as well as a plan for 
meetings, a Google Group for communication, etc. The precise path for each
working group in terms of teleconference meetings and actual in-person
sessions or workshops varied from group to group. Each of the active
working groups has produced a working group report, which is available from 
the HSF CWP webpage~\cite{HSF2017}.
An overall Community White Paper document, synthesizing the information from 
the individual working group white papers, has also been prepared.
As of 18 December, 2017, most of the working groups have prepared final 
drafts of their documents and a final version of the Community White Paper 
has been published.~\cite{CWPDOC} The CWP working group documents are also being 
published in the arXiv (links pending).

The CWP process was intended to assemble the global roadmap for software
and computing for the HL-LHC. In addition, \s2i2-specific activities
were organized to explore which subset of the global roadmap would be
appropriate for a U.S.\ university-based Software Institute and what
role it would play together with other U.S.\ efforts (including
both DOE efforts, the US-ATLAS and US-CMS Operations programs and the Open
Science Grid) and with international efforts. In addition the
\s2i2-HEP conceptualization project investigated how the U.S.\ HEP
community could better collaborate with and leverage the intellectual
capacity of the U.S. Computer Science and NSF Sustainable Software (SI2)~\cite{SI2} communities. Two dedicated
\s2i2 HEP/CS workshops were held as well as a dedicated \s2i2 workshop,
co-located with the ACAT conference. In addition numerous outreach activities
and discussions took place with the U.S.\ HEP community and specifically 
with PIs interested in software and computing R\&D.

\tempnewpage

\section{The HEP Community}

HEP is a global science. The global nature of the community is both
the context and the source of challenges for an \s2i2. A fundamental
characteristic of this community is its globally distributed knowledge
and workforce. The LHC collaborations each comprise thousands of
scientists from close to 200 institutions across more than 40
countries. The large size of these collaborations is due to the complexity of the
scientific endeavor. No one person or small team understands all aspects of
the experimental program. Knowledge is thus collectively obtained,
held, and sustained over the decades long LHC program. Much of that
knowledge is curated in software. Tens of millions of lines of code
are maintained by many hundreds of physicists and engineers.  Software
sustainability is fundamental to the knowledge sustainability
required for a research program that is expected to last 
well into the 2040s.

\subsection{The HEP Software Ecosystem and Computing Environment}


The HEP software landscape itself is quite varied.  Each HEP
experiment requires, at a minimum, ``application'' software for
data acquisition, data handling, data processing, simulation and
analysis, as well as related application frameworks, data persistence
and libraries.  In addition significant ``infrastructure'' software
that spans all aspects of an experiment
is required. The scale of the computing environment itself drives
some of the complexity and requirements for infrastructure tools.
Over the past 20 years, HEP experiments have became large enough 
to require significantly greater resources than the host laboratory can
provide by itself. Collaborating funding agencies typically provide 
in-kind contributions of computing resources rather than send funding
to the host laboratory. This makes a distributed computing infrastructure 
essential, and 
thus HEP research needs have driven the development of sophisticated software
for data management, data access, and workload/workflow management.

\begin{figure}[htbp]
\begin{center}
\includegraphics[width=0.96\textwidth]{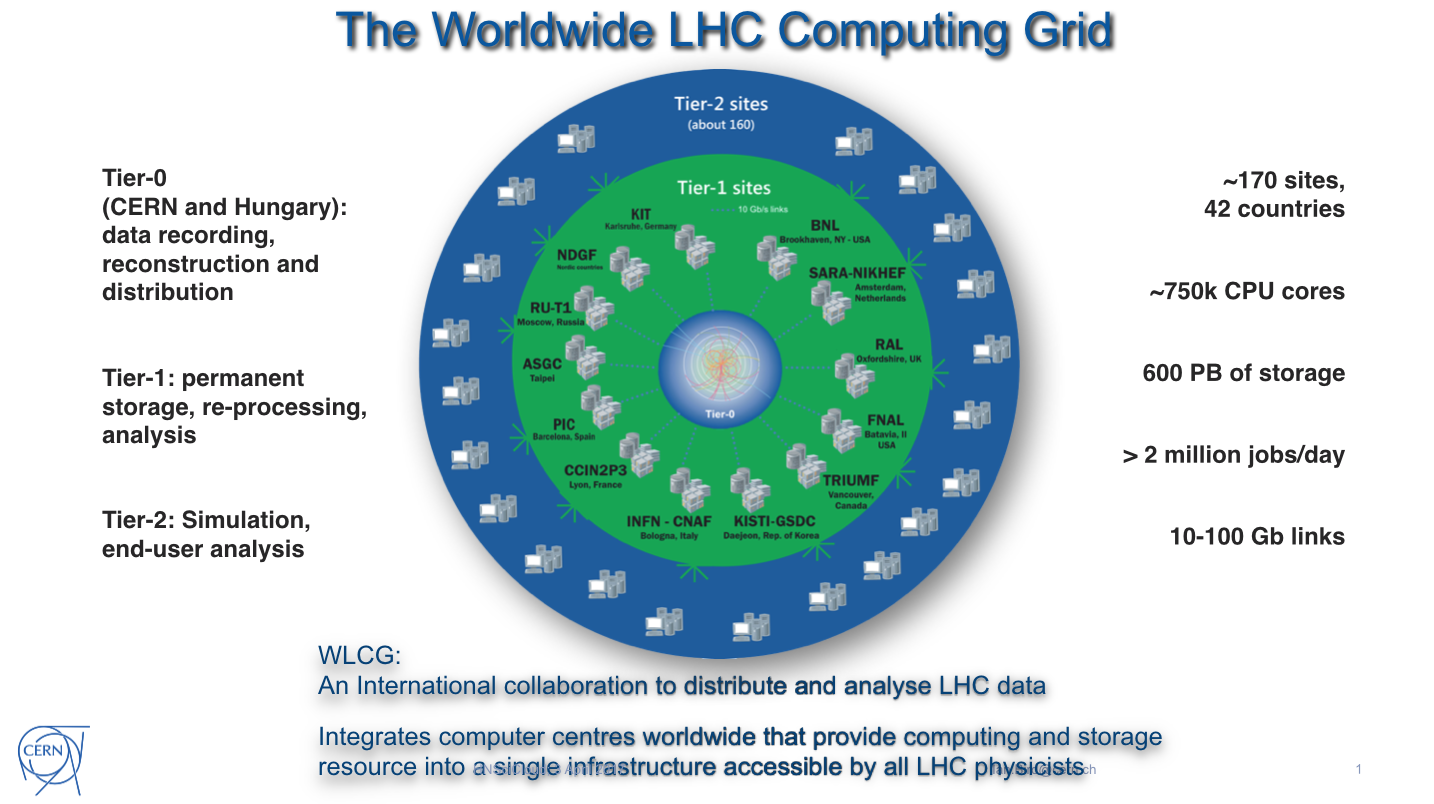}
\caption{The Worldwide LHC Computing Grid (WLCG), which federates national grid infrastructures to provide the computing
resources needed by the four LHC experiments (ALICE, ATLAS, CMS, LHCb). The
numbers shown represent the WLCG resources from 2016.}
\label{fig:wlcg}
\end{center}
\end{figure}

These software elements are in constant use, as computing operations
continues 24 hours a day, 7 days a week throughout the year.
The LHC experiments rely on
$\sim$170 computing centers and national grid infrastructures that
are federated via the Worldwide LHC Computing Grid (shown in
Figure~\ref{fig:wlcg}). The U.S.\ contribution is organized and run
by the Open Science Grid (OSG)~\cite{OSGPAPER,OSGWEB}. The intrinsic
nature of data-intensive collider physics maps very well to the use
of high-throughput computing. The computing use ranges from
``production'' activities that are organized centrally by the
experiment (e.g., basic processing of RAW data from the detector and 
creation of high statistics Monte Carlo simulations) to ``analysis'' activities 
initiated by individuals or small groups of researchers for their specific
research investigations.

\vskip 0.1cm
\noindent{\bf Software Stacks:} In practice much of the actual software and 
infrastructure is implemented {\em independently} by each experiment. This 
includes managing 
the software development and deployment process and the resulting software
stack. Some of this is a natural result of the intrinsic differences
in the actual detectors (scientific instruments) used by each experiment.
Independent software stacks are also the healthy result of
different experiments and groups making independent scientific investigations 
using different algorithmic and 
implementation choices. And last, but not least, each experiment must
have control over its own schedule to insure that it can deliver physics 
results in a competitive environment. This implies sufficient control over 
the software development process and the software itself that the experiment 
uses.

The independence of the software processes in each experiment of course
has some downsides. At times, similar functionalities are implemented
redundantly in multiple experiments, parts as a by-product of the 
physics research program (i.e.\ the result of R\&D by postdocs and graduate 
students). Typically, software is developed and used without with the explicit aim of producing 
sustainable software over the lifetime of an experimental program. Issues of long term software 
sustainability can arise in these cases when the particular functionality is 
not actually mission-critical or specific to the experiment. 
Trivial technical and/or communication issues can prevent
even high quality tools developed in one experiment from being adopted
by another.

The HEP community has nonetheless a developed an ecosystem of common software 
tools that are widely shared in the community. Ideas and experience
with software and computing in the HEP community are shared at 
general dedicated HEP software/computing conferences such as 
the Conference on Computing in High Energy and Nuclear Physics (CHEP)~\cite{CHEP2016},
the Workshop on Advanced Computing and Analysis Techniques (ACAT)~\cite{ACAT2017} and HEPiX~\cite{HEPiX}. 
In addition there are many 
specialized workshops on software and techniques for pattern recognition, 
simulation, data acquisition, use of machine learning, and other 
topics.

An important exception to the organization of software stacks 
by the experiments is the national grid infrastructures, such as the OSG
in the U.S. The federation of computing resources from
separate computing centers which at times support more than one HEP experiment 
or that support HEP and other scientific domains requires and creates
incentives that drive the development and deployment of ``common'' 
solutions.



\vskip 0.1cm
\noindent{\bf Examples of Shared Application Software Toolkits:}
The preparations for the LHC have nonetheless yielded important community 
software tools for data analysis like ROOT~\cite{root} and detector simulation
Geant4~\cite{Geant4Ref1,Geant4Ref2,Geant4Ref3}, both of which have been
critical not only for LHC but in most other areas of HEP and beyond.
Other tools have been shared between some, but not all, experiments.
Examples include
the GAUDI~\cite{GAUDI} event processing framework,
IgProf~\cite{IGPROF} for profiling very large C++ applications like those used
in HEP, 
RooFit~\cite{Verkerke:2003ir} for data modeling and fitting and
the TMVA~\cite{Hocker:2007ht} toolkit for multivariate data analysis.

In addition software is a critical tool for the interaction and knowledge
transfer between experimentalists and theorists. 
Software provide an important physics input from the theory community to 
the LHC experimental program, for example through event 
generators such
as SHERPA~\cite{SHERPA} and ALPGEN~\cite{ALPGEN} and through jet finding 
tools like FastJet~\cite{FASTJET1,FASTJET2}, which is a critical piece of 
software for the LHC experiments RAW data processing applications.


\vskip 0.1cm
\noindent{\bf Infrastructure Software Examples:} 
As noted above, the need for reliable ``infrastructure'' tools which must 
be deployed as services in multiple computer centers creates incentives for the 
development of common tools which can be used by multiple HEP experiments,
and often by other scientific applications.
Examples include
FRONTIER~\cite{FRONTIER} for cached access to databases,
XROOTD~\cite{XROOTD} and dCache~\cite{DCACHE} for distributed access to
bulk file data, 
Rucio~\cite{RUCIO} for distributed data management,
EOS~\cite{EOS1,EOS2} for distributed disk storage cluster 
management,
FTS~\cite{FTS} for data movement across the distributed computing system,
CERNVM-FS~\cite{CERNVMFS} for distributed and cached access to software,
GlideinWMS~\cite{GLIDEINWMS} and PanDA~\cite{PANDA1,PANDA2} for
workload management. 
Although not developed specifically for HEP, HEP has been an important
domain-side partner in the development of tools such as 
HTCondor~\cite{condor-practice} for distributed high throughput computing 
and the Parrot~\cite{PARROT} virtual file system.

Global scientific collaborations need to meet and discuss, and this has 
driven the development of the scalable event organization software 
Indico~\cite{INDICO1,INDICO2}.  Various tools have been developed by
the HEP community to support information exchange and preservation across
the experimental and theoretical communities. Examples include: the
preservation of experimental data samples, analysis results, 
and software developed by experiments (e.g., RECAST~\cite{Cranmer2011} and REANA~\cite{reanaweb}); 
information discovery for HEP papers, authors, and collaborations (e.g., INSPIRE~\cite{inspireweb}); 
and to facilitating technical collaborations (e.g, SWAN analysis service based on 
notebook technologies~\cite{swan}).

In a similar way, the CS and HEP communities have collaborated on the
broader problems of data and software
preservation such as the DASPOS Project~\cite{daspos-project} in
scientific computing, which has led to case studies and software
prototypes applied to LHC software as well as service prototypes such
as the CERN data analysis portal.

\subsection{Software Development and Processes in the HEP Community}
\label{sec:hepsw}
The HEP community has by necessity developed significant experience in 
creating software infrastructure and processes that integrate 
contributions from large, distributed communities of physics researchers. 
To build its software ecosystem, each of the major 
HEP experiments provides a set of
``software architectures and lifecycle processes, development,
testing and deployment methodologies, validation and verification
processes, end usability and interface considerations, and required
infrastructure and technologies'' (to quote the NSF \s2i2 solicitation~\cite{NSF15553}).
Computing hardware to support the development process for the application
software (such as continuous integration, development, and test machines) 
is typically provided by the host laboratory of each experiment, 
e.g., CERN for the LHC experiments. Each experiment
manages software release cycles for its own unique application software code
base, and works to update software elements, such has shared software toolkits,
 that are integrated into its software stack. Release cycles are organized 
to meet goals ranging from physics needs, to bug and performance fixes. The 
software development infrastructure is also designed to allow 
individuals to write, test and contribute software from any laboratory, university,
or personal laptop.  The software development and testing support for 
the ``infrastructure'' part of the software ecosystem, supporting the 
distributed computing environment, is more diverse and not centralized at 
CERN. It relies much more heavily on resources such as the Tier-2 centers 
and the OSG. Given the non-uniformity in computing site infrastructures, 
the integration and testing process for computing infrastructure software elements 
is complex than that for application software. Nevertheless, the 
full set of processes has also been put in place by each LHC experiment, and continues to evolve
as the software stacks change.

\begin{figure}[ht]
\begin{center}
\includegraphics[width=0.96\textwidth]{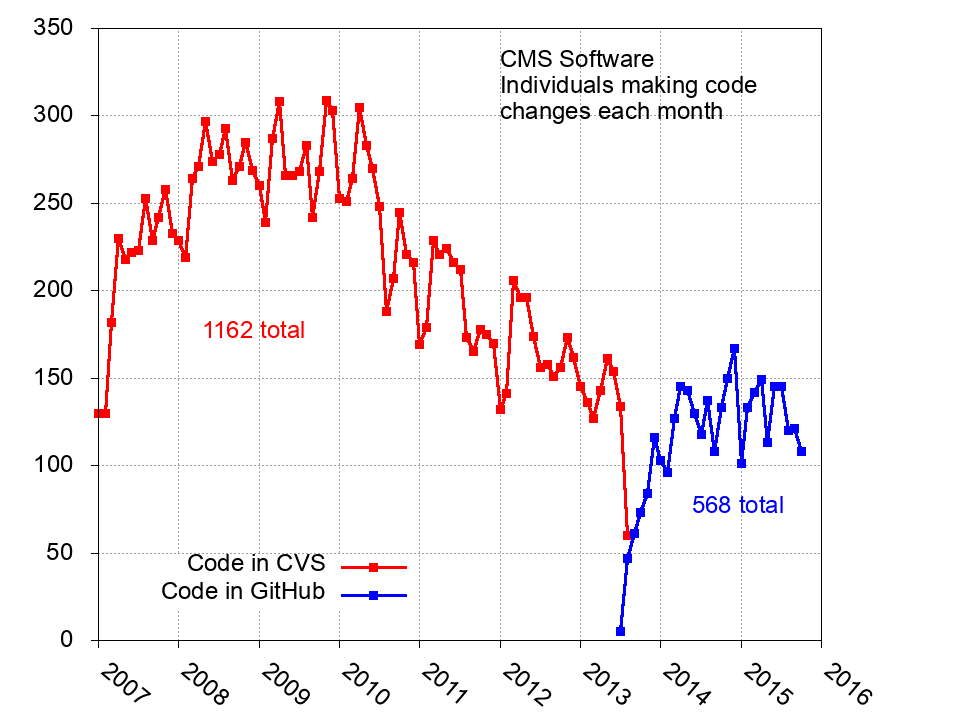}
\caption{Evolution of the number of individuals making contributions to the CMS application software release each month over the period from 2007 to 2016. Also shown is how the
developer community was maintained through large changes to the technical
infrastructure, in this case the evolution of the version control system from 
CVS hosted at CERN to git hosted in GitHub. This plot shows only the application software managed in the experiment-wide software release (CMSSW) and not ``infrastructure'' software (e.g., for data and workflow management) or ``analysis'' software developed by individuals or small groups.
\label{fig:cmsswdevelopers}}
\end{center}
\end{figure}

For the most part, the HEP community has not 
formally adopted any explicit development methodology or model, however the 
de-facto method adopted is very similar to agile software 
development~\cite{AGILE}.
On slightly longer time scales, the software development
efforts within the experiments must respond to various
challenges including evolving physics goals and discoveries, general
infrastructure and technology evolution, as well as the evolution of the
experiments themselves (detector upgrades, accelerator energy, and
luminosity increases, etc.). 
HEP experiments have also maintained these software infrastructures
over time scales ranging from years to decades and in projects
involving hundreds to thousands of developers. 
Figure~\ref{fig:cmsswdevelopers} shows the example
of the application software release (CMSSW) of CMS experiment at the LHC. 
Over a ten year period, up to
300 people were involved in making changes to the software each month.  
The software process shown in the figure results in the integration, testing 
and deployment of tens of releases per year on the global computing 
infrastructure. 
The figure also shows an example of the evolution in the
technical infrastructure, in which the code version control system was
changed from CVS (hosted at CERN) to git (hosted on GitHub~\cite{cmsswgithub}). 
Similar software processes are also in routine
use to develop, integrate, test and deploy the computing infrastructure 
elements in the software ecosystem which support distributed data management
and high throughput computing.

In this section, we described ways in which HEP community develops
its software and manages its computing environment to produce physics
results. In the next section (Section~\ref{sec:role}), we present the
role of the Institute to facilitate a successful HL-LHC physics program
through targeted software development and leadership, more generally,
within the HEP software ecosystem.

\tempnewpage

\section{The Institute Role}
\label{sec:role}

\subsection{Institute Role within the HEP Community}
\label{sec:s2i2role-hep}
\begin{figure}[htbp]
\begin{center}
\includegraphics[width=0.99\textwidth]{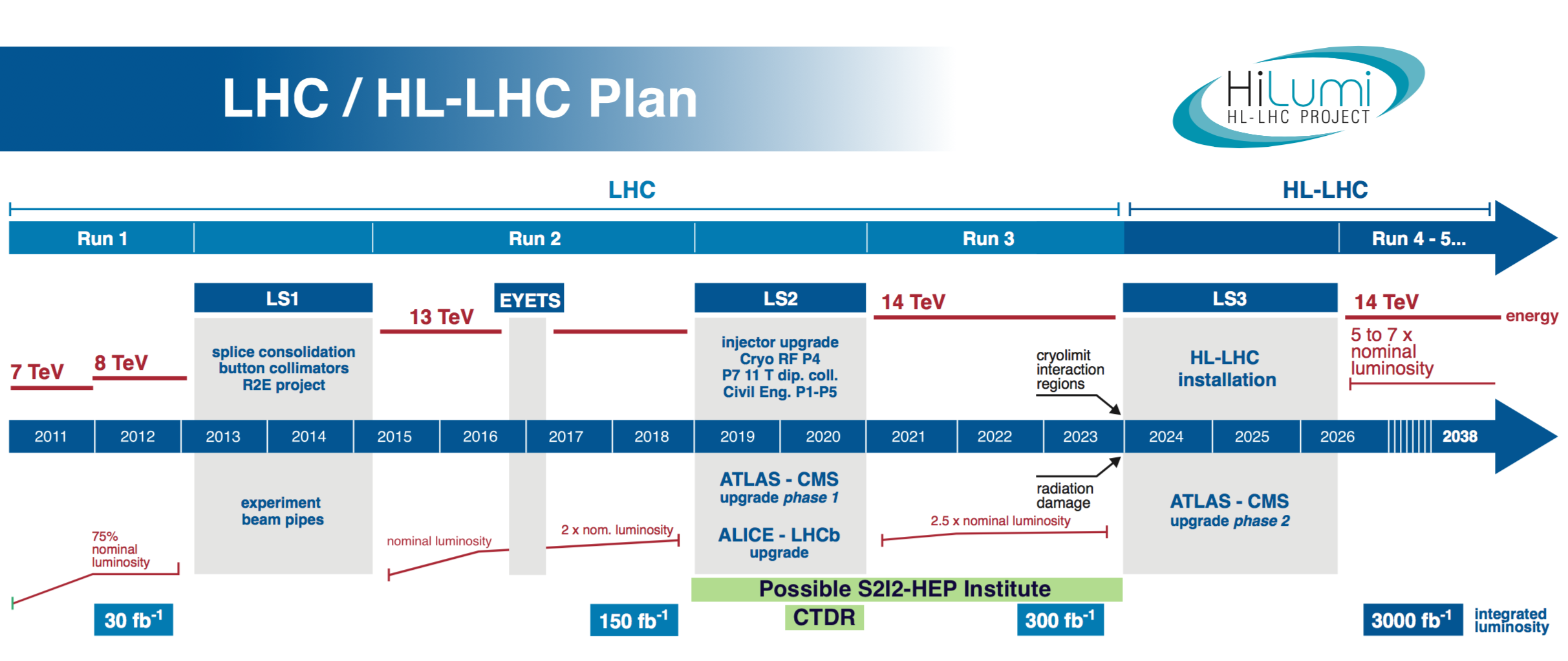}
\caption{Timeline for the LHC and HL-LHC~\cite{HLLHCTIMELINE}, indicating 
both data-taking periods and ``shutdown'' periods which are used for upgrades 
of the accelerator and detectors. Data-taking periods are 
indicated by the (lower) red lines showing the relative luminosity and (upper)
red lines showing the center of mass energy. Shutdowns with no data-taking are 
indicated by blue boxes (LS = Long Shutdown, EYETS = Extended Year End Technical
Stop). The approximate periods of execution for an \s2i2 for HEP and the writing and delivery of the CTDRs are shown in green.}
\label{fig:hllhctimeline}
\end{center}
\end{figure}

The mission of a Scientific Software Innovation Institute
(\s2i2) for HL-LHC physics should be to serve as both an
active software research and development center 
and as an intellectual hub for the larger R\&D effort required
to ensure the success of the HL-LHC scientific program.  The timeline for
the LHC and HL-LHC is shown in Figure~\ref{fig:hllhctimeline}.  A
Software Institute operating roughly in the 5 year period from 2019
to 2023 (inclusive) will coincide with two important steps in the
ramp up to the HL-LHC: the delivery of the Computing Technical Design 
Reports (CTDRs) of ATLAS and CMS in $\sim$2020 and LHC Run 3 in 2021-2023.  
The CTDRs will
describe the experiments' technical blueprints for building software
and computing to maximize the HL-LHC physics reach, given the
financial constraints defined by the funding agencies.  For ATLAS
and CMS, the increased size of the Run 3 data sets relative to Run
2 will not be a major challenge, and changes to the detectors will be
modest compared to the upgrades anticipated for Run 4.  As a
result, ATLAS and CMS will have an opportunity to deploy prototype
elements of the HL-LHC computing model during Run 3 as real road
tests, even if not at full scale.  In contrast, LHCb is making its
major transition in terms of how much data will be processed at the
onset of Run 3.  Some Institute deliverables will be deployed at
full scale to directly maximize LHCb physics and provide valuable
experience the larger experiments can use to prepare for the HL-LHC. 

The Institute will exist within a larger context of international
and national projects that are required for software and computing
to successfully enable science at the LHC, both today, and in the
future. Most importantly at the national level, this includes the U.S.\ LHC 
``Operations Programs'' jointly funded by DOE and NSF, as well
as the Open Science Grid project. In the present section we focus
on the role of the Institute while its relationships to these
national and international partners are elaborated on in
Section~\ref{sec:partnerships}.

The Institute's mission
will be
realized by building a more cooperative, community process
for developing, prototyping, and deploying software.
The Institute itself should be greater than the sum of its
parts, and the larger community efforts it engenders should
produce better and more sustainable software than would be possible
otherwise. Consistent with this mission, the role of the Institute
within the HEP community will be to:
\begin{enumerate}
\item 
  drive the software R\&D process in 
  specific focus areas 
  using its own resources directly,
  and also leveraging them through collaborative efforts 
  (see Section~\ref{sec:focusareas}).
\item
  work closely with the LHC experiments, 
  their U.S.\ Operations Programs, the
  relevant national laboratories, and the greater HEP community
  to identify the highest priority software and computing issues
  and then create collaborative mechanisms to address them.
\item 
  serve as an intellectual hub for the 
  larger community effort
  in HEP software and computing. 
  For example, it will bring together a critical mass of
  experts from HEP, other domain sciences, academic computer science,
  and the private sector to
  advise the HEP community
  on sustainable software development. 
  Similarly, the Institute will serve as a center for
  disseminating knowledge related to the current software and computing
  landscape, emerging technologies, and tools.
  It will  provide critical
  evaluation of new proposed software elements for algorithm essence
  (e.g. to avoid redundant efforts), feasibility and sustainability,
  and provide recommendations to collaborations (both experiment and
  theory) on training, workforce, and software development.
\item 
  deliver value
  through its (a) contributions to 
  the development of the CTDRs for ATLAS and CMS and 
  (b) research, development and deployment of software that is used for physics during Run 3.
\end{enumerate}



\subsection{Institute Role in the Software Lifecycle}
\label{sec:s2i2role-sw}

\begin{figure}[htbp]
\begin{center}
\includegraphics[width=0.55\textwidth]{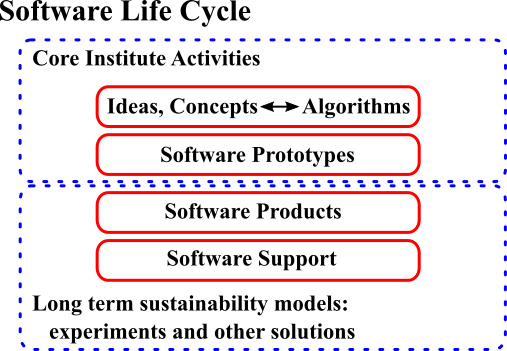}
\caption{Roles of the Institute in the Software Life Cycle}
\label{fig:swprocess}
\end{center}
\end{figure}

Figure~\ref{fig:swprocess} shows the elements of the software life cycle,
from development of {\it core concepts and algorithms}, through 
{\it prototypes} to 
deployment of {\it software products} and {\it long term support}. The 
community vision 
for the Institute is that it will focus its resources on developing innovative 
ideas and concepts through the prototype stage and along the path to become 
software products used by the wider community. It will partner with the
experiments, the U.S.\ LHC Operations Programs and others to transition software 
from the prototype stage to the software product stage. 
As described in Section~\ref{sec:hepsw} the experiments already
provide full integration, testing deployment and lifecycle processes.
The Institute will not duplicate these, but instead
collaborate with the experiments and Operations Programs on the efforts
required for software integration activities and activities associated to 
initial deployments of new software products. This may also include the 
phasing out of older software elements, the transition of existing systems 
to new modes of working and the consolidation of existing redundant software
elements. 

The Institute will have a finite lifetime of 5 years (perhaps extensible in 
a 2nd phase to 10 years), but this is still much shorter than the planned 
lifetime of HL-LHC activities. 
The Institute will thus also provide technical support to the experiments and 
others to identify sustainability and support models for the software products 
developed. It may at times provide technical support for driving transitions in
the HEP software ecosystem which enhance sustainability.
In its role as an intellectual hub for HEP software innovation, it will
provide advice and guidance broadly on software development within
the HEP ecosystem. For example, a new idea or direction under
consideration by an experiment could be critically evaluated by the
Institute in terms of its essence, novelty, sustainability and
impact which would then provide written recommendations for the
proposed activity. This will be achieved through having a critical mass
of experts in scientific software development inside and outside
of HEP and the computer science community who partner with the
Institute.

%

%
%

\subsection{Institute Elements}\label{institute_elements}

\begin{figure}[ht]
\begin{center}
\includegraphics[width=0.9\textwidth]{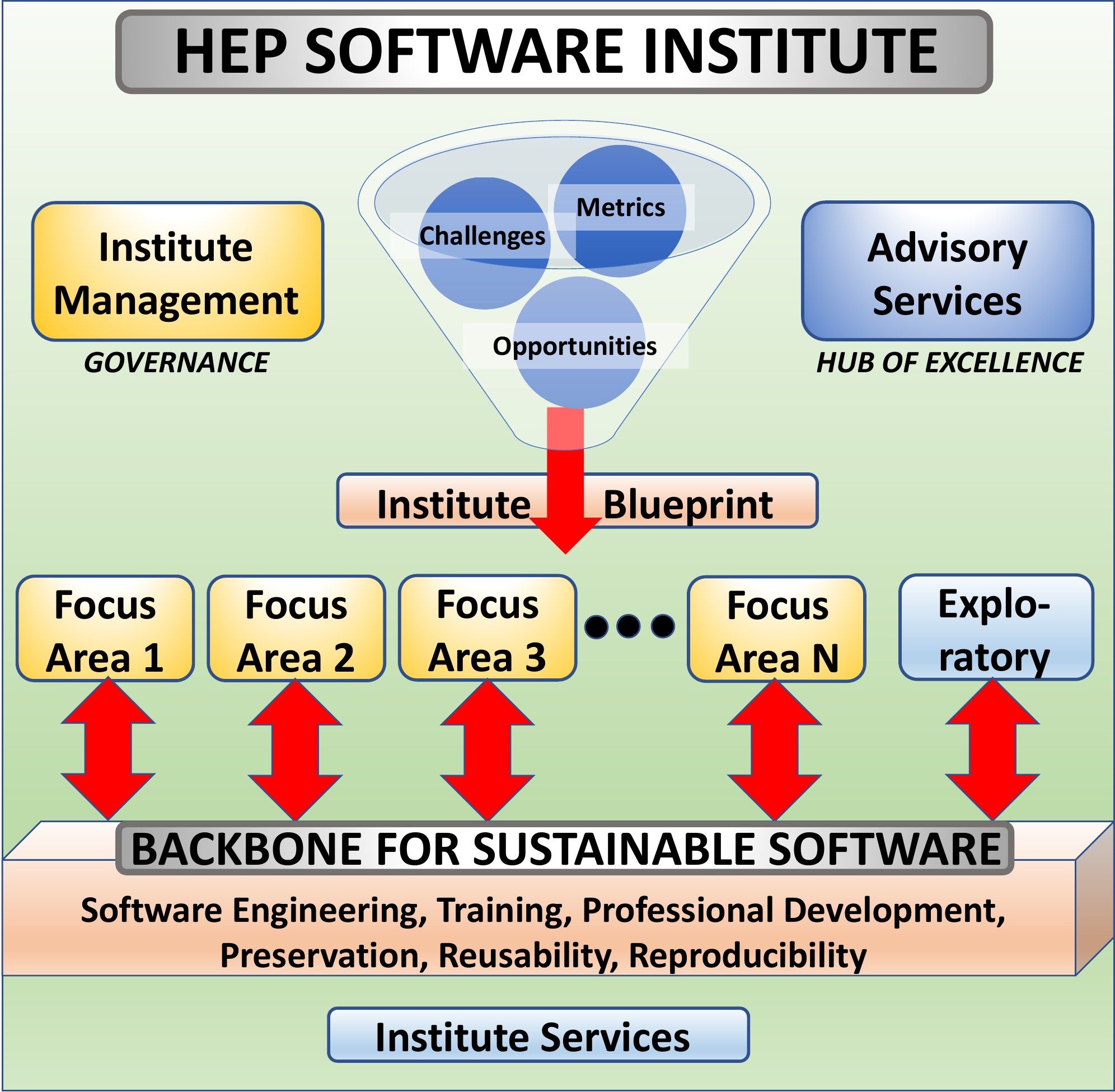}
\caption{Internal elements of the Institute.}
\label{fig:s2i2_elements}
\end{center}
\end{figure}

The Institute will have a number of internal functional elements, as shown 
in Figure~\ref{fig:s2i2_elements}. (External interactions of the institute
will be described in Section~\ref{sec:partnerships}.)

\vskip 0.1in
\noindent{\bf Institute Management:} In order to accomplish its mission, the institute will have a well-defined internal management structure, as well as external governance and advisory structures. Further information on this aspect is provided in Section~\ref{sec:orggov}.
\vskip 0.1in
\noindent{\bf Focus Areas:} The Institute will have $N$ focus areas, which
will pursue the main R\&D goals being pursued by the Institute. High priority 
candidates for these focus areas are described in Section~\ref{sec:focusareas}. How many of these will be implemented in an Institute implementation will 
depend on available funding, as described in Section~\ref{sec:funding}. Each 
focus area will have its own specific plan of work and metrics for evaluation.
\vskip 0.1in
\noindent{\bf Institute Blueprint:} The Institute Blueprint activity 
will maintain the software vision for the Institute and, 3-4 times per 
year, will bring together expertise to answer specific key questions within the 
scope of the Institute's activities or, as needed, within the wider scope 
of HEP software/computing. 
Blueprint activities will be an essential element to build a common vision with other
HEP and HL-LHC R\&D efforts, as described in Section~\ref{sec:partnerships}.
The blueprints will then inform the evolution of both the Institute 
activities and the overall community HL-LHC R\&D objectives in the 
medium and long term.

\vskip 0.1in
\noindent{\bf Exploratory:} From time to time the Institute may deploy modest resources for short term exploratory R\&D projects of relevance to inform the planning and overall mission of the Institute.
\vskip 0.1in
\noindent{\bf Backbone for Sustainable Software:} In addition to the specific
technical advances which will be enabled by the Institute, a dedicated 
``backbone'' activity will focus on how these activities are communicated
to students and researchers, identifying best practices and possible 
incentives, developing and providing training and making data and tools
available to the public. Further information on this activity is included
in Section~\ref{sec:backbone}.
\vskip 0.1in
\noindent{\bf Advisory Services:} The Institute will play a role
in the larger research software community (in HEP and beyond) by
being available to provide technical and planning advice to other
projects and by participating in reviews. The Institute will execute
this functionality both with individuals directly employed by the
Institute and by involving others through its network of partnerships.
\vskip 0.1in
\noindent{\bf Institute Services:} The Institute may
provide other services in support of its software R\&D activities.
Possible examples include access to build platforms
and continuous integration systems; software stack build and packaging
services; technology evaluation services; performance benchmarking services;
access to computing resources and related services required for
testing of prototypes at scale in the distributed computing
environment. 
In most cases, the actual services will not be owned
by the Institute, but instead by one its many partners.
The role of the Institute in this case will be to guarantee 
and coordinate access to 
the services in support of its mission.

%
%
%

\tempnewpage

\section{Strategic Areas for Initial Investment}
\label{sec:focusareas}
A university-based \s2i2 focused on the software 
required to ensure the scientific success of the
HL-LHC will be part of a larger research, development, and deployment
community. 
It will directly fund and lead some of the R\&D efforts;
it will support related deployment efforts by the experiments;
and
it will serve as an intellectual hub for
more diverse efforts.
The process leading to the CWP, discussed in 
Section \ref{sec:Conceptualization},
identified  three {\em impact criteria}
for judging the value of additional investments, regardless of who
makes the investments:
\begin{itemize}
\item {\bf Impact - Physics:} Will efforts in this area enable new approaches to
computing and software that maximize, and potentially radically extend, 
the physics reach of the detectors?
\item {\bf Impact - Resources/Cost:} Will efforts in this area lead to
improvements in software efficiency, scalability and performance and 
make use of the advances in CPU, storage and network technologies, that 
allow the experiments to maximize their physics reach 
within their computing budgets?
\item {\bf Impact - Sustainability:} Will efforts in this area 
significantly improve
the long term sustainability of the software through the 
lifetime of the HL-LHC?
\end{itemize}
These are key questions for HL-LHC software R\&D
projects funded by any 
mechanism,  especially an \s2i2.
During the CWP process, Working Groups (WGs) formed to consider potential
activities in areas spanning the HL-LHC software community:
\begin{itemize}
\item Careers, Staffing and Training
\item Conditions Database
\item Computing Models, Facilities and Distributed Computing
\item Data Access, Organization and Management
\item Data Analysis and Interpretation
\item Data and Software Preservation
\item Detector Simulation
\item Event Processing Frameworks
\item Machine Learning
\item Physics Generators
\item Software Development, Deployment and Validation/Verification
\item Software Trigger and Event Reconstruction
\item Visualization
\item Workflow and Resource Management
\end{itemize}
Each WG was asked to prepare a section of the CWP including the research and development 
topics identified in a roadmap for software and computing R\&D in HEP for the 2020s, and to
evaluate these activities in terms of the impact criteria.

\subsection{Rationale for choices and prioritization of a university-based \s2i2}
\label{sec:focusareas-rationale}

\noindent
The \s2i2 will not be able to solve all of the challenging
software problems for the HL-LHC, and it should not take 
responsibility for deploying and sustaining experiment-specific
software.
It should, instead, focus its efforts in targeted areas where R\&D will have a
high impact on the HL-LHC program. The \s2i2 needs to align its
activities with the expertise of the U.S.\ university program and with the rest of the community.
In addition to identifying areas in which it will lead efforts,
the Institute should clearly identify areas in which it will
not.
These will include some where it will have no significant role
at all,
and others where it might participate with lower priority.

The \s2i2 process was largely community-driven. During this process,
{\em additional \s2i2-specific criteria} were 
developed for identifying Focus Areas for the Institute
and specific initial R\&D topics within each:
\begin{itemize}
\item {\bf Interest/Expertise:} Does the U.S.\ university community have 
      strong interest and expertise in the area?
\item {\bf Leadership:} Are the proposed focus areas complementary to 
      efforts funded by the US-LHC Operations programs, the DOE,
      and international partners?
\item {\bf Value:} Is there potential to provide value to more than one 
    HL-LHC experiment and to the wider HEP community?
\item {\bf Research/Innovation:} Are there opportunities for combining 
       research and innovation as part of partnerships between the HEP and 
       Computer Science/Software Engineering/Data Science communities?
\end{itemize}
At the end of the \s2i2 process, there was a general consensus 
that highest priority Focus Areas where an \s2i2 can 
play a leading role are:
\begin{itemize}
\item {\bf Data Analysis Systems:} Modernize and evolve tools and techniques for analysis of high-energy physics data sets. 
Potential focus areas include adoption of data science tools and approaches, development of analysis systems,
analysis resource management, analysis preservation, and visualization for data analytics. 
\item {\bf Reconstruction Algorithms and Software Triggering:} Develop algorithms able to exploit next-generation detector
technologies and next-generation computing platforms and programming techniques. Potential focus areas include algorithms
for new computing architectures, modernized programming techniques, real-time analysis techniques, and
anomaly detection techniques and other approaches that target 
high precision reconstruction and identification techniques 
enabled by new experimental apparatus and larger data rates.
\item {\bf Applications of Machine Learning :} Exploit Machine Learning approaches to improve the physics reach of HEP data sets. 
Potential focus areas include track and vertex reconstruction, raw data compression, parameterized simulation methods,
and data visualization. 
\item {\bf Data Organization, Management and Access (DOMA):} Modernize the way HEP organizes, manages, and accesses its data. 
Potential focus areas include approaches to data persistence, caching, federated data centers,
and interactions with networking resources.
\end{itemize}
Two additional potential Focus Areas were identified as medium priority
for an \s2i2:
\begin{itemize}
\item Production Workflow, Workload and Resource Management
\item Event Visualization techniques, primarily focusing on collaborative and immersive event displays
\end{itemize}
Production workflow as well as workload and resource management
are absolutely critical software elements for the success of the
HL-LHC that will require sustained investment to keep up with the
increasing demands. However, the existing operations programs plus other DOE-funded
projects are leading the efforts in these areas.
One topic in this area where an \s2i2 may collaborate extensively
is workflows for compute-intensive analysis. Within the \s2i2, this can
be addressed as part of the Data Analysis Systems focus area.
Similarly, there are likely places where the \s2i2 will collaborate with
the visualization community. Specifically, visualization techniques for data analytics 
and ML analytics can be addressed as part of
Data Analysis Systems and ML Applications, respectively.

Although software R\&D efforts in each of the following areas will
be critical for the success of the HL-LHC, 
there was a general consensus that other entities are leading the
efforts, 
and these areas should be low priority for \s2i2
efforts and resources:
\begin{itemize}
\item Conditions Database
\item Event Processing Frameworks
\item Data Acquisition Software
\item General Detector Simulation
\item Physics Generators
\item Network Technology
\end{itemize}
As is evident from our decision to include elements of production
workflow and visualization into higher priority focus areas,
the definitions of focus areas are intentionally fluid.
In addition, some of the proposed activities intentionally
cross nominal boundaries.


\subsection{Data Analysis Systems}

At the heart of experimental HEP is the development of facilities
(e.g.\ particle colliders, underground laboratories) and instrumentation
(e.g.\ detectors) that provide sensitivity to new phenomena. The
analysis and interpretation of data from sophisticated detectors
enables HEP to understand the universe at its most fundamental
level, including the constituents of matter and their interactions,
and the nature of space and time itself.  The final stages of
data analysis are undertaken by small groups, or individual researchers.
The baseline analysis model utilizes successive stages of data
reduction, finally reaching a compact dataset for quick real-time
iterations. This approach aims at exploiting the maximum possible
scientific potential of the data, whilst minimising the 
``time-to-insight'' for a large number of different analyses performed in
parallel. Optimizing analysis systems is a complicated combination of 
diverse constraints, ranging from the need to make efficient use of
computing resources to navigating through specific policies of
experimental collaborations. Any analysis system has to be flexible
enough to handle bursts of activity driven primarily by conference
schedules. Future analysis models must also be nimble enough to adapt 
to new opportunities for discovery (intriguing hints in the data or
new experimental signatures), massive increases in data volume by the
experiments, and potentially significantly more complex analyses,
while still retaining this essential ``time-to-insight''
optimization.

\subsubsection{Challenges and Opportunities}

Over the past 20 years the HEP community has developed and primarily
utilized an analysis ecosystem centered on ROOT~\cite{bib:root}. This
software  ecosystem currently both dominates HEP analysis and impacts
the full event processing chain, providing the core libraries, I/O
services, and analysis tools. This approach has certain advantages for
the HEP community as compared with other scientific disciplines. It
provides an integrated and validated toolkit, lowering the barrier
for analysis productivity and enabling the community to speak in a
common analysis language. It also facilitates improvements and
additions to the toolkit being made available quickly to the community
and therefore benefiting a large number of analyses. More recently,
open source tools for analysis have become widely available from
industry and data science. This newer ecosystem includes data analysis
platforms, machine learning tools, and efficient data storage
protocols. In many cases, these tools are evolving very quickly and
surpass the HEP efforts both in total investment in analysis software
development and the size of communities that use and maintain these
tools.

The maintenance and sustainability of the current analysis ecosystem
is a challenge. The ecosystem supports a number of use cases and
integrates and maintains a wide variety of components. Support for
these components has to be prioritized to fit within the available 
effort, which is provided by a few institutions and not very
distributed across the community. Legacy and less used parts of the
ecosystem are hard to retire and their continued support strains the
available effort. The emergence and abundance of alternative and new
analysis components and techniques coming from industry open source
projects is also a challenge for the HEP analysis software
ecosystem. The community is very interested in using these new
techniques and technologies. This leads to additional support needs in
order to use the new technologies together with established components
of the ecosystem and also be able to interchange old components with
new open source components. 

Reproducibility is the cornerstone of scientific results. It is currently 
difficult to repeat most HEP analyses in the same the manner they were 
originally performed. This difficulty mainly arises due to the number of 
scientists involved, the large number of steps in a typical HEP analysis 
workflow, and the complexity of the analyses themselves. A challenge 
specific to data analysis and interpretation is tracking the evolution 
of relationships between all the different components of an
analysis. Better methods for the preservation of analysis workflows
and reuse of analysis software and data products would improve the
quality of HEP physics results and reduce the time-to-insight because
it would be easier for analyses to progress through increases in data
volume and changes in analyses personnel. 

Robust methods for data reinterpretation are also critical. Collaborations 
typically interpret results in the context of specific models for new
physics searches and sometimes reinterpret those same searches in the
context of alternative theories. However, understanding the full
implications of these searches requires the interpretation of the
experimental results in the context of many more theoretical models
than are currently explored at the time of publication. Analysis
reproducibility and reinterpretation strategies need to be considered
in all new approaches under investigation, so that they become a
fundamental component of the system as a whole.

\subsubsection{Current Approaches}

Methods for analyzing the data at the LHC experiments have been
developed over the years and successfully applied to LHC data to
produce physics results during Run 1 and Run 2. The amount of data
typically used by a LHC Run 2 data analysis at the LHC (hundreds of TB
or PBs) is far too large to be delivered locally to the user. The
baseline analysis model utilizes successive stages of data reduction,
finally analyzing a compact dataset with quick real time
iteration. Experiments and their analysts use a series of processing
steps to reduce large input datasets down to sizes suitable for
laptop-scale analysis. The line between managed production-like
analysis processing and individual analysis, as well as the balance
between harmonized vs.\ individualized analysis data formats differs
by experiment, based on their needs and optimization level and the
maturity of an experiment in its life cycle. 

An evolution of this baseline approach is to produce physics-ready
data right from the output of the high-level trigger of the
experiment, avoiding the need for any further processing of the data
with updated or new software algorithms or detector conditions. The
online calibrations are not of sufficient quality to yet enable this
approach for all types of analysis, however this approach is now in
use across all of the LHC experiments, and will be the primary method
used by LHCb in Run 3.  Referred to as ``real-time analysis'', this
technique could be a key enabler of a simplified analysis model that
allows simple stripping of data and very efficient data reduction.

The technologies to enable both analysis reproducibility and analysis
reinterpretation are evolving quickly. Both require preserving the
data and software used for an analysis in some form. This ``analysis
capture'' is best performed while the analysis is being developed, or
at least before it has been published. Recent progress using workflow
systems and containerization technology have rapidly transformed this
area to provide robust solutions to help analysts adopt techniques
that enable reproducibility and reinterpretation of their work.

The LHC collaborations are pursuing a vast number of searches for new
physics. Interpretation of these analyses sits at the heart of the LHC
physics priorities, and aligns with using the Higgs as a tool for
discovery, identify the new physics of dark matter, and explore the
unknown of new particles, interactions, and physical principles. The
collaborations typically interpret these results in the context of
specific models for new physics searches and sometimes reinterpret
those same searches in the context of alternative theories. However,
understanding the full implications of these searches requires the
interpretation of the experimental results in the context of many more
theoretical models than are currently explored by the
experiments. This is a very active field, with close theory-experiment
interaction and with several public tools in development.

For example, a forum~\cite{LHCInterpretationForum} on the
interpretation of the LHC results for Beyond Standard Model (BSM)
studies was initiated to discuss topics related to the BSM
(re)interpretation of LHC data, including the development of the
necessary public recasting tools~\cite{Cranmer2011} and related
infrastructure, and to provide a platform for a continued interaction
between the theorists and the experiments.

The infrastructure needed for analysis reinterpretation is a focal
point of other cyber infrastructure components including the INSPIRE
literature database~\cite{inspireweb}, the HEPData data
repository~\cite{hepdata,hepdataweb}, the CERN Analysis Preservation
framework~\cite{lhcopendata,cern-data-preservation-web}, and the REANA
cloud-based workflow execution system~\cite{reanaweb}. Critically,
this cyber infrastructure sits at the interface between the
theoretical community and various experimental collaborations. As a
result, this type of infrastructure is not funded through the
experiments and tends to fall through the cracks. Thus, it is the
perfect topic for a community-wide, cross-collaboration effort.

\subsubsection{Research and Development Roadmap and Goals}

The goal for future analysis models is to reduce the ``time-to-insight''
while exploiting the maximum possible scientific potential of the data
within the constraints of computing and human resources. Analysis
models aim to give scientists access to the data in the most 
interactive and reproducible way possible, to enable quick turn-around 
in iteratively learning new insights from the data.

Many analyses have common deadlines defined by conference schedules
and the availability of physics-quality data samples. The increased
analysis activity before these deadlines require the analysis system
to be sufficiently elastic to guarantee a rich physics harvest. Models
must evolve to take advantage of new computing hardware such as GPUs
and new memory as they emerge to reduce the `time-to-insight'
further.

\vskip 0.1in\noindent
\textbf{\textit{Diversification of the Analysis Ecosystem}}. 
ROOT and its ecosystem currently dominate HEP analysis and impact the
full event processing chain in HEP, providing foundation libraries,
I/O services, etc. The analysis tools landscape is now evolving in
ways that will influence on the analysis and core software landscape
for HL-LHC.
\begin{itemize}
\item
Data-intensive analysis is growing in importance in other science
domains as well as the wider world. Powerful tools from data science
and new development initiatives, both within our field and in the
wider open source community, have emerged. These tools include
software and platforms for visualizing large volumes of complex data
and machine learning applications such as TensorFlow, Dask, Pachyderm,
Blaze, Parsl, and Thrill. R\&D into these tools is needed to enable
widespread adoption of these tools in HEP in cases where they can have
a big impact on the efficiency of HEP analysts.
\item
One increasingly important aspect is automation of workflows and the
use of automated analysis pipelines. Technologies behind these often
leverage open source software such as continuous integration
tools. With a lower bar to adoption, these pipeline toolkits could
become much more widespread in HEP, with benefits including reduced
mechanical work by analysts and enabling analysis reproducibility at a
very early stage.
\item
Notebook interfaces have already demonstrated their value for
tutorials and exercises in training sessions and facilitating
reproducibility. Remote services like notebook-based
analysis-as-a-service should be explored and HEP research tool in
addition to education and outreach. 
\item
The HEP community should leverage data formats which are standard
within data science, which is critical for gaining access to non-HEP
tools, technologies and expertise from computer scientists. We should 
investigate optimizing some of the more promising formats for
late-stage HEP analysis workflows.
\end{itemize}

\vskip 0.1in\noindent
\textbf{\textit{Connecting to Modern Cyberinfrastructure}}.
Facilitating easy access and efficient use of modern
cyberinfrastructure for analysis workflows will be very important
during the HL-LHC due to the anticipated proliferation of such
platforms and an increased demand for analysis resources to achieve
the physics goals. These include scalable platforms, campus clusters,
clouds, and HPC systems, which employ modern and evolving
architectures such as GPUs, FPGAs, specialized architectures like
Google's Tensor Processing Units (TPUs)~\cite{TPU}, memory-intensive systems, 
and web services. We should develop mechanisms to instantiate resources
for analysis from shared infrastructure as demand arises and share
them elastically to support easy, efficient use. An approach gaining a
lot of interest for deployment of analysis job payload is containers
on grid, cloud, HPC and local resources. The goal is to develop
approaches to data analysis which make it easy to utilize
heterogeneous resources for analysis workflows. The challenges include
making heterogeneous resources look more uniform to the analyzers and 
adapting to changes in resources not directly controlled by analysts or 
their experiments (both technically and financially). 

\vskip 0.1in\noindent
\textbf{\textit{Functional, Declarative Programming}}.
In a functional approach to programming, an analyst defines what tasks 
she or he would like the computing system to perform, rather than
telling the system how to do it. In this way, scientists express the
intended data transformation as a query on data. Instead of having to
define and control the `how', the analyst would declare the `what' of
their analysis, essentially removing the need to define the event loop 
in an analysis and leave it to underlying services and systems to
optimally iterate over events. This model allows (and gives the
responsibility to) the underlying infrastructure to optimize all
aspects of the application, including data access patterns and
execution concurrency. HEP analysis throughput could be greatly
enhanced by switching to a functional or declarative programming
model.  The HEP community is already investing in R\&D projects to
enable a functional programming approach (for example TDataFrame
in ROOT). Additional R\&D projects are needed to develop functional
programming models, along with the sophisticated algorithms to match
declarative specifications to underlying resources within a convergent
optimization framework.

\vskip 0.1in\noindent
\textbf{\textit{Improved Non-event Data Handling}}. An important area
that has not received sufficient development is the access to
non-event data required for analysis. Example data types include
cross-section values, scale factors, efficiencies and fake rate
tables, and potentially larger data tables produced by methods such
as BDTs or neural networks.  Easy storage of non-event data of all
sorts of different content, during the analysis step is needed to
bring reliable and reproducible access to non-event data just as
it currently exists for event data.  While a number of ways of doing
this have been developed, no commonly accepted and supported way
has yet emerged.

\vskip 0.1in\noindent
\textbf{\textit{High-throughput, Low-latency Analysis Systems}}.
An interesting alternative approach to the current approach to
analysis data reduction via a series of time-intensive processing
steps is a very low-latency analysis system. To be of interest, an
analysis facility would need to provide results, such as histograms,
on time-scales short enough to allow many iterations per day by the
analyzer. Two promising, new approaches to data analysis systems in
this area are:
\begin{itemize}
\item \textbf{Spark-like analysis systems.} A new model of data analysis,
  developed outside of HEP, maintains the concept of sequential ntuple
  reduction but mixes interactivity with batch processing. Spark is
  one such system, but TensorFlow, Dask, Pachyderm, and Thrill are
  others. Distributed processing is either launched as a part of user
  interaction at a command prompt or wrapped up for batch
  submission. The key differences from the above are:
  \begin{enumerate}
  \item parallelization is implicit through map/filter/reduce
    functions.
  \item data are abstracted as remote, distributed datasets, rather
    than files.
  \item computation and storage are mixed for data locality: a
    specialized cluster must be prepared, but can yield higher
    throughput.
  \end{enumerate}
  A Spark-like analysis facility would be a shared resource for
  exploratory data analysis (e.g., making quick plots on data subsets
  through the spark-shell) and batch submission with the same
  interface (e.g., substantial jobs through spark-submit). The primary 
  advantage that software products like Spark introduce is in
  simplifying the user's access to data, lowering the cognitive
  overhead to setting up and running parallel jobs. 

\item \textbf{Query-based analysis systems.} In one vision for a
  query-based analysis approach, a series of analysis cycles, each of
  which provides minimal input (queries of data and code to execute), 
  generates the essential output (histograms, ntuples, etc.) that can
  be retrieved by the user. The analysis workflow should be
  accomplished without focus on persistence of data traditionally
  associated with data reduction, however transient data could be
  generated in order to efficiently accomplish this workflow and
  optionally could be retained to a facilitate an analysis
  `checkpoint' for subsequent execution. In this approach, the focus
  is on obtaining the analysis end-products in a way that does not
  necessitate a data reduction campaign and associated provisioning of
  resources. Advantages of a query-based analysis system and its key
  components include:
  \begin{enumerate}
    \item \emph{Reduced resource needs for Analysis}. A critical
      consideration of the currently ntuple-driven analysis method is
      the large CPU and storage requirements for the intermediate data
      samples. The query-based system provides only the final outcomes
      of interest (histograms, etc).
    \item \emph{Sharing resources with traditional systems.} Unlike a
      traditional batch system, access to this query system is
      intermittent and extremely bursty, so it would be hard to
      justify allocating exclusive resources to it. The query system
      must share resources with a traditional batch system in such a
      way that it could elastically scale in response to load.
      \item \emph {Fast Columnar Data Caching.} Presenting column
        partitions (``Columnar cache'') to an analysis system as the
        fundamental unit of data management as opposed to files is an
        essential feature of the query system. It facilitates retaining
        input data between queries, which are usually repeated with
        small modifications (intentionally as part of a systematics
        study or unplanned as part of normal data exploration). 
      \item \emph {Provenance}. The query system should also attach
        enough provenance to each dataset that it could be recreated
        from the original source data, which is considered
        immutable. User datasets, while they can't be modified
        in-place, can be deleted, so a dataset's paper trail must
        extend all the way back to source data.
  \end{enumerate}
\end{itemize}

\vskip 0.1in\noindent
\textbf{\textit{Data Interpretation}}. 
The LHC provides a large increase in center-of-mass energy over
previous collider experiments, starting from 7-8 TeV in Run 1 to 13
TeV during Run 2. The associated large increase in gluon luminosity
provided the necessary conditions for discovery of the Higgs boson by
the ATLAS and CMS
collaborations~\cite{HIGG-2012-27,Chatrchyan:2012xdj}. Searches for
other new particles at high mass has been a primary focus at the LHC,
with lower limits on new particle masses reaching several TeV in many
new physics models. The HL-LHC will be an era of increased integrated
luminosity rather than increased collision energy. It is conceivable, 
maybe even likely, that the focus of many analyses during the HL-LHC
will shift from direct searches for new particle production to
indirect searches for new states with masses beyond the direct reach
of the experiments. In this scenario, many LHC analyses will be
searching for virtual effects from particles at high-scale, evident
only through a detailed study of the kinematics of many events and
correlations among many observables. Given its general
parameterization of new physics at high-scale, a central framework
for this type of analysis is the Effective Field Theory (EFT) 
extension of the SM~\cite{Goertz:2017gor}. Constraining possible
higher-dimensional operators within the context of EFT or other
model parameter estimation in high-dimensional spaces using the large
datasets afforded by the HL-LHC will be both a challenge and an
opportunity for HEP, demanding improvements in analysis techniques,
software and computing. An Institute could bring together HEP
theorists and experimentalists and computer scientists to tackle the
challenges associated with these kinds of generalized interpretations
of LHC data. Examples include developing better high-dimensional
minimization methods and machine learning approaches to approximate
event probability densities~\cite{sustainableMEM} and provide
likelihood-free inference~\cite{Cranmer2016}.

\vskip 0.1in\noindent
\textbf{\textit{Analysis Reproducibility and Reinterpretation}}. 
To be successful, analysis reproducibility and reinterpretation
need to be considered in all new approaches under investigation and
needs to be a fundamental component of the analysis ecosystem as a
whole.  These considerations become even more critical as we explore
analysis models with more heterogeneous hardware and analysis
techniques. One specific piece of infrastructure that is currently
missing is an analysis database able to represent the many-to-many
mapping between publications, logical labels for the event selection
defining signal and control regions, data products associated to
the application of those event selections to specific datasets, the
theoretical models associated to simulated datasets, the multiple
implementations of those analyses from the experiments and theoretical
community created for the purpose of analysis interpretation, and
the results of those interpretations.  The protocol for analysis
(re)interpretation is clear and narrowly scoped, which makes it
possible to offer it as a service.  This type of activity lends
itself to an ``Interpretation Gateway'' concept, whose goal is to
facility access to shared data, software, computing services,
instruments, and related educational materials~\cite{Sciencegateways2017}.
Much of the necessary infrastructure is in place to create
it~\cite{Cranmer2011,Akopov:2012bm,AtlasDataAccessPolicy}.  Such
an interpretation service would greatly enhance the physics impact
of the LHC and also enhance the legacy of the LHC well into the
future. An Institute could potentially drive the integration analysis 
facilities, analysis preservation infrastructure, data repositories, and 
recasting tools.

\subsubsection{Impact and Relevance for \s2i2}

\vskip 0.1in
\noindent{\bf Physics Impact:} The very fast turnaround of analysis
results that could be possible with new approaches to data access and
organization would lead to rapid turnaround for new science. 

\vskip 0.1in
\noindent{\bf Resources Impact:} Optimized data access for analysis will lead 
to more efficient use of both CPU and (especially) storage resources. This 
is essential holding down the overall costs of computing. 

\vskip 0.1in
\noindent{\bf Sustainability Impact:} This effort would improve the
reproducibility and provenance tracking for workflows (especially
analysis workflows), making physics analyses more sustainable through
the lifetime of the HL-LHC. 

\vskip 0.1in
\noindent{\bf Interest/Expertise:} University groups have already
pioneered significant changes to the data access model for the LHC
through the development of federated storage systems, and are prepared
to take this further. Other groups are currently exploring the
features of modern storage systems and their possible implementation
in experiments.

\vskip 0.1in
\noindent{\bf Leadership:} Universities are where data analyses for
Ph.D. theses are done, together with postdocs and professors. There is
also much to be gained within the US physics effort for HL-LHC by
focusing on improving the last-mile of analysis computing. Therefore,
it is natural for the US Universities to lead in the development of
data analysis systems, especially considering the potential for
computer science colleagues to collaborate on innovative approaches
to such systems. This is also an area where partnership with
national labs can be very productive, since the much of technical
infrastructure to develop these systems at required scales are at the
labs.

\vskip 0.1in
\noindent{\bf Value:} All LHC experiments will benefit from new
methods of data access and organization, although the implementations
may vary due to the different data formats and computing models of
each experiment.

\vskip 0.1in
\noindent{\bf Research/Innovation:} This effort would rely on
partnerships with data storage and access experts in the CS community,
some of whom are already providing consultation in this area.

\subsection{Reconstruction and Trigger Algorithms}
\label{sec:recotrigger}

The real-time processing in the trigger and reconstruction of raw
detector data (real and simulated) represent major components of
today's computing requirements in HEP. A recent
projection~\cite{Campana2016} of the ATLAS 2016 computing model
results in $>$85\% of the HL-LHC CPU resources being spent on the
reconstruction of data or simulated events. 
Several types of software algorithms are essential to the
transformation of raw detector data into analysis-level objects.
Specifically, these algorithms can be broadly grouped:

\begin{enumerate}
\item Online: Algorithms, or sequences of algorithms, executed on
  events read out from the detector in near-real-time as part of the
  software trigger, typically on a computing facility located close to
  the detector itself.
\item Offline: As distinguished from online, any algorithm or sequence
  of algorithms executed on the subset of events preselected by the
  trigger system, or generated by a Monte Carlo simulation
  application, typically in a distributed computing system. 
\item Reconstruction: The transformation of raw detector information
  into higher level objects used in physics analysis. A defining
  characteristic of `reconstruction' that separates it from `analysis'
  is that the quality criteria used in the reconstruction to, for
  example, minimize the number of fake tracks, are independent of how
  those tracks will be used later on. This usually implies that
  reconstruction algorithms use the entirety of the detector
  information to attempt to create a full picture of each interaction
  in the detector. Reconstruction algorithms are also typically run as
  part of the processing carried out by centralized computing
  facilities.
\item Trigger: the online classification of events which reduces
  either the number of events which are kept for further `offline'
  analysis, the size of such events, or both. Software triggers, whose
  defining characteristic is that they process data without a fixed
  latency, are part of the real-time processing path and must make
  decisions quickly enough to keep up with the incoming data, possibly
  using substantial disk buffers.
\item Real-time analysis: Data processing that goes beyond object
  reconstruction, and is performed online within the trigger
  system. The typical goal of real-time analysis is to combine the
  products of the reconstruction algorithms (tracks, clusters,
  jets...) into complex objects (hadrons, gauge bosons, new physics
  candidates...) which can then be used directly in analysis without
  an intermediate reconstruction step.
\end{enumerate}

\subsubsection{Challenges}

Software trigger and event reconstruction techniques in HEP face a
number of new challenges in the next decade. These are broadly
categorized into 1) those from new and upgraded accelerator
facilities, 2) from detector upgrades and new detector technologies,
3) increases in anticipated event rates to be processed by algorithms
(both online and offline), and 4) from evolutions in software
development practices.

Advances in facilities and future experiments bring a dramatic
increase in physics reach, as well as increased event complexity and
rates. At the HL-LHC, the central challenge for object reconstruction
is thus to maintain excellent efficiency and resolution in the face of
high pileup values, especially at low object $p_T$. Detector upgrades
such as increases in channel density, high precision timing and
improved detector geometric layouts are essential to mitigate these
problems. For software, particularly for triggering and event
reconstruction algorithms, there is a critical need not to
dramatically increase the processing time per event.

A number of new detector concepts are proposed on the 5-10 year
timescale  to help in overcoming the challenges identified above. In
many cases, these new technologies bring novel requirements to
software trigger and event reconstruction algorithms or require new
algorithms to be developed. Ones of particular importance at the
HL-LHC include high-granularity calorimetry, high precision timing
detectors, and hardware triggers based on tracking information which
may seed later software trigger and reconstruction algorithms.

Trigger systems for next-generation experiments are evolving to be
more capable, both in their ability to select a wider range of events
of interest for the physics program of their experiment, and their
ability to stream a larger rate of events for further
processing. ATLAS and CMS both target systems where the output of the
hardware trigger system is increased to $ 10 \times $ the current
capability, up to 1 MHz~\cite{ATLAS2015,CMS2015}. In other cases, such
as LHCb~\cite{LHCb2014} and ALICE~\cite{ALICE2015}, the full collision
rate (between 30 to 40 MHz for typical LHC operations) will be
streamed to real-time or quasi-realtime software trigger systems
starting in Run 3. The increase in event complexity also brings a
`problem' of overabundance of signal to the experiments, and
specifically the software trigger algorithms. The evolution towards a
genuine real-time analysis of data has been driven by the need to
analyze more signal than can be written out for traditional
processing, and technological developments which make it possible to
do this without reducing the analysis sensitivity or introducing
biases.

The evolution of computing technologies presents both opportunities 
and challenges. It is an opportunity to move beyond commodity x86
technologies, which HEP has used very effectively over the past 20
years, to performance-driven architectures and therefore software
designs. It is also a significant challenge to derive sufficient event
processing throughput per cost to reasonably enable our physics
programs~\cite{Bird2014}.  Specific items identified included 1) the
increase of SIMD capabilities (processors capable of running a single
instruction set simultaneously over multiple data), 2) the evolution
towards multi- or many-core architectures, 3) the slow increase in
memory bandwidth relative to CPU capabilities, 4) the rise of
heterogeneous hardware, and 5) the possible evolution in facilities
available to HEP production systems.

The move towards open source software development and continuous
integration systems brings opportunities to assist developers of
software trigger and event reconstruction algorithms. Continuous
integration systems have already allowed automated code quality and
performance checks, both for algorithm developers and code integration
teams. Scaling these up to allow for sufficiently high statistics
checks is still among the outstanding challenges. As the timescale for
recording and analyzing  data increases, maintaining and supporting
legacy code will become more challenging. Code quality demands
increase as traditional offline analysis components migrate into
trigger systems, and, more generally, into algorithms that are run
only once.

\subsubsection{Current Approaches}

Substantial computing facilities are in use for both online and
offline event processing across all experiments surveyed. Online
facilities are dedicated to the operation of the software trigger,
while offline facilities are shared for operational needs including
event reconstruction, simulation (often the dominant component) and
analysis. CPU use by experiments is typically at the scale of
tens or hundreds of thousands of x86 processing cores. The projections
of CPU requirements discussed in
Section~\ref{sec:computing_challenges} clearly demonstrate the need
for either much larger facilities than anticipated or correspondingly
more performant algorithms. 

The CPU time needed for event reconstruction tends to be dominated by
that used by charged particle reconstruction (tracking), especially as
the need for efficiently reconstructing low $p_T$ particles is
considered. Calorimetric reconstruction, particle flow reconstruction
and particle identification algorithms also make up significant parts
of the CPU budget in some experiments.

Disk storage is currently 10s to 100s of PB per experiment. It is
dominantly used to make the output of the event reconstruction, for
both real data and simulated data, available for analysis.

Current generation experiments have moved towards smaller, but still
flexible, data tiers for analysis. These tiers are typically based on
the ROOT~\cite{bib:root} file format and constructed to facilitate
both skimming of interesting events and the selection of interesting
pieces of events by individual analysis groups or through centralized
analysis processing systems. Initial implementations of real-time 
analysis systems are in use within several experiments. These
approaches remove the detector data that typically makes up the raw
data tier kept for offline reconstruction, and to keep only final
analysis objects~\cite{Aaij2016,Abreu2014,CMS2016}.

Detector calibration and alignment requirements were
surveyed. Generally a high level of automation is in place across
experiments, both for very frequently updated measurements and more
rarely updated measurements. Often, automated procedures are
integrated as part of the data taking and data reconstruction
processing chain. Some longer term measurements, requiring significant
data samples to be analyzed together remain as critical pieces of
calibration and alignment work. These techniques are often most
critical for a subset of high precision measurements rather than for
the entire physics program of an experiment.

\subsubsection{Research and Development Roadmap and Goals}

The CWP identified seven broad areas which will be critical for software 
trigger and event reconstruction work over the next decade. These are:

\vskip 0.1in
\noindent{\bf Roadmap area 1: Enhanced vectorization programming techniques -}
HEP-developed toolkits and algorithms typically make poor use of
vector processors on commodity computing systems. Improving this will
bring speedups to applications running on both current computing
systems and most future architectures. The goal for work in this
area is to evolve current toolkit and algorithm implementations,
and best programming techniques to better use SIMD capabilities of
current and future computing architectures.

\vskip 0.1in
\noindent{\bf Roadmap area 2: Algorithms and data structures to efficiently exploit many-core architectures -} 
Computing platforms are generally evolving towards having more cores
to increase processing capability. This evolution has
resulted in multi-threaded frameworks in use, or in development,
across HEP. Algorithm developers can improve throughput by being
thread safe and enabling the use of fine-grained parallelism. The
goal is to evolve current event models, toolkits and algorithm
implementations, and best programming techniques to improve the
throughput of multi-threaded software trigger and event reconstruction
applications.

\vskip 0.1in
\noindent{\bf Roadmap area 3: Algorithms and data structures for non-x86 computing architectures (e.g.\ GPUs, FPGAs) -}
Computing architectures using technologies beyond CPUs offer an
interesting alternative for increasing throughput of the most time
consuming trigger or reconstruction algorithms. Such architectures
(e.g.\ GPUs, FPGAs) could be easily integrated into dedicated trigger
or specialized reconstruction processing facilities (e.g.\ online
computing farms). The goal is to demonstrate how the throughput of
toolkits or algorithms can be improved through the use of new
computing architectures in a production environment.
The adoption of these technologies will particularly affect the research
and development needed in other roadmap areas.

\vskip 0.1in
\noindent{\bf Roadmap area 4: Enhanced QA/QC for reconstruction techniques -}
HEP experiments have extensive continuous integration systems,
including varying code regression checks that have enhanced the
quality assurance (QA) and quality control (QC) procedures for
software development in recent years. These are typically maintained
by individual experiments and have not yet reached the scale where
statistical regression, technical, and physics performance checks
can be performed for each proposed software change. The goal is to
enable the development, automation, and deployment of extended QA
and QC tools and facilities for software trigger and event
reconstruction algorithms.

\vskip 0.1in
\noindent{\bf Roadmap area 5: Real-time analysis -} 
Real-time analysis techniques are being adopted to enable a wider
range of physics signals to be saved by the trigger for final
analysis. As rates increase, these techniques can become more
important and widespread by enabling only the parts of an event
associated with the signal candidates to be saved, reducing the
required disk space. The goal is to evaluate and demonstrate the
tools needed to facilitate real-time analysis techniques. Research
topics include compression and custom data formats; toolkits for
real-time detector calibration and validation which will enable
full offline analysis chains to be ported into real-time; and
frameworks which will enable non-expert offline analysts to design
and deploy real-time analyses without compromising data taking
quality.

\vskip 0.1in
\noindent{\bf Roadmap area 6: High precision physics-object reconstruction, identification and measurement techniques -}
The central challenge for object reconstruction at the HL-LHC is 
to maintain excellent efficiency and resolution in the face of high
pileup values, especially for low object $p_T$. Both trigger and
reconstruction algorithms must exploit new techniques and higher
granularity detectors to maintain, or even improve,
future  physics measurements.
It is also becoming clear that
reconstruction in very high pileup environments at the HL-LHC
will only be possible by adding timing information
to our detectors.
Designing appropriate detectors requires that 
the corresponding reconstruction algorithms 
be developed and demonstrated to work well in complex environments.

\vskip 0.1in
\noindent{\bf Roadmap area 7: Fast software trigger and reconstruction algorithms for high-density environments -}
Future experimental facilities will bring a large increase in event
complexity. The scaling of current-generation algorithms with this
complexity must be improved to avoid a large increase in resource
needs. Where possible, toolkits and algorithms will be evolved or
rewritten, focusing on their physics and technical performance at high
event complexity (e.g.\ high pileup). It is likely also necessary to
deploy new algorithms and new approaches, including advanced machine
learning techniques developed in other fields, in order to solve these
problems. One possible approach is that of anomaly detection, where
events not consistent with known processes or signatures are
identified and retained. The most important targets are those which
limit expected throughput performance (e.g.\ charged-particle
tracking).

\subsubsection{Impact and Relevance for \s2i2}

Reconstruction algorithms are projected to be the biggest CPU
consumers at the HL-LHC. 
Code modernization and new approaches are needed
to address the large increases in pileup (4x) and trigger output rates (5-10x).
Trigger/Reco algorithm
enhancements (and new approaches) enable extended physics reach
even in more challenging detection environments (e.g., pileup).
Moreover, Trigger/Reco algorithm development is needed to take full
advantage of enhanced detector capabilities (e.g., timing detectors,
high-granularity calorimeters). 
`Real time analysis' promises to
effectively increase achievable trigger rates (for fixed budgets)
through making reduced-size, analysis-ready output from online
trigger(-less) systems.

\vskip 0.1in
\noindent{\bf Physics Impact:}  Effectively selecting datasets to be persisted,
and processing them sufficiently rapidly while maintaining the quality
of the reconstructed objects, will allow analysts to use the higher
collision rates in the more complex environments to address the broadest
range of physics questions.

\vskip 0.1in
\noindent{\bf Resources Impact:} 
Technical improvements achieved in trigger or reconstruction 
algorithms directly reduce the computing resources needed for 
HL-LHC computing. 
In addition,
targeted optimizations of existing code will allow HL-LHC experiments 
to fully take advantage of the 
significant computing resources at HPC centers that may become
available at little direct cost.

\vskip 0.1in
\noindent{\bf Sustainability Impact:} 
University personnel, including graduate students and post-docs
working in the research program, frequently develop and maintain
trigger, reconstruction, and real-time analysis
 algorithms and implementations.
Doing so in the context of an \s2i2 will focus efforts on
best practices related to reproducible research, 
including design and documentation.

\vskip 0.1in
\noindent{\bf Interest/Expertise:} U.S.\ university groups are already
leading many of the efforts in these areas.
They are also working on designs of detector upgrades that require
improved algorithms to take advantage of new features such as
high precision timing.
Similarly, they are already studying the use of more specialized
chipsets, such as FPGAs and GPUs, for HEP-specific applications
such as track pattern recognition and parameter estimation.

\vskip 0.1in
\noindent{\bf Leadership:}
As in the bullet above.

\vskip 0.1in
\noindent{\bf Value:} All LHC experiments will benefit from these techniques, 
although detailed implementations will be experiment-specific given 
the differing detector configurations.

\vskip 0.1in
\noindent{\bf Research/Innovation:} 
The CPU evolution requirements described in 
Section~\ref{sec:computing_challenges} are about  $ 6 \times $ greater 
than those promised by Moore's Law.
Achieving this level of performance will require significant algorithmic
innovation and software engineering research to take advantage
of vector processors and other emerging technologies.
Machine learning also promises the ability to replace some of the
most CPU-intensive algorithms with fast inference engines trained on mixtures
of simulated and real data.
These efforts will require collaboration of domain experts 
with software engineers,
computer scientists, and data scientists with complementary experience.






\subsection{Applications of Machine Learning}

Machine Learning (ML) is a rapidly evolving area of computer science, with close ties to statistics, 
aimed at algorithmic approaches for solving a wide variety of tasks based on data. 
These tasks include classification, regression, clustering, density estimation, 
data compression, anomaly detection, statistical inference, and various forms of prediction. 
Each of these tasks have applications in HEP.
The high-dimensional and highly structured
data resulting from the complex sensor arrays of modern particle detectors
provides an environment where ML methods can reasonably be 
expected to radically change how data is reduced and analyzed. 
The presence of high-fidelity simulations makes supervised learning approaches particularly 
powerful for HEP; however, unsupervised learning based on unlabeled collision data is also promising. 
Some applications of ML will qualitatively
improve the physics reach of HEP data sets.  
Others will allow much more efficient use 
of processing and storage resources, allowing the HL-LHC experiments to
achieve their goals within cost limitations.  
It is anticipated that ML will become ubiquitous in HEP, thus, 
many of the activities in this focus area will explicitly intersect with activities
in the other focus areas.

%


\subsubsection{Challenges and Opportunities}

Clearly, HEP can profit by leveraging the developments in ML methodology and 
software solutions being developed by computer scientists, data scientists, 
and scientific software developers from outside the HEP world. 
There are enormous financial and intellectual investments going into
the development of modern ML software and methodology. 
Harnessing these developments is a challenge as there is a 
technical gap between most industrial ML platforms, the software used
for ML research, and the software frameworks used by the HEP experiments. 

Another challenge is that several HEP problems have unique 
considerations that do not map nicely onto existing problems with well established
solutions. For instance, we often deal with  
steeply falling distributions, sparse data or data with very large dynamic range, 
extreme real-time demands, complex detector geometries with non-uniform cell sizes, and we are 
very concerned with systematic uncertainties and calibrated statistical statements.
Similarly, while labeled training data can be produced with simulations,  these simulations
are computationally intensive and not completely accurate.
However, experience has shown that
recasting HEP problems into abstract formulations reveals that they are often 
of more general interest. For example, the treatment of systematics uncertainties can be related to 
ML topics of domain adaptation and fairness~\cite{Louppe:2016ylz}. Challenges posed by scientific simulators 
appear in a wide range of scientific disciplines including systems and population biology, computational neuroscience,
epidemiology, cosmology, astrophysics, and personalized health~\cite{abc, implicitModels,Cranmer:2015bka,Louppe:2017pay}. 
This provides an opportunity to engage with the ML community and provide broader 
impacts beyond HEP.


A number of ML approaches have been used productively in HEP for more than 20 years;
others have been introduced relatively recently or are still in the research and development
phase. 
HEP now has the opportunity to exploit these developments
to make substantial improvements over traditional techniques with
effects on both physics and technical performance. Broad research
and development programs are needed to leverage these capabilities for HEP. Example
applications where ML software could have a large impact on HEP:
\begin{itemize}
  \item
   Replace the most computationally expensive
   parts of pattern recognition algorithms and 
   algorithms that extract parameters
   characterizing reconstructed objects;
 \item
  Optimize the real-time decision making in the trigger and data acquisition systems;
  \item
   Compress data significantly with negligible loss of fidelity
   in terms of physics utility;
 \item 
  Provide fast, high-fidelity  simulation; 
 \item 
   Extend the physics reach of experiments by qualitatively
   changing the types of analyses that can be done.
\end{itemize}

The fast pace of ML research and the plethora of algorithms and implementations presents 
both opportunities and challenges for HEP. The community
needs to understand which resources are most appropriate for our use,
tradeoffs for using one approach compared to another, and 
the tradeoffs of using ML algorithms compared to more traditional approaches.
These issues are intermixed, and a key goal of an Institute will
be to streamline the integration of knowledge and solutions 
to the greater HEP community. The Institute would complement other community 
efforts around using ML in HEP, serve as a hub of expertise 
on techniques and tools, and extend existing training programs. 
The Institute's university presence would accelerate the participation of 
academics in computer science, machine learning, data science, applied mathematics, and statistics. 
The Institute could also provide the missing effort that is key to organize and manage successful challenges 
around specific topics like jet tagging and tracking. Well organized challenges of various forms have been key to 
the rapid advance of deep learning. The ImageNet Challenge~\cite{imagenet-challenge,imagenetweb}, for example, is often associated to the rise of deep learning and convolutional neural networks. 
Beyond the R\&D projects it sponsors directly,
the Institute would help teams develop and deploy experiment-specific 
ML-based algorithms in their software stacks. It will work with industry 
as standards such as the Open Neural Network Exchange (\textsc{ONNX})~\cite{onnx} develop.

\subsubsection{Current Approaches}

The use of ML in HEP analyses has become commonplace over the past 
two decades. Many analyses  use the HEP-specific software package  
\textpack{TMVA}~\cite{Hocker:2007ht} included in the
CERN \textpack{ROOT}~\cite{root} project. These tools have mainly been used for classification and regression either at the level of individual reconstructed objects or at the event-level.
Recently, many HEP analysts have begun migrating to ML packages developed outside of HEP,
such as \textpack{SciKit-Learn}~\cite{sklearn},
\textpack{Keras}~\cite{keras},  \textpack{TensorFlow}~\cite{tensorflow}, \textpack{MXNet}~\cite{mxnet}, and \textpack{PyTorch}~\cite{pytorch}.
Data scientists at  Yandex created a Python package that
provides a consistent API to most ML packages used in
HEP~\cite{rep}, and another that provides some HEP-specific
ML algorithms~\cite{hepml}.  Unfortunately, integrating modern ML algorithms for reconstruction and bulk data processing into the HEP software frameworks is currently inefficient. Moreover, the use of machine learning in higher-level analyses often involves a transition from the HEP software ecosystem to the ML software ecosystem and back. These software issues are currently barriers to integrating ML deeply into the bulk data processing and higher-level data analysis.

%
%
%

\subsubsection{Research and Development Roadmap and Goals}

The possible scope of applications where ML techniques can be applied is
broad and spans most HEP technical areas, from trigger algorithms up through
analysis. Examples that the HEP community believes to be of primary 
interest include:
 
\vskip 0.1in
\noindent\textbf{Track and vertex reconstruction}. 
  Charged track and vertex reconstruction is one of the most CPU intensive 
  elements of the software stack.  
  The algorithms are typically iterative, alternating between selecting hits 
  associated with tracks and characterizing the trajectory of a track (a collection of hits).  
  Similarly, vertices are built from collections of tracks, and then characterized 
  quantitatively. 
  ML algorithms  have been used extensively outside HEP to recognize, 
  classify, and quantitatively describe objects.  
  We will investigate how to replace components of the
  pattern recognition algorithms and the `fitting' algorithms that extract 
  parameters characterizing the reconstructed objects.  
  As existing algorithms already produce high-quality physics, the primary 
  goal of this activity will be developing replacement algorithms that 
  execute much more quickly while maintaining sufficient fidelity.

\vskip 0.1in
\noindent\textbf{Jet tagging}. 
   ML algorithms can often discover patterns and correlations more powerfully than 
   human analysts alone.  
   This allows qualitatively better analysis of recorded data sets.  
   For example, ML algorithms can be used to characterize the substructure of 
   ``jets" observed in terms of their underlying physics processes~\cite{Larkoski:2017jix,deOliveira:2015xxd,Louppe:2017ipp}.  
   ATLAS, CMS, and LHCb already use ML algorithms to separate jets into those 
   associated with b-quark, c-quarks, or lighter quarks~\cite{Guest:2016iqz,CMS:2017xxp,Scodellaro:2017wli,Sirunyan:2017obz}.  
   Areas where new ML approaches will have a big impact include 
   exploiting sub-jets associated with quarks or gluons, and how calorimetric imaging
   techniques can be used as the basis for jet tagging.  
   If this can be done with both good efficiency and accurate understanding of 
   efficiency, the physics reach of the experiments will be significantly enhanced.

\vskip 0.1in
\noindent\textbf{Particle classification and regression}. 
   ML algorithms offer a promising avenue to improve the performance of
   particle classification algorithms as well as the resolution of
   particle energies and directions measured using regression
   techniques. Recent studies have demonstrated promising improvements 
   in classification and energy resolution of electrons and photons
   using energy depositions in calorimeter cells produced in
   electromagnetic showers~\cite{PCR1,PCR2}. Additional possible
   applications include the use of charged particle tracks for the
   identification of prompt electrons, muons and hadronically-decaying
   $\tau$ leptons. These improvements have the potential to benefit a
   wide variety of BSM searches and SM measurements in signatures with
   leptons and photons.
   
\vskip 0.1in
\noindent\textbf{Real-time Event classification}.
  The ATLAS, CMS, and LHCb detectors all produce much more data than can be 
  moved to permanent storage. The hardware and software components in the  
  trigger are responsible for reducing this data volume to what can be kept
  for analysis.  
  Electronics sparsify the data stream using zero suppression and they do some 
  basic data compression. 
  While this reduces the data rate by a factor of 100 
  (or more, depending on the experiment)  to about 1 terabyte per second, 
  another factor of order 1500 is required before the data can be 
  written to tape (or other long-term storage). 
  ML algorithms have already been used very successfully to rapidly
  characterize which events should be selected for additional consideration
  and eventually saved to long-term storage.
  The challenge will increase both quantitatively and qualitatively
  as the number of proton-proton collisions per bunch crossing increases.

\vskip 0.1in
\noindent\textbf{Tuning Monte Carlo simulations and fast simulation methods}
  All HEP experiments rely on simulated data sets to accurately compare 
  observed detector response data with expectations
  based on the hypotheses of the Standard Model or models of new physics. 
  While the high-energy processes of subatomic
  particle interactions are well modeled, the modeling of the parton shower, fragmentation, and hadronization 
  involve phenomenological models with several free parameters. Furthermore, Monte Carlo simulation tools, such as GEANT4~\cite{Geant4Ref1,Geant4Ref2,Geant4Ref3}, 
  have been developed to simulate the propagation
  of particles through detectors.  These simulations are computationally expensive and have many free parameters, and tuning them to 
  accurately describe the data is a challenge. 
   HEP physicists have begun using packages like {\sc Spearmint}~\cite{spearmint} and {\sc Scikit-optimize}~\cite{skopt} for Bayesian optimization to tune HEP simulations~\cite{Ilten:2016csi,TuneMC}. Recently, HEP use cases motivated a novel Adversarial Variational Optimization algorithm~\cite{Louppe:2017pay}, which is now being used for problems as diverse as computational topography and cardiac simulation for personal health~\cite{cardiac}.
   Once tuned, HEP simulators accurately model the complex interactions of particles and the subsequent detector response. Unfortunately, simulating a single LHC proton-proton
  collision takes on the order of minutes, corresponding to a significant part of 
  the computing needs for LHC since tens of billions of events are generated each year.
  {\em Fast simulation} techniques are an interesting option for replacing 
  the slowest components of the simulation chain 
  with computationally efficient approximations.
  Often, this is  done using simplified parameterizations or 
  look-up tables which don't reproduce detector response with the required
  level of precision.
  A variety of ML tools, such as Generative Adversarial Networks and Variational
  Auto-encoders, promise better fidelity and comparable executions 
  speeds (after training)~\cite{Paganini:2017hrr,nyugan}. The Institute could play a critical role in defining
  and developing reliable fast simulation algorithms based on ML tools that
  execute much more quickly than full simulation
  while maintaining sufficient fidelity for most physics studies. 

\vskip 0.1in
\noindent\textbf{Simulation-based inference} 
A fundamental challenge for statistical inference in HEP experiments arises 
from the fact that predictions are made using complicated simulations
of both the quantum mechanical scatterings as well as the complex interactions 
within the detector. The simulation implicitly defines a probability distribution
over the data, but evaluating this probability is intractable. These types of problems
appear in a wide range of scientific disciplines including systems and population biology, 
computational neuroscience, epidemiology, cosmology, astrophysics, 
and personalized health~\cite{abc, implicitModels}. 
In population biology, and other fields,
 Approximate Bayesian Computation (ABC) has  been used 
for simulation-based inference~\cite{abc}. 
Traditionally, HEP has 
approached this problem by approximating the probability distributions for a single variable with histogram templates 
or ad hoc analytic functions.
For instance, many searches are based on the invariant mass distribution.
However, in many cases, such as an Effective Field Theory analysis of the Higgs boson, an analysis based on a single observable sacrifices physics reach compared to an analysis based on a higher-dimensional representation of the data~\cite{Brehmer:2017lrt,Brehmer:2016nyr}. Another example is the calculation of approximate event probability densities using the Matrix Element Method~\cite{Kondo:1988yd,Fiedler:2010sg,2011arXiv1101.2259V,Elahi:2017ppe}.
More recently, a number of approximate inference and calculation techniques based on 
machine learning have been developed or proposed, which have the potential to extract great promise~\cite{Cranmer:2015bka,Louppe:2017pay,NIPS2017_7136,Papamakarios:2017:maf,DBLP:journals/corr/LeBW16,sustainableMEM}.  Tools such as \textsc{carl}~\cite{carl}, \textsc{edward}~\cite{tran2016edward,edward}, and \textsc{pyro}~\cite{pyro} are being developed (by HEP physicists, computer scientists, statisticians, and  industry researchers) to enable this deep integration of machine learning and statistical inference. 
This is a major shift in the analysis practices of HEP, and to realize it 
will require training and a deep integration of ML with HEP statistical software, a task 
that is well-suited for the Institute.

\subsubsection{Impact and Relevance for \s2i2}

\vskip 0.1in
\noindent{\bf Physics Impact:}  
Machine learning can enable qualitatively new types of data analysis and provide substantial gains in 
physics reach through improved data acquisition, particle identification, object reconstruction,
and event selection.
%

\vskip 0.1in
\noindent{\bf Resources Impact:} 
Replacing the most computationally expensive parts of simulation and reconstruction
will allow the experiments to use computing resources more efficiently.
Optimizing data compression will allow the experiments to use data
storage and networking resources more efficiently.

\vskip 0.1in
\noindent{\bf Sustainability Impact:} 
Building our domain-specific software on top of ML tools from
industry and the larger scientific software community should reduce the need to 
maintain (or build)  equivalent tools ourselves, but it will
require that we maintain components needed for interoperability.

\vskip 0.1in
\noindent{\bf Interest/Expertise:} 
U.S.\ university personnel are already leading significant efforts in
using ML for reconstruction, data-acquisition, jet tagging, event selection,
and inference. Some personnel are actively developing novel ML 
methodology.

\vskip 0.1in
\noindent{\bf Leadership:} 
There is a natural area for Institute leadership: in addition to the
existing interest and expertise in the university HEP community,
this is an area where engaging academics from other disciplines
will be a critical element in making the greatest possible progress.
Although specific software implementations of algorithms will differ, 
much of the R\&D program can be common.

\vskip 0.1in
\noindent{\bf Value:} All LHC experiments will benefit from using ML
to enable more performant data acquisition, particle identification, 
and event selection software that directly extends the physics
reach of HEP experiments like the HL-LHC.  
Experience has shown that solutions to HEP ML problems often translate to
other scientific disciplines.

\vskip 0.1in
\noindent{\bf Research/Innovation:} 
ML is evolving very rapidly, so there are many opportunities for
basic and applied research as well as innovation.
As most of the work developing ML algorithms and implementing them in
software (as distinct from 
the applications software built using them) is done by experts in the 
computer science and data science communities,
HEP needs to learn how to effectively use toolkits provided by the
open scientific software and industrial research community.
At the same time, some of the HL-LHC problems may be of special interest
to these other communities, either because the sizes of our data sets 
are large or because they have unique features.
Solutions to HEP problems lead to innovations that have historically had 
broader impact.


\subsection{Data Organization, Management and Access (DOMA)}

Experimental HEP has long been a data intensive science and it will 
continue to be through the HL-LHC era. The success of HEP experiments is 
built on their ability to reduce the tremendous amounts of data produced by 
HEP detectors to physics measurements. 
The reach of these data-intensive experiments is limited
by how quickly data can be accessed and digested by the computational
resources. Both changes in technology and large increases in data volume require
new computational models~\cite{SMSTORAGE}. HL-LHC and the HEP experiments
of the 2020s will be no exception.

Extending the current data handling methods and methodologies is expected
to be intractable in the HL-LHC era. The development and adoption of new
data analysis paradigms gives the field, as a whole, a window in
which to adapt our data access and data management schemes to ones
which are more suited and optimally matched to a wide range of
advanced computing models and analysis applications. This type of
shift has the potential to enable new analysis methods and
allow for an increase in scientific output.

\subsubsection{Challenges and Opportunities}

The LHC experiments currently provision and manage about an exabyte
of storage, approximately half of which is archival, and half is
traditional disk storage. The storage requirements per year are
expected to jump a factor of 20 or more for the HL-LHC. This itself is faster 
than projected Moore's Law gains and will present major challenges. 
Storage will remain one of the visible cost drivers for HEP computing,
however the projected growth and cost of the computational resources
needed to analyze the data is also expected to grow even faster than the
base storage costs. The combination of storage and analysis computing
costs may restrict scientific output and potential physics reach
of the experiments, thus new techniques and algorithms are likely to be
required. 

These three main challenges for data in the HL-LHC era can thus be summarized:

\begin{enumerate}
\item {\bf Big Data:} The HL-LHC will bring significant increases to both the data 
rate and the data volume. The computing systems will need to handle this 
without significant cost increases and within evolving storage technology limitations.
\item {\bf Dynamic Distributed Computing:} The significantly increased computational 
requirements for the HL-LHC era will also place new requirements on data. Specifically the use of 
new types of compute resources (cloud, HPC, and hybrid platforms) with different dynamic 
availability and characteristics will require more dynamic DOMA systems.
\item {\bf New Applications:} New applications such as machine learning training or high rate 
data query systems for analysis will likely be employed to meet the computational constraints 
and to extend the physics reach of the HL-LHC. These new applications will place new requirements on how 
and where data is accessed and produced. For example, specific applications (e.g.\ training for 
machine learning) may require use of specialized processor resources such as GPUs, placing 
further requirements on data formats.
\end{enumerate}

The projected event complexity of data from future LHC runs will require advanced reconstruction 
algorithms and analysis tools. The
precursors of these tools, in the form of new machine learning
paradigms, pattern recognition algorithms, and fast simulations, already show promise in reducing CPU needs 
for HEP experiments.
As these techniques continue to grow and blossom, they will place new
requirements on the computational resources that need to be leveraged
by all of HEP. The storage systems that are developed, and the data
management techniques that are employed will need to directly support
this wide range of computational facilities, and will need to be
matched to the changes in the computational work, so as not to
impede the improvements that they are bringing.

As with CPU, the landscape of storage protocols accessible to us
is trending towards heterogeneity. Thus, the ability to leverage
new storage technologies as they become available into existing
data delivery models becomes a challenge for which we must be prepared.
In part, this also means HEP experiments should be prepared to separate storage 
abstractions from the underlying storage resource systems~\cite{thain2005separating}.
Like opportunistic CPU, opportunistic storage resources available for limited duration 
(e.g. from a cloud provider) require data management and provisioning systems that 
can exploit them on short notice.  Much of this change can be aided by 
active R\&D of our own IO patterns which to date have not been 
well characterized.

On the hardware side, R\&D is needed in alternative approaches to
data archiving to determine the possible cost/performance tradeoffs.
Currently, tape is extensively used to hold data that cannot be
economically made available online. While the data is still accessible,
it comes with a high latency penalty, restricting its use in analysis and 
many processing pipelines. 
We need to do R\&D on both separate direct access-based archives
(e.g. disk or optical) and new models that overlay online direct
access volumes with archive space. This is especially relevant when
access latency is proportional to storage density. 

Closely related is research into splitting files or datasets into objects 
that are always kept together (i.e. the ``atomic size'') which can have implications 
at all levels in the software, storage and network infrastructure. 
In storage systems, as the atomic size increases so does memory pressure, 
CPU cycles spent on copying/moving/compressing data, and the likelihood of
hot spots developing on data servers. As atomic size decreases, the CPU cycles spent on requesting data,
round-trip times, and metadata overhead increase while locality is reduced. Luckily, modern storage systems 
such as Ceph~\cite{Weil:2006:CSH:1298455.1298485} have a number of effective knobs to navigate these trade-offs, including sizing of objects, partitioning and striping of data to objects, and co-location of objects. However, these currently must
be manually tuned for the workflow being optimized. Research in automating and ``learning'' which sets 
of storage system parameters yield optimal access performance is needed.

In the end, the results have to be weighed against the storage
deployment models that, currently, differ among the various
experiments. In the near term, this offers an opportunity to evalute the 
effectiveness of chosen approaches at scale. The lessons drawn will provide 
guidance going forward into the HL-LHC era.


Finally, any and all changes undertaken must improve the ease of
access to data current computing models offer while achieving greater scales.
We must also be prepared to accept the fact that the best possible
solution may require significant changes in the way data is handled
and analyzed. Simple extrapolations make clear that existing solutions
will not scale to meet the needs of HL-LHC experiments~\cite{BIRD2016}. 

\subsubsection{Current Approaches}

The original LHC computing models (circa 2005) were derived from the
simpler models used before distributed computing was a central part
of HEP computing. This allowed for a reasonably clean separation between
three different aspects of interacting with data: organization, management
and access. We define these terms in context here:

\vskip 0.1in
\noindent{\bf Data Organization:} This is essentially how data is structured as it is written. Most data is written in flat files, in ROOT~\cite{bib:root} format, typically with a column-wise organization of the data. The records corresponding to these columns are compressed. The internal details of this organization are typically visible only to individual software applications.

\vskip 0.1in
\noindent{\bf Data Management:} The key challenge here was the transition to the use of distributed 
computing in the form of the grid. The experiments developed dedicated data transfer and placement systems, 
along with catalogs, to move data between computing centers. To first order the computing models were rather static: 
data was placed at sites and the relevant compute jobs were sent to the right locations. Applications might interact 
with catalogs or, at times, the workflow management systems does this on behalf of the 
applications.
\vskip 0.1in
\noindent{\bf Data Access:} Various protocols are used for direct reads (rfio, dcap, xrootd, https, etc.) with a given computer center and/or explicit local stagein and caching for read by jobs. Application access may use different protocols than those used by the data transfers between site.
\vskip 0.1in

Before the LHC turn-on and in the first years of the LHC, these
three areas were to first order optimized independently. Many of the
challenges were in the area of ``Data Management (DM)'' as the Worldwide
LHC Computing Grid was commissioned. As the LHC computing matured through
Run 1 and Run 2, the interest has turned to optimizations spanning these
three areas. For example, the recent use of ``data federations''~\cite{CMSDATAFED,AAA} 
couples data management and data access aspects. As we will see below, some of 
the foreseen opportunities towards HL-LHC may require global optimizations.

We thus take a broader view than traditional ``DM'', and 
consider the combination of ``Data Organization, Management and Access 
(DOMA)'' together. We believe that such an integrated
view of all aspects of how an experiment interacts with and uses data
in HEP will provide important opportunities for efficiency and 
scalability as we enter the many-Exabyte era.



\subsubsection{Research and Development Roadmap and Goals}

First and foremost, the Institute should develop and maintain an
overarching, integrated vision for how experiments interact with thier
data and help them to articulate
a coherent strategy in their computing models. It should strive to
understand and document how 
any changes in one area of DOMA would affect all of the elements of 
the experiment's computing models. Historically, HEP experiments making
major changes in the DOMA area have stumbled when technical
investigations and deployments were done in a fragmented fashion
without a complete vision. Clear examples where this has happened
include BaBar at SLAC and the LHC experiments in the early phases
of the preparations for the LHC. The Institute should work closely
with the experiments and the US LHC Operations programs to build
a coherent strategy for data organization, management, and access and
understand how to integrate and test at-scale the key elements to validate
this strategy.

In the following, we identify DOMA specific task areas and goals that address the increased volume and complexity of data expected over the coming decade.

\vskip 0.1in
\noindent{\bf Atomic Size of Data:} An important goal is to create abstractions that make questions like the atomic size of data go away because that size is determined automatically. In higher layers of abstraction, we generally mean sub-file granularity, e.g. event-based.  This should be studied to see whether it can be implemented efficiently and in a scalable, cost-effective manner. Applications making use of event selection can be assessed as to whether it offers an advantage over current file-based granularity.  Example tasks in this area for the
early years of the Institute include:

\begin{itemize}
	\item Quantify the impact on performance and resource utilization (CPU, storage, network) for the main types of access 
	patterns (simulation, reconstruction, analysis).
	\item Assess impact of different access patterns on catalogs and on data distribution.
	\item Assess whether event-granularity makes sense in object stores that tend to require large chunks of data for efficiency.
	\item Test for improvement in recoverability from job or task preemption, in particular when using cloud spot resources and/or dynamic HPC resources.
\end{itemize}

\vskip 0.1in
\noindent{\bf Data Organization Paradigms:} We will seek to derive benefits from data organization and analysis technologies adopted by other big data users. A proof-of-concept that involves the following tasks needs to be established in the early years of the Institute to allow full implementations to be made in the years that follow. 

\begin{itemize}
	\item Study the impact of column-wise, versus row-wise, organization of data on the performance of each kind of access, including the associated storage format.
	\item Investigate efficient data storage and access solutions that support the use of MapReduce or Spark-like analysis services.
	\item Evaluation of declarative interfaces and in-situ processing.
	\item Evaluate just-in-time decompression schemes and mappings onto hardware architectures considering the flow of data, from spinning disk to memory and application.
	\item Investigate the long term replacement of Gridftp as the primary data transfer protocol.  Define metrics (performance, etc.) for evaluation.
	\item Benchmark end-to-end data delivery for the main use cases.  Identify impediments to efficient data delivery to/from CPU and storage.  What are the necessary storage hierarchies, and how are they mapped to available technologies.
\end{itemize}

\vskip 0.1in
\noindent{\bf Data Placement and Caching:}
Discover the role that data placement optimizations can play, including the use of data caches distributed over the WAN, in order to use computing resources more effectively. Investigate and or develop the technologies that can be used to achieve this. The following tasks would be appropriate for the early years of the Institute:

\begin{itemize}
	\item Quantify the benefit of placement optimization for the main use cases i.e. reconstruction, analysis, and simulation.
	\item Assess the benefit of caching for machine learning-based applications (in particular for the learning phase) and follow-up the evolution of technology outside HEP itself.
	\item Assess the potential benefit of content delivery networks in the HEP data context.
	\item Assess the feasibility and potential benefit of a named data network component in a HL-LHC data management system, in both medium and long-term as this new technology matures~\cite{ndn}.
\end{itemize}

\vskip 0.1in
\noindent{\bf Federated Data Centers (prototyping ``Data Lakes'')}

As storage operational costs are significant, models which consolidate storage into a smaller number of data centers with high capacity, well-managed networks, i.e. so-called ``Data Lakes'', should be prototyped. This would include the necessary qualities of service, and options for regionally distributed implementations, including the ability to flexibly respond to model changes in the balance between disk and tape. 

\begin{itemize}
\item Understanding the needed functionalities, including policies for managing data and replications, availability, quality of service, service levels, etc.
\item Understand how to interface a data-lake federation with heterogeneous storage systems in different sites.
\item Investigate how to define and manage the interconnects, network performance and bandwidth, monitoring, service quality etc.  Integration of networking information and testing of advanced networking infrastructure.
\item Investigate policies for managing and serving derived data sets, lifetimes, re-creation (on-demand?), caching of data, etc.
\end{itemize}

In the longer term, the benefits that can be derived from using different approaches to the way HEP is currently managing its data delivery systems should be investigated. Different content delivery methods should be studied for their utility in the HL-LHC context, namely Content Delivery Networks (CDN) and Named Data Networking (NDN).

\vskip 0.1in
\noindent{\bf Support for Query-based analysis techniques:} Data analysis is currently tied in with ROOT-based formats. 
In many currently-used paradigms, physicists consider all events at an equivalent level of detail and in the format 
offering the highest level of detail that needs to be considered in an analysis.  
However, not every event considered in analysis requires the same level of detail. 
One consideration to improve event access throughput is to design event tiers with different abstractions, 
and thus data sizes. All events can be considered at a lighter-weight tier while events of interest only 
can be accessed with a more information-rich tier.

For more scalable analysis, another opportunity to evaluate is how much work can be offloaded to a storage system, 
for example caching uncompressed or reordered data for fast access. The idea can be extended to virtual data 
and to query interfaces which would perform some of the transformation logic currently executed on CPU workers. 
Interactive querying of large datasets is an active field in the Big Data industry; examples include Spark-SQL, Impala, 
Kudu, Hawq, Apache Drill, and Google Dremel/BigQuery. A key question is about the usability of these techniques in 
HEP and we need to assess if our data transformations are not too complex for the SQL-based query languages used by 
these products. We also need to take into account that the adoption of these techniques, if they 
prove to be beneficial, would represent a disruptive change which directly impacts the end user and therefore 
promoting acceptance through intermediate solutions would be desirable.

\subsubsection{Impact and Relevance for \s2i2}

\vskip 0.1in
\noindent{\bf Physics Impact:} The very fast turnaround of analysis
results that could be possible with new approaches to data access
and organization would lead to rapid turnaround for new science.

\vskip 0.1in
\noindent{\bf Resources Impact:} Optimized data access will lead
to more efficient use of resources. In addition, by changing the
analysis models, and by reducing the number of data replicas required, 
the overall costs of storage can be reduced.

\vskip 0.1in
\noindent{\bf Sustainability Impact:} This effort would improve the
reproducibility and provenance tracking for workflows (especially
analysis workflows), making physics analyses more sustainable through
the lifetime of the HL-LHC.

\vskip 0.1in
\noindent{\bf Interest/Expertise:} University groups have already
pioneered significant changes to the data access model for the LHC
through the development of federated storage systems, and are
prepared to take this further.  Other groups are currently exploring
the features of modern storage systems and their possible implementation
in experiments.

\vskip 0.1in
\noindent{\bf Leadership:} CS research and technology innovation in several pertinent 
areas are being carried out by university groups, including research on methods for large 
scale adaptive and elastic database systems that support intensive mixed workloads 
(e.g.\ high data ingest, online analytics, and transactional updates).  Also universities 
are leading centers for work addressing critical emerging problems across many science domains, 
including analytical systems that benefit from column-oriented storage, where data is organized 
by attributes instead of records, thus enabling efficient disk I/O. As many teams perform 
data analytics in a collaborative way, where several users contribute to cleaning, modeling, 
analyzing, and integrating new data. To allow users to work on these tasks in isolation and 
selectively share the results, research at universities is actively developing systems to 
support lightweight dataset versioning, that is similar to software control systems like Git, 
but for structured data.

\vskip 0.1in
\noindent{\bf Value:} All LHC experiments will benefit from new
methods of data access and organization, although the implementations
may vary due to the different data formats and computing models of
each experiment.

\vskip 0.1in
\noindent{\bf Research/Innovation:}  This effort would rely on
partnerships with data storage and access experts in the CS community,
some of whom are already providing consultation in this area.


\subsection{Fabric of distributed high-throughput computing services
  (OSG)}
\label{sec:fabric}

Since its inception, the Open Science Grid (OSG) has evolved into an
internationally-recognized element of the U.S.\ national
cyberinfrastructure, enabling scientific discovery across a broad
range of disciplines. This has been accomplished by a unique
partnership that cuts across science disciplines, technical expertise,
and institutions. Building on novel software and shared hardware
capabilities, the OSG has been expanding the reach of high-throughput
computing (HTC) to a growing number of communities. Most importantly,
in terms of the HL-LHC, it provides essential services to US-ATLAS and
US-CMS.

The importance of the fabric of distributed high-throughput computing
(DHTC) services was identified by the National Academies of Science
(NAS)  2016 report on NSF Advanced Computing Infrastructure: {\em
  Increased advanced computing capability has historically enabled new
  science, and many fields today rely on high-throughput computing for
  discovery}~\cite{NAP21886}. HEP in general, and the HL-LHC science
program in particular, already relies on DHTC for discovery; we expect
this to become even more true in the future. While we will continue to
use existing facilities for HTC, and similar future resources, we must
be prepared to take advantage of new methods for accessing
both ``traditional" and newer types of resources.

The OSG provides the infrastructure for accessing all different types
of resources as transparently as possible. Traditional HTC resources
include dedicated facilities at national laboratories and
universities. The LHC is also beginning to use allocations at 
national HPC facilities, (e.g., NSF- and DOE- funded leadership class
computing centers) and  elastic, on-demand access to commercial
clouds. It is sharing facilities with collaborating institutions in
the wider national and international community. Moving beyond
traditional, single-threaded applications running on x86
architectures, the HEP community is writing software to take advantage
of emerging architectures. These include vectorized versions of x86
architectures (including Xeon, Xeon Phi and AMD) and various types of
GPU-based accelerator computing. The types of resources being
requested are becoming more varied in other ways. Deep learning is
currently most efficient on specialized GPUs and similar
architectures. Containers are being used to run software reliably and
reproducibly moving from one computing environment to
another. Providing the software and operations infrastructure to
access scalable, elastic, and heterogeneous resources is an essential
challenge for LHC and HL-LHC computing and the OSG is helping to
address that challenge.

The software and computing leaders of the U.S. LHC Operations Program,
together with input from the OSG Executive Team, have defined a
minimal set of services needed for the next several years. These
services and their expected continued FTE levels are 
listed in Table~\ref{tab:osg} below. They are orthogonal to the \s2i2
R\&D program for HL-LHC era software, including prototyping. Their
focus is on operating the currently needed services. They include R\&D
and prototyping only to the extent that this is essential to support
the software lifecycle of the distributed DHTC infrastructure. The
types of operations services supported by the OSG for US-LHC fall into
six categories, plus coordination. \katznoteoff{maybe add
  numbers to the table and to the paragraphs that should match each
  row?} 

\begin{table}[ht]
\centering
\begin{tabular}{|l|l|l|l|l|}
\hline
Category & ATLAS-only & Shared ATLAS & CMS only & Total\\
         &            & and CMS      &     &     \\
\hline
Infrastructure software & 0.85 & 2.9 & 1.7 & 5.45\\
maintenance and integration & & & & \\ \hline
CVMFS service & 0.2 & 0.1 & 0.4 & 0.7\\ 
operation  & & & & \\ \hline
Accounting, registration, & 0.35& 0.3& 0.2& 0.85\\
monitoring & & & & \\ \hline
Job submission & 1.5 & 0.0 & 1.0 & 2.5 \\ 
infrastructure operations & & & & \\ \hline
Cybersecurity & 0.0 & 0.3 & 0.0 & 0.3 \\
infrastructure & & & & \\ \hline
Ticketing and & 1.0 & 1.2 & 1.0 & 3.2 \\
front-line support & & & & \\ \hline
Coordination & 0.0 & 0.5 & 0.0 & 0.5 \\ \hline\hline
{\bf Services Total} & 3.9 & 5.2 & 4.2 & 13.3\\ \hline\hline
Technology evaluation &  & 3.0 &  & 3.0 \\ \hline
\end{tabular}
\caption{OSG LHC Services (in FTEs), organized into six categories
  that are described in the text. Also shown at the bottom is the FTE
  level for the OSG technology evaluation area.}
\label{tab:osg}
\vspace{-1.0em}
\end{table}

\vskip 0.05in
\noindent{\bf Infrastructure software maintenance and integration}
includes creating, maintaining, and supporting an integrated software
stack that is used to deploy production services at compute and
storage clusters that support the HL-LHC science program in the U.S.\ and
South America. The entire software lifecycle needs to be supported,
from introducing a new product into the stack, to including updated 
versions in future releases that are fully integrated with all other
relevant software to build production services, to retirement of
software from the stack. The retirement process typically includes a
multi-year ``orphanage'' during which OSG has to assume responsibility
for a software package between the time the original developer
abandons support for it, and the time it can be retired from the
integrated stack  This is because the software has been replaced with
a different product or is otherwise no longer needed.
\vskip 0.05in
\noindent{\bf CVMFS service operations} includes operating three types
of software library infrastructures. Those that are specific to the
two experiments, and the one that both experiments share. As the bulk
of the application level software presently is not shared between the
experiments, the effort for the shared instance is smallest in
Table~\ref{tab:osg}. The shared service instance is also shared with
most, but not all other user communities on OSG.
\vskip 0.05in
\noindent{\bf Accounting, registration, and monitoring} includes
any and all production services that allow U.S.\ institutions to
contribute resources to WLCG.
\vskip 0.05in
\noindent{\bf Job submission infrastructure operations} is presently
not shared between ATLAS and CMS because both have chosen radically
different solutions. CMS shares its job submission infrastructure with
all other communities on OSG, while ATLAS uses its own set of
dedicated services. Both types of services need to be operated.
\vskip 0.05in
\noindent {\bf Cybersecurity infrastructure} US-ATLAS and US-CMS
depend on a shared cybersecurity infrastructure that includes software
and processes, as well as a shared {\it coordination with the
Worldwide LHC Computing Grid (WLCG)}. Both of these are also shared
with all other communities on OSG.
\vskip 0.05in
\noindent {\bf Ticketing and front-line support} The OSG operates a
ticketing system to provide support for users and individual
sites, including feature requests and handling issues related to
security, wide-area networking, and installation and configuration of the
software. The OSG also actively tracks and pushes to resolution issues
reported by the WLCG community by synchronizing their respective
problem ticket systems.

\katznoteoff{there were 7 rows in the table, but only 4 paragraphs here,
  and then there's the next paragraph that doesn't match the table.}

\vskip 0.05in
\noindent {\bf Technology evaluation} In addition to these production
services, the OSG presently includes a {\it technology evaluation}
area that comprises 3 FTE. 
\katznoteoff{why isn't this in the table?}
This area provides OSG with a mechanism for medium- to long-term
technology evaluation, planning and evolution of the OSG software
stack. It includes a blueprint activity that OSG uses to engage
with computer scientists on longer term architectural discussions
that sometimes lead to new projects that address functionality or
performance gaps in the software stack. Given the planned role of
the \s2i2 as an intellectual hub for software and computing
(see Section~\ref{sec:role}), it could be natural for this part of 
the current OSG activities to  reside within a new Institute. Given
the operational nature of the remainder of current OSG activities, and
their focus on the present and the near future, it may be more
appropriate for the remaining 13.3 FTE to be housed in an independent
{\em but} collaborating project.

The full scope of whatever project houses OSG-like services in support
of the LHC experiments moving forward, in terms of domain sciences,
remains ill-defined. The OSG project has demonstrated that a single
organization whose users span many different domains and
experiments provides a valuable set of synergies and
cross-fertilization of tools, technologies and ideas. The DHTC
paradigm serves science communities beyond the LHC experiments,
communities even more diverse than those of HEP. As clearly identified
in the NAS NSF Advanced Computing Infrastructure
report~\cite{NAP21886}, {\em many fields today rely on
  high-throughput computing for discovery}.
\katznoteoff{though many of them use commercial clouds, not OSG...} We
encourage the NSF to develop a funding mechanism to deploy and
maintain a common DHTC infrastructure for HL-LHC as well as LIGO, DES,
IceCube, and other current and future science programs.

\tempnewpage

\subsection{Backbone for Sustainable Software}
\label{sec:backbone}


In addition to enabling technical advances, the Institute must also focus 
on how these software advances are communicated to
and taken up by students, researchers developing software (both
within the HEP experiments and outside), and members of the general
public with scientific interests in HEP and big data and software. 
The Institute will
play a central role in elevating the recognition of software as a critical 
research cyberinfrastructure within the HEP community and beyond. 
To enable this elevation,
we envision a ``backbone'' activity of the Institute that focuses
on finding, improving, and disseminating best practices; determining
and applying incentives around software; developing, coordinating
and providing training; and making data and tools accessible by and
useful to the public.

The experimental HEP community is unique in that the organization of 
its researchers into very large experiments results in significant community
structure on a global scale. It is possible within this structure to explore 
the impact of changes to the software development processes with concrete 
metrics, as much of the software development is an open part of the 
collaborative process. This makes it a fertile ground both for study and for 
concretely exploring the nature and impact of best practices. 
This large community also provides the infrastructure to conduct surveys and 
interviews of project personnel to supplement the metrics with subjective and 
qualitative evaluations of the need for and the changes to software process. An
Institute Backbone for Sustainable Software, with a mandate to pursue
these activities broadly within and beyond the HEP community, would be
well placed to leverage this community structure.

\vskip 0.1in
\noindent{\bf Best Practices:} The Institute should document, 
disseminate, and work towards community adoption of the best practices (from 
HEP and beyond) in the areas of software sustainability, including topics
in software engineering, data/software preservation, and reproducibility. 
Of particular importance are best practices surrounding the modernization of 
the software development process for scientists. Individual experts can 
improve the technical performance of software significantly (sometimes by more than an order of magnitude) by understanding the 
algorithms and intended optimizations and 
providing advice on how to achieve the best performance.
Surveys and interviews of HEP scientists can provide the information need to elicit and document best practices as well as to identify the area still in need of improvement.
The Institute can improve the overall process so that the quality of 
software written by the original scientist author is already optimized. In some cases
tool support, including packaging and distribution, may be be an integral
part of the best practices.
Best practices should also include the use of testbeds for validation
and scaling. This is a natural area for collaboration between the
Institute and the LHC Ops programs: the Institute can provide the
effort for R\&D and capabilities while the Ops programs can provide
the actual hardware testbeds.  
The practices can be disseminated in general outreach to the HEP 
software development
community and integrated into training activities. The Backbone can
also engage in planning exercises and modest, collaborative efforts
with the experiments to lower the barrier to adoption of these
practices.

The Institute should also leverage the experience of the wider
research community interested in sustainable software issues,
including the NSF SI2 community and other \s2i2 institutes
including the recently recommended SI2-S2I2 Conceptualization: Conceptualizing a US Research Software Sustainability Institute (URSSI),
the Software Sustainability Institute
in the UK~\cite{SSIUK}, the HPC centers, industry, open source software communities, and other organizations and 
adopt this experience for the HEP community.
In particular, URSSI and the UK SSI work in the wider research space and seek to work with focused domain communities that bring in particular challenges. These wider institutes can then generalize these challenges and attempt to solve them (or at least make progress in solving them). The solutions can then be applied back to the domain communities to both help them and allow them to learn from these initial applications.  

The Institute should also collaborate 
with empirical software engineers and external
experts to {\it (a)} study HEP processes and suggest changes and improvements
and {\it (b)} develop activities to deploy and study the implementation
of these best practices in the HEP community. 
These external
collaborations may involve a
combination of unfunded collaborations, official partnerships, (funded) 
Institute activities,
and potentially even the pursuit of dedicated proposals and projects.
The Institute should provide the fertile ground in which all of
these possibilities can grow.

\vskip 0.1in
\noindent{\bf Incentives:} The Institute should also play a role in developing 
incentives within the HEP community for {\it (a)} sharing software 
and for having your software used (in discoveries, by others building off it), 
{\it (b)} implementing best practices (as above) and {\it (c)} valuing research software 
development as a career path. This may include defining metrics regarding
HEP research software (including metrics related to software productivity 
and scientific productivity) and publicizing them within the HEP community. It
could involve the use of blogs, webinars, talks at conferences, or
dedicated workshops to raise awareness and to publicize useful software
development practices used within the institute. Most importantly, the Institute
can advocate for use of these metrics in hiring, promotion, and tenure 
decisions at Universities and laboratories. To support this, the Institute 
should create sample language and circulate these to departments and to
relevant individuals.


\tempnewpage

\section{Institute Organizational Structure and Evolutionary Process}
\label{sec:orggov}

During the \s2i2 conceptualization process, the U.S.\ community had a 
number of discussions regarding possible management and governance structures.
In order to organize these discussions, it was agreed that the
management and governance structures chosen for the Institute should be
guided by answers the following questions: 

\begin{enumerate}
\item {\bf Goals:} What are the goals of the Institute?
\item {\bf Interactions:} Who are the primary clients/beneficiaries of the Institute? How are their interests represented? How can the Institute align its priorities with those of the LHC experiments? 
\item {\bf Operations:} How does the Institute execute its plan with the resources it directly controls? How does the Institute leverage and collaborate with other organizations? How does the Institute maintain transparency?
\item {\bf Metrics:} How is the impact of the Institute evaluated? And by whom?
\item {\bf Evolution:} What are the processes by which the Institute’s areas of focus and activities evolve? 
\end{enumerate}

\begin{figure}[ht]
\begin{center}
\includegraphics[width=0.9\textwidth]{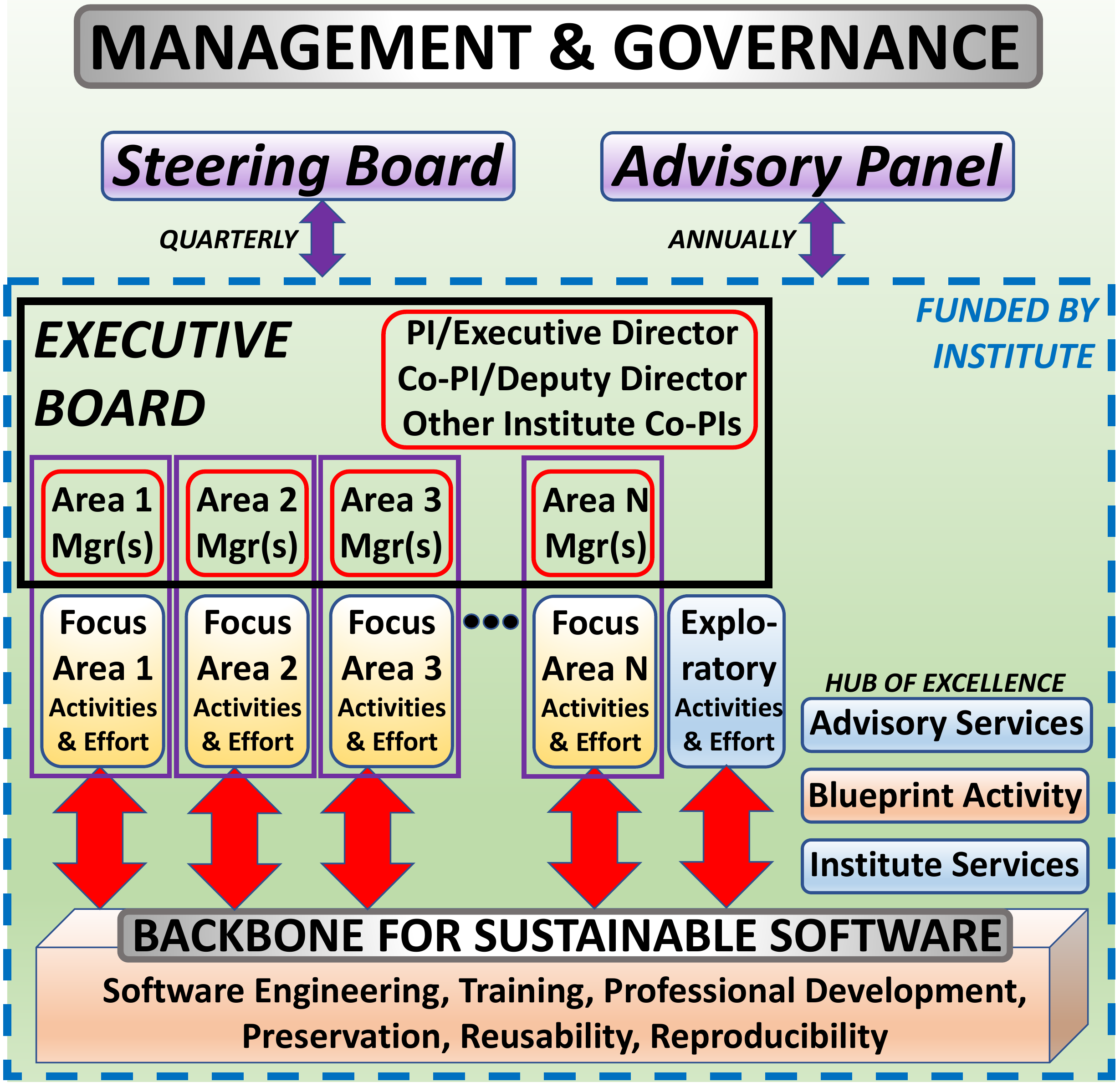}
\caption{Baseline Model for Institute Management and Governance.
}
\label{fig:orgmgmt}
\end{center}
\end{figure}

The \s2i2 discussions converged on the baseline model as described in 
Figure~\ref{fig:orgmgmt}. The specific choices may evolve in  an
eventual implementation phase depending on funding levels, specific
project participants, etc., but the basic functions here are expected
to be relevant and important.

\noindent The main elements in this organizational structure and their
roles within the Institute are:

\vskip 0.10in
\noindent{\bf PI/co-PIs:} The PI/co-PIs on an eventual Institute
implementation proposal will have project responsibilities as defined
by NSF.

\vskip 0.10in
\noindent{\bf Focus Areas:} A number of Focus Areas will be defined
for the Institute at any given point in time. These areas will
represent the main priorities of the Institute in terms of activities
aimed at developing the software infrastructure to achieve the mission
of the Institute. The \s2i2-HEP conceptualization process has
identified a initial set of high impact focus areas. These are
described in Section~\ref{sec:focusareas} of this document. The number
and size of focus areas which will be included in an Institute
implementation will depend on funding available and resources needed
to achieve the goals. The areas could also evolve over the course of
the Institute, but it is expected to be typically between three and
five. Each focus area within an Institute will have a written set of
goals for the year and corresponding Institute resources. The active
focus areas will be reviewed together with the Advisory Panel
once/year and decisions will be taken on updating the list of areas an
their yearly goals, with input from the Steering Board.

\vskip 0.10in
\noindent{\bf Area Manager(s):} Each Area Manager will manage the
day to day activities within a focus area. It is for the moment
undefined whether there will be an Area Manager plus a deputy,
co-managers or a single manager. An appropriate mix of HEP, Computer
Science and representation from different experiments will be a
goal. 

\vskip 0.10in
\noindent{\bf Executive Board:} The Executive Board will manage the
day to day activities of the Institute. It will consist of the PI,
co-PIs, and the managers of the focus areas. A weekly meeting will be
used to manage the general activities of the Institute and make
shorter term plans. In many cases, a liaison from other organizations
(e.g. the US LHC Ops programs) would be invited as an ``observer'' to
weekly Executive Board meetings in order to facilitate transparency
and collaboration (e.g.\ on shared services or resources).

\vskip 0.10in
\noindent{\bf Steering Board:} A Steering Board will be defined to
meet with the executive board approximately quarterly to review the
large scale priorities and strategy of the Institute. (Areas of
focus will also be reviewed, but less frequently.) The steering
board will consist of two representatives for each participating
experiment, representatives of the US-LHC Operations programs, plus 
representatives of CERN, FNAL, etc. Members of the
Steering Board will be proposed by their respective organizations
and accepted by the Executive Director in consultation with the
Executive Board.

\vskip 0.10in
\noindent{\bf Executive Director:} An Executive Director will manage
the overall activities of the Institute and its interactions with
external entities. In general day-to-day decisions will be taken
by achieving consensus in the Executive Board and strategy and
priority decisions based on advice and recommendations by the
Steering and Executive Boards. In cases where consensus cannot be
reached, the Executive Director will take a final decision. A Deputy
Director will be included in the Institute organization, to assume
duties of the Executive Director during periods of unavailability to
ensure continuity of Institute operations.

\vskip 0.10in
\noindent{\bf Advisory Panel:} An Advisory Panel will be convened
to conduct an internal review of the project once per year. The
members of the panel will be selected by the PI/co-PIs with input
from the Steering Board. The panel will include experts not otherwise
involved with the Institute in the areas of physics, computational
physics, sustainable software development and computer science.

\tempnewpage

\section{Building Partnerships}
\label{sec:partnerships}


\begin{figure}[htbp]
\begin{center}
\includegraphics[width=0.70\textwidth]{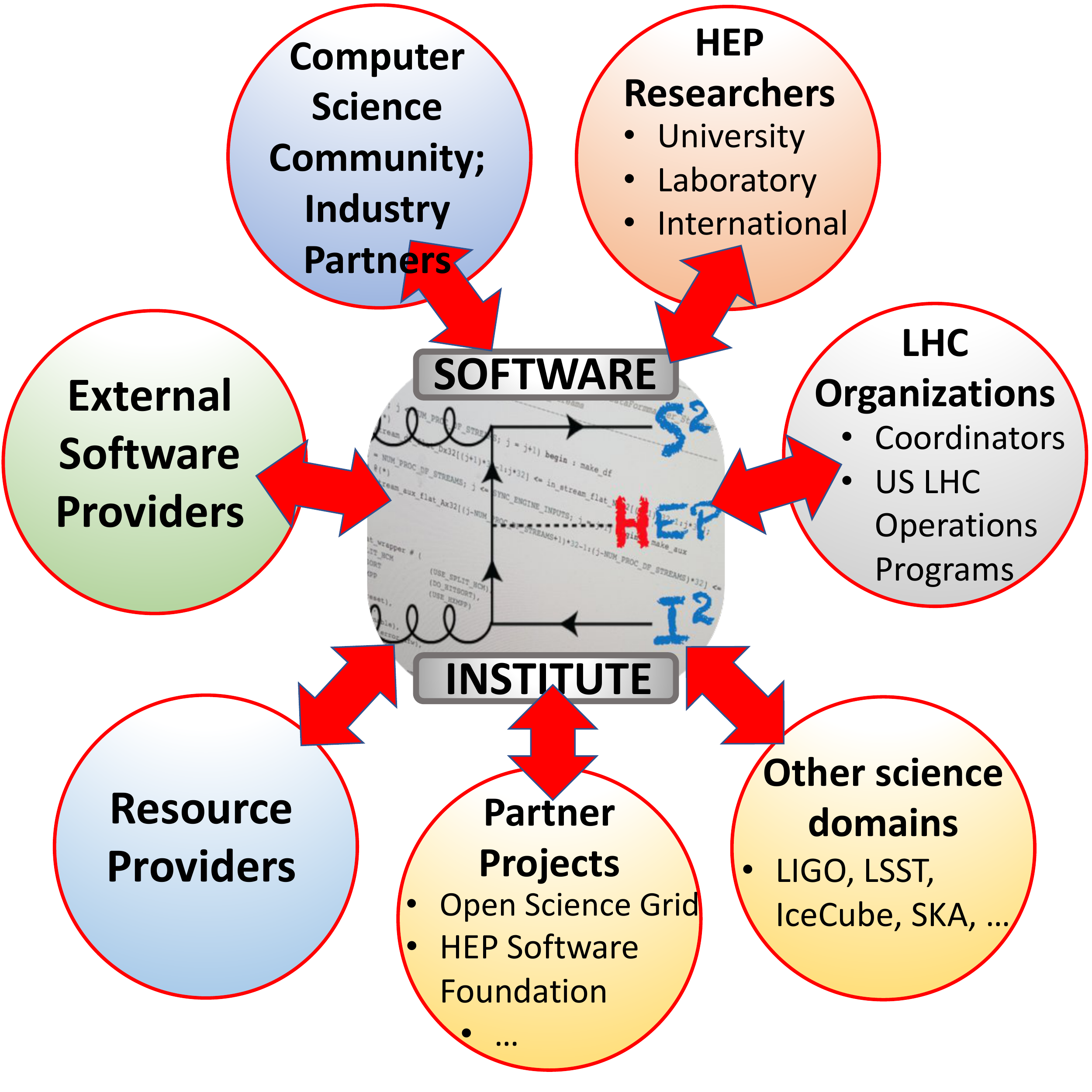}
\caption{Partners of the Institute.} 
\label{fig:s2i2_interactions}
\end{center}
\end{figure}

\subsection{Partners}
  The roles envisioned for the Institute in Section~\ref{sec:role} will require
collaborations and partnerships with external entities, as illustrated
in Figure~\ref{fig:s2i2_interactions}. These  include:

\vskip 0.1in
\noindent{\bf HEP Researchers (University, Lab, International):} LHC researchers are 
the primary repository of expertise related to all of the domain-specific software
to be developed and deployed; they also define many of the goals for domain-specific
implementations of more general types of software such as workflow management.
Areas in which collaboration with HEP researchers will be especially close
include
technical aspects of the detectors and their performance, algorithms for 
reconstruction and simulation, analysis techniques and, ultimately, the 
potential physics reach of the experiments. 
These researchers will define the
detailed and evolving physics goals of the experiments. 
They will participate in many of the roles described in 
Section~\ref{sec:orggov},
and some will be co-funded by the Institute. 
In addition, the Institute should
identify, engage and build collaborations with other non-LHC HEP researchers  
whose interests and expertise align with the Institute's 
focus areas.

\vskip 0.1in
\noindent{\bf LHC Experiments:} 
The LHC experiments, and especially the US-ATLAS and US-CMS collaborations,
are key partners.
In large measure, the success of an Institute will be judged in terms of
its impact on their Computing Technical Design Reports (CTDRs), to be submitted
in 2020, and its impact on software deployed (at least as a test for HL-LHC) 
during LHC's Run 3.
The experiments will play leading roles in understanding and defining
software requirements and how the pieces fit together
into coherent computing models with maximum impact on 
cost/resources, physics reach, and sustainability. 
As described in 
Section~\ref{sec:orggov}, representatives of the experiments will participate 
explicitly in the Institute Steering Board to help provide big-picture
guidance and oversight.
In terms of daily work, the engagement will be 
deeper. 
Many people directly supported by the Institute will be 
collaborators on the LHC experiments,
and some will have complementary roles in the
physics or software \& computing organizations of their experiments. 
Building on these
natural connections will provide visibility
for Institute activities within the LHC experiments, foster collaboration across experiments,
and provide a 
feedback mechanism from the experiments to the Institute at the 
level of individual researchers. 
The experiments will also be integral
to developing sustainability paths for software products 
they deploy that emerge  from
R\&D performed by the Institute;
therefore, they  must be partners starting early in the
software lifecycle.

\vskip 0.1in
\noindent{\bf US LHC Operations Programs:} 
As described in Section~\ref{sec:role}, the Institute will be 
an R\&D engine in the earlier phases of the software life cycle.
The Operations Programs will be one of the primary partners within the U.S.\ for 
integration activities, testing ``at-scale" on real facilities, and eventual 
deployment. 
In addition
they will provide a long run sustainability path for some elements of the 
software products. 
Ultimately, much of the software emerging from Institute efforts
will be essential for the LHC Operations Programs or run in facilities
they operate. 
The Institute will address many of
the issues that the Operations Programs expect to encounter
in the HL-LHC era. 
Thus, the Institute must have, within
the U.S., a close relationship to the Operations Programs. 
Their representatives
will serve as members of the Steering Board,
and  they will be invited to  participate in Executive Board
meetings as observers.

\vskip 0.1in
\noindent{\bf Computer Science (CS) Community:} During the \s2i2-HEP 
conceptualization process we ran two workshops focused on 
how the HEP and CS communities could work to their mutual benefit
in the context of an Institute, and, also, more generally.
We identified some specific areas for collaboration, and others
where the work in one field can inform the other.
Several joint efforts have started as results of these conversations.
More importantly, we discussed the challenges, as well as the opportunities,
in such collaborations, and established a framework for continued exchanges.
For example, we discussed the fact that the computer science research 
interest in a problem is often to map specific concrete problems to more 
abstract problems whose solutions are of research interest, as opposed 
to simply providing software engineering solutions to the concrete problems.
This can nonetheless bring intellectual rigor and new points of view
to the resolution of the specific HEP problem, and the HEP domain
can provide realistic environments for exploring CS solutions at-scale, but 
it is very important to keep in mind the differing incentives of the two 
communities for collaboration.
We anticipate direct CS participation in preparing a proposal if there
is a solicitation, and collaboration in Institute
R\&D projects if it comes to fruition.
Continued dialogue and engagement with
the CS community will help assure the success of the Institute.
This may take the form of targeted workshops focused on
specific software and computing issues in HEP and their 
relevance for CS, or involvement of CS researchers in blueprint activities
(see below).
It may also take the form of joint
exploratory projects. 
Identified topics of common interest include:
science practices \& policies, sociology
and community issues; machine learning; software life cycle; software
engineering; parallelism and performance on modern processor
architectures, software/data/workflow preservation \& reproducibility,
scalable platforms; data organization, management and access; data
storage; data intensive analysis tools and techniques; visualization;
data streaming; training and education; and professional development
and advancement.
We also expect that one or two members of the CS and Cyberinfrastructure communities
will serve on  the
Institute Advisory Panel, as described in Section~\ref{sec:orggov}, to provide
a broad view of CS research.

\vskip 0.1in
\noindent{\bf External Software Providers:} The LHC experiments depend on
numerous software packages developed by external providers, both within
HEP and from the wider open source software community. For the non-HEP
software packages, the HEP community interactions are often a bit diffuse
and unorganized. The Institute could play a role in developing the 
collaborations with these software providers, as needed, including
engaging them for relevant planning, discussions regarding
interoperability with other HEP packages, and software packaging and performance
issues. For non-HEP packages the Institute can also play a role in
introducing key developers of these external software packages to
the HEP community. This can be done through invited seminars or sponsored 
visits to work at HEP institutions or by raising the visibility of 
HEP use cases in the external development communities. Examples of these
types of activity
can be found in the ``topical meetings'' being organized by the DIANA-HEP
project~\cite{diana,dianaindico}.

\vskip 0.1in
\noindent{\bf Open Science Grid (OSG):} The strength of the OSG
project is its fabric of services that allows the integration
of an at-scale globally distributed computing infrastructure for
HTC that is fundamentally elastic in nature, and can scale out
across many different types of hardware, software, and business
models. It is the natural partner for the Institute on all aspects
of ``productizing'' prototypes, or testing prototypes at scale. For example,
the OSG already supports machine learning environments across a range of
hardware and software environments. New environments
could be added in support of the ML focus area. It is also a natural
partner to facilitate discussions with IT infrastructure providers,
and deployment experts, e.g. in the context of the DOMA and Data
Analysis Systems focus areas. Because of its strong connections to
the computer science community, the OSG also may also provide
opportunities for engaging computer scientists (as described
above) in other areas of interest to the Institute.

\vskip 0.1in
\noindent{\bf DOE and the National Labs:} The R\&D roadmap outlined in
the Community White Paper~\cite{HSF2017} is much broader than what will
be possible within an Institute. 
The DOE labs will necessarily
engage in related R\&D activities both for the HL-LHC and for 
the broader U.S.\ HEP program in the 2020s. 
Many DOE lab personnel participated in 
both the CWP and \s2i2-HEP processes. 
In addition, a dedicated workshop was held
in November 2017 to discuss how {\s2i2}- and DOE-funded efforts
related to 
HL-LHC upgrade software R\&D might be aligned to provide for
maximum coherence (see Appendix~\ref{workshoplist}). 
Collaborations between university personnel and national laboratory personnel
will be critical, as will be collaborations with foreign partners.
In particular, the 
HEP Center for Computational Excellence (HEP-CCE)~\cite{HEPCCE}, a DOE 
cross-cutting initiative
focused on preparations for 
effectively utilizing DOE's future high performance computing (HPC) facilities, 
and the R\&D projects funded as
part of DOE's SciDAC program are critical elements of the HL-LHC software
upgrade effort.
While \s2i2 R\&D efforts will tend to be complementary, 
the Institute will establish
contacts with all of these projects 
and will use the blueprint process (described below) to establish
a common vision of how the various efforts align into a coherent set
of U.S.\ activities.

\vskip 0.1in
\noindent{\bf CERN:} As the host lab for the LHC experiments, CERN must
be an important collaborator for the Institute. 
Two entities within 
CERN are involved with software and computing activities. 
The IT department
is  focused on computing infrastructure and hosts
CERN openlab (for partnerships with industry, see below). 
The Software (SFT)
group in the CERN Physics Department develops and supports critical software
application libraries relevant for both the LHC experiments and the HEP 
community at large, most notably the ROOT analysis framework and the
Geant4 Monte Carlo detector simulation package. There are currently
many ongoing collaborations between the experiments and U.S.\ projects
and institutions with the CERN software efforts. CERN staff from these
organizations were heavily involved the CWP process. 
The Institute will
naturally build on these existing relationships with CERN. A representative
of CERN will be invited to serve as a member of the Institute Steering Board, as described
in Section~\ref{sec:orggov}.

\vskip 0.1in
\noindent{\bf The HEP Software Foundation (HSF):} The HSF was established 
in 2014 to facilitate international coordination and common efforts in 
high energy physics 
(HEP) software and computing. Although a relatively
new entity, it has already demonstrated its value.
Especially relevant for the \s2i2 conceptualization project,
it organized the broader roadmap process leading to the
parallel preparation of  the Community White Paper. 
This was a collaboration
with our conceptualization project, and we expect that 
the Institute will naturally partner with the HSF in future roadmap
activities.
Similarly, it will work under the HSF umbrella to sponsor relevant workshops
and coordinate community efforts to share information and code.

\vskip 0.1in
\noindent{\bf Industry:} Partnerships with industry are particularly important.
They allow R\&D activities to be informed by technology developments in the 
wider world and, through dedicated projects, to inform and provide feedback
to industry on their products. HEP has a long history of such collaborations
in many technological areas including software and computing. Prior experience
indicates that involving industry partners in actual 
collaborative projects is far more effective than simply inviting them for 
occasional one-way presentations or training sessions. 
There are a number of projects underway today with industry partners.
Examples include collaboration with Intel like the Big Data Reduction Facility~\cite{CMSBIGDATA}, through an Intel Parallel Computing Center~\cite{IPCCROOT}, with Google~\cite{HEPCLOUD,HEPCLOUDDPF} and AWS~\cite{HEPCLOUD,HEPCLOUDAWS,HEPCLOUDDPF} for cloud computing, etc. A variety of areas will be of interest going forward,
including processor, storage and networking technologies, tools for data
management at the Exabyte scale, machine learning and data analytics, computing
facilities infrastructure and management, cloud computing and software
development tools and support for software performance.

In 2001 CERN created a framework for such public-private partnerships with
industry called CERN openlab~\cite{OPENLAB}. Initially this was used to build
projects between CERN staff and industry on HEP projects, however in recent 
years the framework has been broadened to include other research 
institutions and scientific disciplines. Fermilab recently joined the
CERN openlab collaboration and Princeton University is currently finishing the
process to join. Others may follow. 
CERN openlab can also be leveraged by the Institute to build 
partnerships with industry and to make them maximally effective. This can
be done in addition to direct partnerships with industry.

\vskip 0.1in
\noindent{\bf Other domain science areas and projects:} The Institute should 
also play an active role in building relationships with key individuals and
groups in the software \& computing areas of other scientific
domains. These relationships can help identify commonalities and
possibilities for integration across the different fields, as well
as opportunities for common research and development activities.
For example, one successful workshop that the \s2i2 project ran
together with the Flatiron Institute focused on ``Data Organisation,
Management and Access (DOMA) in Astronomy, Genomics and High Energy
Physics'' (see Appendix~\ref{workshoplist}). 
The Institute can
foster interactions of the LHC software \& computing community
with those of other large U.S.\ and international big science projects
such as IceCube, LIGO and LSST.
Similarly, it 
can help connect  the other NSF SI2
and OSG domain science communities (not only Computer Science) to HEP. 

\subsection{The Blueprint Process}
\label{sec:partnerships_blueprint}

\begin{figure}[htbp]
\begin{center}
\includegraphics[width=0.95\textwidth]{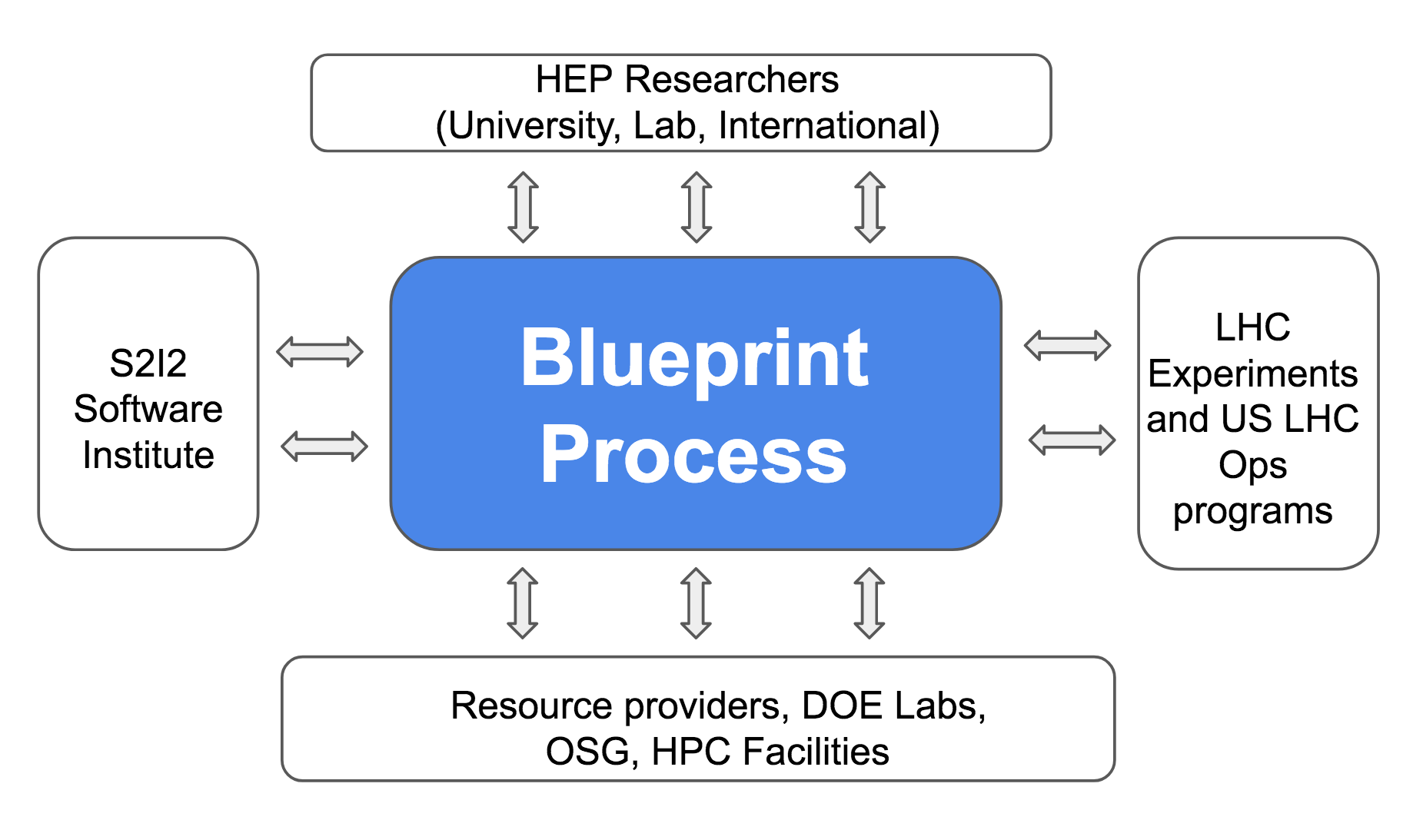}
\caption{The Blueprint Process will be a primary means of developing a common 
vision with the major partners.}
\label{fig:blueprint}
\end{center}
\end{figure}

To facilitate the development of effective collaborations with the 
various partners described above, the Institute should proactively engage and 
bring together key personnel for small ``blueprint'' workshops on specific 
aspects of the full R\&D effort. 
During these blueprint workshops the various partners will not only inform 
each other about the status and goals of various projects, but actively 
{\em articulate and document}
a common vision for how the various activities fit together into a coherent
R\&D picture. The scope of each blueprint workshop should be sized in
a pragmatic fashion to allow for convergence on the common vision, 
and some of the key personnel involved should have the means of realigning 
efforts within the individual projects if necessary.
The ensemble of these small blueprint workshops will be the process by which
the Institute can establish its role within the full HL-LHC R\&D effort. The
blueprint process will also be the mechanism by which
the Institute and its various partners can drive the evolution of the
R\&D efforts over time, as shown in Figure~\ref{fig:blueprint}.

Following the discussions at the November 2017 \s2i2-DOE workshop on 
HL-LHC R\&D, we expect that jointly sponsored blueprint activities between 
NSF and DOE activities relevant for HL-LHC, the US LHC Operations Programs
and resource providers like OSG will likely be possible.
All parties felt strongly that an active blueprint process would contribute 
significantly to the coherence of the combined U.S.\ efforts. 
The Institute could also play a leading role to bring other parties into
specific blueprint activities, where a formal joint sponsorship is 
less likely to be possible. This may include specific HEP and CS researchers,
other relevant national R\&D efforts (non-HEP, non-DOE, other NSF), 
international efforts and 
other external software providers, as required for the specific blueprint
topic. 

Blueprint activities will likely happen 3-4 times per year, typically with a
focus on a different specific topics each time. The topics will be chosen 
based on recognizing areas where a common vision is  
required for the coordination between partners. Input from the Institute 
managment, the Institute Steering Board and the management of various partner 
projects and key personnel will be explicitly solicited to identify potential
blueprint activities. 
The Institute will take an active role in organizing blueprint activities
by itself and jointly with its partners based on this input.
From year to year specific topics may be revisited.

\tempnewpage

\section{Metrics for Success (Physics, Software, Community Engagement)}
\label{sec:metrics}

The primary goal of the proposed Institute is doing,
and fostering, research and
development that leads to deployment of high impact software, as defined in 
Section~\ref{sec:focusareas}.
It should significantly advance the physics reach of the
HL-LHC experiments.
Because the Institute will exist within a larger context of international 
and national projects, a second major goal will be to
build a more cooperative, community process 
for developing, prototyping, and deploying sustainable software.
Metrics for success must address both of these goals.

As the HL-LHC experiments will start taking data almost a decade
from now, it will be impossible to directly determine how 
R\&D done in the next few years enables transformative 
science later.
Instead, it will be necessary to judge year-to-year progress
in achieving more concrete goals.

Within each Focus Area, specific metrics for funded activities
will be defined annually in terms of
projected impact.
For example, the first roadmap item identified in the 
Reconstruction and Trigger Algorithms Section (\ref{sec:recotrigger})
is enhanced vectorization programming techniques.
Appropriate questions to judge technical progress might include:
\begin{itemize}
 \item
   Has a targeted algorithm been made faster up by a factor of two
   (or four, or more)?
 \item
   Has a prototype event data model been designed to enable non-expert
   developers to write effective SIMD implementations of
   existing algorithms?
 \item
   Have new or substantially modified
   algorithms been developed to take advantage of
   SIMD capabilities of new architectures?
\end{itemize}
Metrics like these will need to be reviewed and revised regularly
as the state-of-the-art advances.
It is equally important to evaluate the level at which the Institute is 
engaging with and makes a significant impact on the LHC experiments. 
Appropriate questions to judge activities in a given Focus Area might include:
\begin{itemize}
 \item
  To what extent are the software development activities of the Focus Area aligned with priorities of the LHC experiments in that area to be ready for HL-LHC physics? How many experiments are impacted by the activities? What quantitative impact on cost/resource issues has been enabled by Institute activities?
 \item
  To what extent are the experiments directly collaborating in the 
  Focus Area projects (most importantly, U.S.\ university
  groups and the U.S.\ LHC Operations Programs)?
 \item
  To what extent are the results being disseminated to the experiments?
  Is the Institute conducting workshops to teach developers from the
  individual experiments to write faster, more performant code? How many
  people have participated in these workshops?
 \item
   To what extent are the larger LHC, HEP, and scientific 
   software worlds engaged?
   How are results being disseminated publicly?
   How many individuals not funded by the Institute are contributing to the 
   software?   
 \item 
   Which software developed by the Institute has been deployed in the
   experiments and how many users are using or affected by the software?
 \item 
   To what extent has the Institute improved the overall sustainability
   of the software ecosystem through the reduction of redundant community
   solutions and/or adoption of tools used by a wider community? 
\end{itemize}

\noindent And more generally for the Institute across development activities:
\begin{itemize}
 \item
  To what extent are the software development activities of the Institute aligned with priorities of the LHC experiments to be ready for HL-LHC physics, taking into consideration the criterion described in Section~\ref{sec:focusareas-rationale}? 
\item What fraction of the software development activities are impacting multiple experiments? Lead to utilizing common software? How is this fraction evolving over time?

\item What fraction of exploratory projects are sufficiently promising to lead to follow-on activities within the institute or beyond? How is this fraction evolving over time?
\end{itemize}

\noindent In terms of its role as an intellectual hub for  the
larger community effort in HEP software and computing,
it should define goals for each year in terms of
activities such as:
\begin{itemize}
 \item
   How many ``blueprint" workshops were organized for
   aligning and coordinating community efforts, and what were the outcomes? How many partners participated in the blueprint activity (and endorsed the 
outcomes)?
 \item
   Did the Institute help evaluate outside projects? How?
 \item
   Did the Institute provide outside projects with significant
   support?
   This could include help with software engineering, packaging,
   access to resources, etc.
\end{itemize}

\noindent In terms of training, education and outreach, it should define goals and activities as the Institute evolves with consideration of metrics such as:
\begin{itemize}
 \item
   How many training sessions such as summer schools did the Institute sponsor, and how many 
   students participated in each?
   Especially in this case, it will be important to report
   on the diversity of participants in terms of under-represented
   populations and level of seniority.
 \item
   How many HEP software developers are being actively mentored by experts partnering with the Institute?
 \item
   Are those developing software within the Institute visible within the experiments and receiving credit (e.g. citation, conference talks) for their work?
 \item
   Are early-career scientists working with the Institute on solid trajectories toward more permanent positions in academia or industry? What factors are behind success stories in term of professional development? Factors behind less-than-sucessful cases?
\end{itemize}

In each case, the Steering Board should work closely with the Executive
Board to define goals for the forthcoming year, and should review
progress at a fine-grained level on a rolling basis.
In addition, the Advisory Panel should review and evaluate progress
on a coarser-grained basis annually.
It should also judge whether proposed goals and metrics for the
forthcoming year are appropriate.

\tempnewpage

\section{Training, Workforce Development and Outreach}

\label{sec:training}

\subsection{The HEP Workforce}

People are the key to successful software.  Computing hardware
becomes obsolete after 3 -- 5 years.  Specific software implementations
of algorithms can have somewhat longer (or shorter) lifetimes.
Developing, maintaining, and evolving algorithms and implementations
for HEP experiments can continue for many decades.  Using the LEP
tunnel at CERN for a hadron collider like the LHC was first considered at a
workshop in 1984; the ATLAS and CMS collaborations submitted letters
of intent in 1992; the CERN Council approved construction of the
LHC in late 1994, and it first delivered beams in 2009.  A decade
later, the accelerator and the detectors are exceeding their design
specifications, producing transformative science.  The community
is building hardware upgrades and planning for an HL-LHC
era which will {\em start} collecting data circa 10 years from
now, and then acquire data for at least another decade.  People,
working together, across disciplines and experiments, over several
generations, are the real cyberinfrastructure underlying sustainable
software. Investing in people through training over the course of
their careers is a vital part of supporting this human facet of scientific
research. Training should include scientists and engineers at all stages 
of their careers, but should take particular care to invest in the young 
students and postdocs who will be faculty leaders driving the research
agenda in the HL-LHC era.

HEP algorithms and their implementations are designed and written
by individuals with a broad spectrum of expertise in the underlying
technologies, be it physics, or data science, or principles or
computing, or software engineering.  Almost all Ph.D. students write
analysis software, as do most post-docs.  Many students and post-docs
write software to acquire data, calibrate and reconstruct it, and
reduce data sets to sizes manageable for analysis by teams and
individuals.  Some of these people have very high levels of domain
and software engineering expertise, and some are raw recruits.  For
example, most experiments have dedicated teams for developing and
maintaining code for tracking charged particles.  The most senior
members of these teams generally have many years of experience and
have developed deep understandings of the current algorithms and
their performances, both in terms of physics performance and resource
usage.  This wisdom in passed along in a somewhat unorganized way
through what amounts to an unofficial apprenticeship program.  

In addition, teams of ``core'' developers are responsible for designing
and implementing infrastructure software such as application
frameworks and tools for workflow and workload management.
These individuals are often responsible for managing use of these tools
to run what are often commonly ``central productions" of reconstruction,
data reduction, and simulation campaigns.  Members of these teams are
considered software professionals, although many have been formally
trained in HEP rather than computer science or software engineering.
Matching the educational and training opportunities to the needs
of the various levels of software developers across the full spectrum
of the community will require carefully assessing what skills and
expertise will have the biggest impact on physics.  In addition,
as most people earning Ph.D.'s in experimental particle physics
eventually leave the field, providing educational and training
opportunities that prepare them for other career trajectories must
be a consideration in setting priorities.

\subsection{Current Practices}

Training support for software-related activities in HEP is uneven and made 
up of a patchwork of training activities with some significant holes. Although 
most universities do provide some relevant computer science and software
engineering courses, and many are starting to provide introductory ``data 
science'' courses, many HEP graduate students and postdocs are not
required to take these classes as part of the curriculum. As students
enter the research phase of the graduate student training, many
recognize the value of such classes, but are no longer in a position
to easily take the classes. No ``standard'' recommendations exist
for incoming students, either for HEP experiments or the HEP field
as a whole. Some universities are developing curricula for STEM
training in general and/or ``certificate'' programs for basic data
science and/or software training, but these are by no means yet universal.
The result is that the graduate student and postdoc population has a
very diverse spectrum of  relevant skills.

HEP collaborations do typically provide opportunities for members to
learn the software tools developed by and/or used within the experiments. 
For example, the week-long CMS Data Analysis School (CMSDAS)~\cite{CMSDAS,CMSDAS2,CMSDAS3} 
pairs software experts with new collaborators to build and run end-to-end 
examples of real analysis applications. Similarly, LHCb has a training
program and workshops called the ``Starter Kit''~\cite{lhcb-starter-kit}.
Since the beginning of LHC data taking, the ATLAS collaboration has
maintained an ``ATLAS Analysis Workbook'' designed to provide
information and examples for new (and experienced) ATLAS scientists
doing physics analysis. 
Other collaborations have similar programs. 
The goals of these programs
are primarily to make new collaborators effective {\em users} of the
complex experiment software ecosystems, rather than effective developers of 
that ecosystem, even if the latter will be often an important part of their
eventual research contribution. In addition these programs need to train
collaborators with very uneven backgrounds in basic ideas of computer
science and software engineering, as described above.

A number of summer schools focused on more advanced software and computing 
topics also exist in the global HEP community. These include, among others,
the CERN School 
of Computing~\cite{CERN-CSC}, the GridKa school~\cite{GridKa-School} in
Germany organized by the Karlsruhe Institute of Technology, the ``Developing
Efficient Large Scale Scientific Applications (ESC)''~\cite{ESC17},
school organized by the Istituto Nazionale di Fisica Nucleare (INFN)
in Italy and (more recently) the ``Computational and Data Science for High
Energy Physics (CoDaS-HEP)'' school~\cite{CODASHEP} in the U.S. 
Similarly, the laboratories also organize some ``short-course'' training
activities. For example, the LHC Physics Center (LPC) at Fermilab also offers 
half-day targeted training on specific topics.

Despite the universal need for computational skills for nearly all
HEP researchers, little exists to bring together all of the pieces of
an end-to-end training program accessible to all HEP students and postdocs, 
as well as more advanced training for more senior HEP researchers or 
specialists. In addition many of the individual HEP training efforts suffer 
from sustainability issues, even when they are successful.
In practice the Institute should not aim to create such a end-to-end
training program by itself, but rather to focus resources in two areas. First,
it can build alliances between the existing successful training HEP
and non-HEP programs and schools as well as the HEP experiment-specific 
training efforts. Within the U.S.\ it can also augment this by documenting and 
promoting a vision for how university courses, certifications and programs
can build the necessary base of skills. Second, it can fill in some of
the gaps between those efforts.

\subsection{Training Roadmap}

The highest impact role for the Institute regarding training will
be to {\em coordinate} training related activities and to assemble and
communicate a {\em coherent vision} of a training program for HEP
graduate students, postdocs and more senior researchers in software
and computing. It should not do this in isolation, but instead
develop a process for creating and updating that vision over time with 
the community. It can build a ``federated'' view over the possible
training opportunities in the experiments, at the labs, in dedicated
summer schools and from other sources (HEP and non-HEP). It can
bring together the people organizing those training activities not
only to articulate the vision, but also to develop plans to enhance
the sustainability, reusability and impact of the training activities.
The Institute itself does not need to organize all training activities 
directly, but could devote some of its resources to fill any gaps and in 
particular to help make a complete training program accessible for all 
U.S.\ graduate students and postdocs.
The training could be organized in three broad areas:


\vskip 0.1in
\noindent{\bf Basic skills training:} This should include topics such
as Unix, version control systems, basic software engineering and
programming (Python, C++, etc.). It may include an introduction to
basic elements of data science and related software tools. The Institute can 
work with University groups to document more clearly course possibilities for 
Ph.D.  students at the beginning of their graduate career. The Institute 
should also work with Software Carpentry~\cite{Software-Carpentry} and Data 
Carpentry~\cite{Data-Carpentry} to customize general and basic software 
training for new HEP students. Training examples can be adapted and
made HEP-specific when appropriate and the curriculum can be adapted to
what is needed in HEP. 

\vskip 0.1in
\noindent{\bf Training for active research:} This should include many
topics that are required for active research in HEP, both as users of, and
contributors to, to HEP software ecosystem. This may include more
sophisticated topics such as general computational and data science, 
introduction to the distributed computing cyberinfrastructure, and also 
HEP specific tools (ROOT, Geant4, etc.) This training should dovetail 
with the experiment-specific training offered on by each HEP experiment 
regarding its own software tools and software ecosystems.

\vskip 0.1in
\noindent{\bf Custom and Specialized Training:} No training curriculum will
ever be complete nor perfectly adapted to a given individual's research
needs. When relevant, the Institute should identify and promote opportunities 
for custom and/or specialized training on specific topics, technologies
and/or applications to fill any gaps. The natural targets of this training
will be advanced users and developers and software professionals working in 
the HEP environment, whether funded by the Institute or other entities.

The Institute should organize and/or enable its partners to provide training
via a variety of means, including in-person schools and short-courses, 
webinars and virtual training. It should also explore how new partners such as
the HPC centers, industry and the wider data science communities might
contribute to the training. Last, but not least, it should also organize
a community process for gathering feedback from users and developers on
the impact of ensemble of the training activities.

\subsection{Workforce Development in HEP}

Large HEP experiments organize themselves for global efforts in
many areas, including large detector construction projects, globally
distributed computing systems, software development involving hundreds
of researchers and, of course, extremely complex multi-faceted data
analysis activities. Actively developing the workforce needed to support
this endeavor is critical.

The Institute could also play a leadership role in HEP in workforce
development for software and computing.
One important aspect of workforce 
development is attracting young talent within the experiments to work
on the most important software and computing challenges. There is a
real opportunity to attract top young talent for
HEP software development given industry trends (especially rapidly increasing
interest in data analytics, machine learning and artificial
intelligence) and the manner in which HEP research provides a
recognized means for developing expertise in these areas.  

The training roadmap just described provides the foundation for
scientists to develop sustainable software for HEP that is impactful for
the HL-LHC, which is of course a core element of the \s2i2-HEP
mission. As previously mentioned, a related challenge is to bring in
new effort in the form of early-career scientists from universities
that bring in new ideas and talents. To achieve this, they (and their
advisors) need to believe that developing software within the Institute
provides some mobility toward their career goals. More succinctly, the
Institute needs to put in place a strong program of {\it professional
  development} to complement the training in software just described
(and also the training in physics research that students need to
complete their degrees and postdocs need to move on in their
careers). Elements of professional development within the Institute
could be organized in broad areas as follows:
\vskip 0.1in
\noindent{\bf Establishing expert mentors:} It will be important for
the Institute to recruit experts in scientific software and computing
not only for software development but also for professional
development through direct mentoring of early career scientists
producing software. This would involve the training aspects previously
described but additionally monitoring the early-career scientist's
progress, getting them to work with tools and techniques known within
the data science community, and helping them to develop contacts
both inside and outside of HEP, for example with Computer Science,
Data Science and/or industry partners (see Section~\ref{sec:partnerships}).
Early partnership
with Industry mentors could lead to better job prospects for some
after they move on in their career from the HEP domain, since they
would better align with Industry in terms of joint projects and 
the tools that are commonplace within Data Science circles. 
This would
help attract young talent and benefit HEP in addressing the HL-LHC
challenges. The same could be said of mentoring in the context of
academia, and faculty in HEP and Computer Science could partner with
the Institute in a number of ways, for example through sabbatical
support. There are also research faculty and scientific staff at
Universities and Centers (e.g. NSF-funded supercomputing centers such
as NCSA and UCSD) that are experts in advanced software and
computation who could provide strong mentors to HEP software
developers within the Institute.

\vskip 0.1in
\noindent{\bf Establishing a fellowship program:} The Institute could
provide named fellowships for young scientists, raising the
profile and visibility of software development activities in HEP. 
Institute Fellows
would have visibility within the broader HEP community as developing
software important for enabling HL-LHC physics. They would receive
active mentoring by one or more experts within the Institute with a
strong emphasis on professional development.

\vskip 0.1in
\noindent{\bf Developing methods for visibility of excellence:} An
important element of professional development for early-career
scientists aside from training and mentoring is ensuring that
excellent work they do is visible and receives proper credit for
facilitating the frontier science. To the end, the Institute should work
with the experiments in terms of policy and process on visibility and
credit and the CS community in current approaches to software citation. 
As a hub of excellence for HEP
scientific software, the Institute should also strive attract and support
HEP software developers by creating the conditions necessary for a
vibrant ecosystem with activities such as sponsored
seminars/colloquia, workshops, summer schools, newsletters, media
communications, etc. that highlight the work being done and inform
opportunities for future directions (see
Section~\ref{sec:partnerships_blueprint} for the Institute Blueprint Process).

\subsection{Outreach}

The expression ``outreach" in the context of scientific projects
is often used as an umbrella term for educational activities
aimed at targeted communities and various activities aimed to engage the general public. 
The LHC experiments and the U.S.\ university HEP groups
already have good track records in these activities,
and the Institute will work with them to expand efforts
that have strong software components.
In addition, we interpret outreach as other activities aimed at 
building bridges with other academic communities that may lead to broader impacts.
For instance, it may encompass elements 
similar to those described in Section~\ref{sec:partnerships}
aimed at building partnerships with Computer Science.
As a specific example, the Institute will work with the LHC
experiments to provide software, datasets and documentation for
challenging HEP applications as a bridge to direct collaboration with
the Computer Science and Data Science 
communities. They can then use this data for their own research in
specific areas -- an approach which was advocated by both HEP and
non-HEP domain scientists in a number of \s2i2 workshops on building
partnerships between HEP and Computer Science communities.
While the core team will undertake some of the outreach
activities itself (especially those related to packaging software or datasets,
as an example),
most will be undertaken in cooperation with partners in 
our Focus Area research.

Some of the specific outreach activities and considerations for the
Institute are: 
\begin{itemize}
\item
The LHC experiments introduce high school students throughout
the world to particle physics through the International
Masterclass program~\cite{int-masterclass}.
Many U.S.\ university groups participate.
As part of their activities, the students analyze data from
the LHC experiments.
The Institute should work with them to provide better tools to prepare,
process, and analyze the data. Tools like Jupyter notebooks and 
\textsc{Binder}~\cite{mybinder} allow for web-based analysis tools and remove
the burden of software installation. DIANA-HEP has initiated projects
using \textsc{Binder} for LIGO and Jupyter notebooks for particle physics~\cite{binderpress, jupyterClang}. 
\item 
Challenges are a successful format for engaging the data science and computer science communities. 
Challenges also engage citizen scientists.  The Higgs Kaggle~\cite{HiggsKaggle} challenge drew 
1,785 teams from around the world. These challenges require
substantial resources to organize, execute, and maximize impact
(e.g. through assessments). The Institute could provide the missing effort that is key to organize and manage successful challenges 
around specific topics of important for HL-LHC physics such as jet
tagging and charged-particle tracking. 

\item
Many U.S.\ university groups have QuarkNet~\cite{quarknet} programs that 
provide high school students with paid summer internships.
Most mentors and students judge these programs to be highly
successful.
The Institute should fund similar internships focused on activities
related to the Institute's research, especially in the areas of data analysis
and machine learning.
\item
Similarly, U.S.\ university groups often hire their own 
undergraduates for
summer research projects, and these sometimes continue into
the academic year.
The Institute should encourage our Focus Area research partners to provide
such opportunities in conjunction with our common projects,
and the Institute will provide funds to support these efforts.
\item
The DIANA-HEP project~\cite{diana} has an undergraduate and graduate fellowship program
that provides stipends, travel support, and subsistence payments
to students
for up to three months to enable them to work on projects
outside their home institutions.
Mentors can come from the immediate DIANA-HEP community
or from other institutions developing software related to
data-intensive analysis.
The Institute should provide similar undergraduate fellowships.
These will provide experiential learning opportunities 
for the students and advance our research program concurrently.
\item
The Institute should provide similar short-term fellowship opportunities to Masters and
Ph.D. students from fields outside HEP, but
with common interests (such as Computer Science, Data Science,
and Software Engineering) to work on projects of mutual interest.
\item
The Institute should work with the LHC experiments and CERN to develop gateways
for open data and {\em also} software that can be used by
non-experts to explore the types of measurements made in 
high energy physics. DIANA-HEP and DASPOS~\cite{daspos-project} 
have made great progress in
the development of these gateways for the core scientific community. 
The Institute could extend this effort towards outreach activities and help assure the
the gateways are designed with sustainability and continued availability of these resources
in mind.
\item
During the \s2i2 conceptualization process, we identified a number
of areas where HEP has unique datasets that might be interest to 
the Computer Science and Data Science communities, 
as well as of interest to individual researchers with whom we may partner.
Making such datasets available for general use through well-designed
portals (built in collaboration with the Science Gateways \s2i2~\cite{Sciencegateways2017})
will be done in conjunction with the individual experiments and
in conjunction with our Focus Area research partners.
The types of datasets identified so far include the
meta-data describing jobs executed on the grid
and input for machine learning problems where qualitatively
new approaches are necessary.
\end{itemize}

This list is not meant to be inclusive, but rather an illustration
of the types of outreach activities that would be appropriate.
During its annual review of the project, the Advisory Panel will
be asked to evaluate the previous  year's activities
in this area
and suggest ways to improve them in the forthcoming year.

\newpage

\section{Broadening Participation}
\label{sec:broadeningparticipation}

The participation of women and ethnic minorities is generally low
in the HEP world, and 
fractionally, it is even lower in the HEP Software and Computing (S\&C) 
world.
We estimate that fewer than 10\% of people in HEP S\&C are women 
while (from LHC
experiment statistics) between 13\% and 20\%
of the LHC experiments' collaborators are women. 
Nationally, 7.4\% of high-tech employment is black~\cite{GuardianSV}, 
while in HEP S\&C the fraction is negligibly small.
Looking forward, increasing the diversity of the HEP S\&C
workforce promises two types of benefits.
From first principles, the top 5\% of a larger pool should 
always be better than
the top 10\% of a pool half as large.
In addition, studies show~\cite{diversity1,diversity2,diversity3} that teams of people from diverse backgrounds are more innovative when 
crafting solutions to complex problems and can make better and more profitable 
decisions.

An \s2i2 will not significantly increase the fraction of under-represented
populations in HEP or in S\&C;
it will be too small a player.
However, the Institute must be sensitive to diversity in building its own teams,
and it can help build the pipeline by partnering with institutions
actively working to do this.
At the high school level, programs like QuarkNet engage diverse
groups of students.
At the undergraduate level, ``Women in Science and Engineering" programs
like those at the University of Michigan,
the University of Arizona, and the University of Cincinnati (as examples)
provide Research Experience for Undergraduates (REU) opportunities
targeting women. At University of Illinois at Urbana-Champaign's
National Center for Supercomputing Applications, the {\bf I}ncubating
a {\bf N}ew {\bf C}ommunity of {\bf L}eaders {\bf U}sing {\bf
  S}oftware, {\bf I}nclusion, Innovation, Interdisciplinary and {\bf
  O}pe{\bf N}-Science (INCLUSION)~\cite{INCLUSION} is a 10-week summer
REU program. INCLUSION provides an opportunity for 10 undergraduate
students from underrepresented communities and Minority Serving
Institutions to work in pairs with pairs of mentors on
interdisciplinary socially-impactful INCLUSION research projects that
develop and use open source software.

At the transition from undergraduate to Ph.D. student level, the 
American Physical Society's Bridge Program targets 
under-represented populations (self-identified,
so including first generation college, as an interesting example).
The Institute's outreach and education program can include 
supporting programs like these by providing both financial support
and opportunities to work with \s2i2 teams and their collaborators.
One model would be sponsoring undergraduate and graduate student
Fellowships, similar to those offered by the DIANA-HEP project~\cite{diana}.
In this case, one of the three Fellows who has already
completed their projects was a woman,
and one of the two lined up for early 2018 Fellowships is a woman.
In addition, the \s2i2 can work with groups like {\tt Data Carpentry} to 
organize workshops using HEP data to introduce high school and college students
to data science.

If there is an \s2i2 solicitation, the proposal should identify specific
models and partners for encouraging participation of under-represented
populations in its outreach and education program.
The proponents will need to reach out to institutions with
programs with track records of increasing diversity to
find out what works.
The NYU Material Research Science and Engineering Center
runs an REU program with 40\% minority student participation and
50\% women.
The University of Maryland, Baltimore Campus has become a center
for cultivating underrepresented minority scholarship and awareness 
in the math, science, and engineering disciplines.
Florida International University, which has a CMS group,
serves a student population which is predominantly Hispanic and
 85\% minority.
While increasing diversity will not be the primary goal of
the envisioned \s2i2,
devoting 2\% - 3\% of its resources to outreach and education
efforts  targeting the pipeline can have a
beneficial impact on diversity in the HL-LHC era.

\tempnewpage

\section{Long Term Sustainability}
\label{sec:sustainability}

Long term sustainability of the software ecosystem is particularly important 
for HEP, given that the HL-LHC and other facilities of the 2020s will be
relevant through at least the 2030s. The Institute should foster improved 
sustainability models not only for the software products it is involved in 
generating, but also more generally for the software ecosystem used by 
the LHC experiments and HEP in general. 

As described in Section~\ref{sec:s2i2role-sw}, the Institute should play a 
driving role in particular in the earlier stages of the software lifecycle. 
It should partner with other organizations (the experiments, the US LHC
operations programs, specific institutions) for the later elements of
the lifecycle, in particular with an eye to developing sustainability 
paths for the long run. The Institute Backbone (Section~\ref{sec:backbone})
and its Training activities (Section~\ref{sec:training}) will be key 
elements in working with the community to develop more sustainable 
software practices and skills from the ground up.

In addition, we can look more globally at the existing software ecosystem
and ask more generally which paths to greater sustainability might exist. 
Given the nature of the current LHC and HEP software ecosystem, two possible 
paths stand out as particularly relevant to the mission of the Institute:

\begin{itemize}
\item {\bf Identification and consolidation of redundant HEP-specific solutions:} for a number of historical and organizational reasons, many HEP software solutions are developed within the context of single experiments. In cases where the experiments actually have similar needs, this has led to multiple solutions to the same problem.
\item {\bf Adoption of solutions used by a wider scientific or open source community:} by moving to more widely used solutions the base of support for sustainability issues typically also becomes larger.
\end{itemize}

Both of these paths effectively boil down to increasing the size of the 
community using a given software element. Most software products cannot 
survive and thrive without {\em some} level of dedicated effort and 
``ownership'' by some institution or long running project. In cases
where increasing the size of the community does not significantly increase
the scope of the software, the increase effectively increases the impact
of effort invested. Concentrating available community effort
on a single solution may ultimately lead to better, more sustainable 
solutions.

Figure~\ref{fig:cmsswdevelopers} illustrates one such example. The CMS
experiment developed a number of scripts and web interfaces to build
a software integration workflow around the CVS source code version control
system to integrate software contributions from many distributed 
collaborators. At the time of their adoption, no general open source tool
provided this functionality. Other experiments (including ATLAS, LHCb 
and ALICE) faced similar problems and developed their own similar
tools, driven by the particular collaboration dynamics and evolving
needs over time. In practice these solutions implemented workflows are
not dissimilar from many software projects. Over time newer source
code version control systems like git appeared, along with tools like GitHub 
or gitlab to manage the relevant workflows. In 2013 CMS transitioned
its software development environment from CVS and the CMS-specific tools
to git and GitHub and the workflow tolls provided by the latter. The net
effect of this was to reduce the CMS- (and HEP-) dedicated effort 
required and to leverage efforts serving a much larger open-source
community. Because those tools serve a large community, the solutions
are both better (more feature rich) and exhibit better sustainability.
For example, adapting to the latest versions of the underlying web software 
(e.g. javascript, browsers) or operating system versions will happen 
without CMS or HEP intervention. 

Even if such transitions in the software ecosystem are ultimately beneficial
to the HEP community and particular experiments (by reducing required effort,
providing better solutions and/or improving sustainability) it should be
noted that the transition itself is not cost-free. There is often an 
``activation energy'' associated with the transition. For example, the
CMS ``CVS to GitHub'' required about 0.5 FTE of dedicated effort over 6-9 
months to prepare the change and orchestrate the transition with the 
community. 

The partnership between HEP and Computer Science can play a big part
in the identification and consolidation (and ultimately, reduction) of
redundant solutions to HEP-specific challenges. Recall the {\it
  Advisory Services} role described in Section~\ref{sec:role} where the
Institute will play an advisory role within the larger research software community
by providing technical and planning advice to other
projects and by participating in reviews. As new software projects are
being proposed or developed by individuals or experiments, the
Institute, with its critical mass of expertise through a network of
partnerships, could evaluate proposed software for algorithmic
essence, scalability, and redundancy with existing software and advise
on the best course of action that will lead to sustainable software
over the long run and make most efficient use of limited resources. 

The Institute should play a key role in the LHC and HEP community to drive
the overall software ecosystem towards more sustainable solutions. It can do 
this by {\em (a)} developing better sustainability models for software it is
involved with and {\em (b)} working with the community to evolve the
existing software ecosystem towards more sustainable solutions. In both
cases, explicit effort will be required.

\tempnewpage

\section{Risks and Mitigation}
\label{sec:RisksAndMitigation}

The primary goal of the envisaged \s2i2 is to enable the science
goals of the HL-LHC through software R\&D leading to deployment
of the requisite software by the experiments.
The risks are  {\em social}, {\em technical}, and {\em contextual}.
Those in the {\em social} category include risks related to:
(i) building and maintaining the \s2i2 team,
(ii) fully engaging in a coherent fashion with the larger HL-LHC software community, and
(iii) executing the R\&D plan successfully.
Those in the {\em technical} category include:
(i) slower improvement of hardware performance than anticipated,
(ii) less benefit from new features like parallelization and SIMD vectorization
     than anticipated, and
(iii) less benefit from Machine Learning than anticipated.
Those in the {\em contextual} category include:
(i) substantial changes to the hardware upgrade plans
for the accelerator and detector,
(ii) substantial changes to the upgrade software R\&D 
funding profiles by other agencies,
and
(iii) major scientific discoveries at the LHC, before the HL-LHC
era begins, that significantly change the physics priorities of
the experiments.
Each of these requires different specific mitigations, but all
require regular review of progress by \s2i2 management, 
the outside stakeholders,
and ``disinterested" external advisors 
coupled with the agility
to redirect resources.
\vskip 0.05in
{\em Building the Institute team} will be the first major challenge.
Subsequently maintaining an effective team will be a continuing challenge
that requires careful thought in advance, as well as continuing 
attention.
An Executive Director, and probably a Deputy Executive Director,
will lead the Executive Board.
The initial choices for these positions will be the responsibility of
the lead PI and co-PIs, probably taken in consultation with the NSF
while negotiating a Cooperative Agreement before an award is made.
The individuals selected for these roles will need to devote substantial
fractions of their professional effort to the Institute.
The Executive Director will almost certainly need to devote at least
50\% of his/her effort to the position.
In general, the Deputy Executive Director must be willing to serve
as Interim Executive Director should the occasion arise, and must
be willing to devote enough effort to the Institute to be ready
assume this role on short notice.
An initial team of Area Managers should be identified while negotiating
a Cooperative Agreement.  
The specific individuals can be identified only when the number
of Focus Areas to be supported is known.
In addition to their domain expertise, members of the Executive Board
should broadly represent the interests of the LHC experiments.
They should all have track records of collaborating effectively.
The activities of the core team and in each Focus Area will be formally
reviewed each year to prepare  annual Statements of Work (SOWs) to
be done. 
At this time, it will be appropriate to consider whether new Area
Managers (plus Co-Managers or Deputy-Managers) should be appointed.
These decisions will be taken by the Executive Board, in consultation
with the Steering Board.
Should the Executive Director or Deputy Executive Director step down
before the five-year term of the award ends, the lead PI and co-PIs
will select a replacement, in consultation with the Executive Board
and the Steering Board.

Building a team of approximately 20 FTE physicists and software professionals
to undertake the support and R\&D responsibilities of the Institute
will take time.
The number of highly qualified personnel with the requisite domain and
software expertise is limited.
The Institute will initially build on the existing software development
infrastructure by co-funding individuals whose other activities
complement those being undertaken by the Institute.
For example, it would be appropriate for someone already working
in DOMA as part of the ATLAS or
CMS operations team to continue that work half time and begin
to work on the software upgrade R\&D as a member of the Institute.
Similarly, someone who already provides continuous integration
services, packaging, etc., to an ongoing project could provide
similar services in support of Institute projects on a part-time,
co-funded basis.
While a certain level of finesse will be required to ensure that
individuals funded for different projects are splitting their
efforts appropriately, co-funding these people will provide
opportunities to build a sense of community across experiments
and help keep the Institute focussed on efforts of interest to
the experiments.
The highest priority for the first year will be hiring members of
the core team who are anticipated to continue as members of the
Institute for the full term of the award.
In parallel, the Area Managers for the R\&D Focus Areas
will identify a mix of post-docs, more senior physicists, and
software professionals with the expertise and interest to
advance their research programs.
As appropriate, these individuals can be co-funded or
hired directly by a University group, in conjunction with an
SOW.

Building a team is more complicated than hiring individuals to
do well-defined jobs -- in this case we want people to collaborate
with each other effectively, and also with the larger community.
This will require defining expectations for collaborative work,
and rewarding it meaningfully.
Code and documentation will be reviewed as part of the engineering
process;
we have observed that this both improves software products {\em and}
tends to build a sense of community, perhaps because it creates
joint responsibility and ownership.
Similarly, developers will be expected to present their work within
the Institute and also to the larger community.
Team progress and individual performance will be formally reviewed on an
annual basis.
Individuals will also be asked to prepared written 3-, 6-,
and 12-month goals and plans on a rolling basis, for less formal
discussions with their immediate supervisors.
This process will provide opportunities to laud excellent work
(which is generally expected) and identify the need for remediation,
when indicated.
Where appropriate, individuals will be members of LHC experiments
as well as members of the Institute.
In these cases, \s2i2 management will work with the experiments'
managements to make sure that \s2i2 efforts are explicitly recognized as
service work to the  experiments.
This will be especially important for students, post-docs, and other
more junior members of experiments who are expected to engage is a 
mix of service work and physics analysis as part of their professional
development.  

Some of the approaches to building and maintaining the team, discussed
above, also address a second key issue: 
{\em fully engaging with the larger HL-LHC software community}.
The Steering Board (discussed in Section \ref{sec:orggov})
will explicitly have representatives of all of the LHC experiments,
the US-LHC Operations programs as well as Fermilab and CERN.
They will review the large scale priorities and strategy of the Institute
quarterly, and provide advice on any changes of direction that
should be considered.
Just as this Strategic Plan has emerged from a community process, executed
in parallel with the broader CWP process,
the Institute will sponsor 
continuing blueprint activities, in conjunction with the
HEP Software Foundation, to update the roadmap
on a rolling basis and identify any changes in
priorities.
Additionally, members of the Institute will make 
presentations to the
individual experiments at different levels of technical detail.
At the finest level of granularity, presentations of specific
algorithms and implementations will presented to tools Working Groups.
At a coarser level of granularity,
projects will be presented and discussed during Software and Computing Weeks.
When appropriate, overview presentations will be made at general 
Collaboration Meetings.
In all cases, the goal will be two-way communication.

{\em Executing the R\&D plan successfully} will require 
that developers be technically strong, that they work
together collaboratively, and that they adhere to
good software engineering practices.
It will also require
careful attention
to short-term (and longer-term) goals by members of the Executive Board,
some of whom may be developers themselves.
Many of the software engineering practices 
described in Section~\ref{sec:backbone} are meant to help keep
projects on track and assure high quality.
As an example, requiring that all code be reviewed by
a second developer before a merge request is accepted
will help assure good documentation and correct algorithmic implementation.
As discussed in Section \ref{sec:orggov}, the EB will meet weekly to 
make short term plans to keep efforts properly focussed.
\vskip 0.05in
The magnitudes of the 
computing challenges described in Section \ref{sec:computing_challenges}
{\em assume} that CPU and mass storage performance per unit price
will  continue to grow at a rate equivalent to 20\% per year,
about a factor of 6 over a decade,
and that the funding profile will remain flat.
The experiments' needs are {\em another} factor of 5 greater,
given current algorithms' use of resources.
The purpose of the \s2i2 is to undertake a software
upgrade to provide the enhanced performance required use the
anticipated resources that much more effectively.
Should effective hardware costs drop more slowly than estimated,
or the hardware acquisition budgets drop,
the software goals may need to be revised.
At the moment, we want to find better algorithms to 
reconstruct, process, and analyze data with essentially
the same fidelity as is done today.
If this is not possible, the experiments will need to 
process as much data as possible with lower fidelity or less data
with greater fidelity.
The Institute, with advice from the Steering Committee
will need to adjust its goals and priorities accordingly.
Somewhat similarly, if new algorithms taking advantage
of vectorization and machine learning do not deliver
the anticipated improvements in performance in the next 
five years,
the HL-LHC experiments will need adapt their plans
to live with what is possible.
\vskip 0.05in
The  American philosopher Yogi Berra warned that
{\em It's tough to make predictions, especially about the
future}, 
and the Scottish poet Robert Burns observed 
that {\em The best laid schemes of mice and men
Go often askew} (as translated into modern English).
We recognize that both these insights apply to preparing
a software upgrade for deployment almost 10 years from
now.
Nonetheless, the plan for R\&D over the next 5 years
should be relatively robust.
Changes in the accelerator and detector upgrade
plans are most likely to produce quantitative,
not qualitative, changes in the computing and
software models for the HL-LHC.
Similarly, changes in physics priorities resulting
from discoveries made  before the HL-LHC
turns on may require  re-balancing the computing
resources --  perhaps  more reconstruction and
less simulation, or vice versa.
But the key problems will remain the same.
In general, the time scales for these contextual 
changes should be long enough that the Institute's
regular reviews will permit it to adapt its efforts without
significant disruption.

\tempnewpage

\section{Funding Scenarios}
\label{sec:funding}

The costs of an \s2i2 will depend on its scope and its relationships
to other entities. Most are estimated in terms of nominal
full-time-equivalent (FTE) professionals. Approximately a third of the
funding will support core personnel and other backbone activities. The
remaining funding will primarily support personnel, affiliated with
other university groups, to lead and contribute to software R\&D in
the identified focus areas.

Some of the Institute personnel may be working  only on \s2i2
projects. However, most effort will be done by a mixture of software
professionals working part-time on \s2i2 projects and part-time on
complementary projects, funded through other mechanisms, plus
post-docs and graduate students supported partly by the \s2i2 for
their work on its projects and supported partly by other funds for
related and complementary activities. Co-funding individuals with
relevant expertise will be a key method of ensuring significant
community buy-in and engagement. The Institute may undertake some
projects on its own, but {\em most} should be of sufficient interest
to attract support from elements of the community who want to
collaborate. For example, one of the topics in the {\em Reconstruction
  and Trigger Algorithms} focus area, identified as important by all of
the experiments, is learning to use vectorization programming
techniques effectively. An individual might develop generic toolkits
(or algorithms), funded by the Institute, and test them (or deploy
them) in experiment-specific software, funded by a partner. In such a
case, the Institute is leveraging its resources {\em and} ensuring
that its work is relevant to at least one experiment.

As a first approximation, we estimate that the fully loaded cost of a
software professional FTE will average \$200K/year.  Typically, this
will include salary, fringe benefits, travel, materials and supplies,
plus overhead.  Based on the experience of the OSG, we estimate that
operations personnel will average \$160K/year.

We expect that the core team will include an Executive Director
/ Deputy Director and project/administrative support plus a core set
of software professionals who will 
(i) engage directly in R\&D
projects related to established focus areas and exploratory studies,
(ii) provide software engineering support across the program,
(iii) provide the effort for the Institute ``backbone'' focused on
     developing, documenting and disseminating best practices and
     developing incentives,
(iv) provide some services (e.g., packaging and infrastructure support across
the program),
(v) lead the education and outreach effort,
(vi) lead the blueprint effort,
(vii) coordinate efforts to build bridges beyond the \s2i2 itself
     to the larger HEP, Computer Science, Software Engineering,
     and Data Science communities 
     and to establish the Institute as an intellectual hub for HL-LHC
     software and computing R\&D.
Depending on the funding available, and the overall scope of the
project, we anticipate that the team will consist of
the Executive Director / Deputy Director plus 5 -- 7  FTEs.
As a first approximation, the bottom lines for what be deemed
``central" expenses range from \$1200K/year to \$1800/year. 

An essential element of building a software R\&D will be sponsoring
workshops and supporting participation in other relevant workshops.
Based on our experience with the \s2i2 conceptualization process,
a Participant Costs budget of \$200K/year will prove sufficient, in large
measure because these funds can be used to supplement those from
other sources for many people.  Similarly, we estimate that a \$200K/year
Participant Costs budget reserved for summer schools and other
explicitly pedagogic activities will make a significant impact.  In
the tighter budget scenarios, these last two items could be reduced
stepwise to half in the lowest scenario.

Beyond the core efforts and backbone team, we anticipate funding 
an average of 4 FTE lines for each of four focus areas in the fully
funded scenario, about \$800K/year each. This level of effort would
provide {\it critical mass} to guarantee a significant leading impact
on a focus areas, given previous experience in smaller (NSF-funded)
projects such as DIANA-HEP~\cite{diana}, DASPOS~\cite{daspos-project},
the Parallel Kalman Filter Tracking Project~\cite{PARKAL} and the
``Any Data, Any Time, Anywhere: Global Data Access for
Science''~\cite{AAA} project. Almost none of the personnel funded by
these lines would be fully funded by the \s2i2 -- the projects they
will work on should be of sufficient interest to the community that
collaborators will co-fund  individuals whose other projects are
closely aligned with their Institute projects.  The total expense of
these activities in a fully funded project would be \$3200K/year.  If
sufficient funding is not available, the number of focus areas would
be reduced, rather than trying to fund all at  insufficient levels.
The bare minimum number of focus areas to have a significant impact on
HL-LHC software development would be 2, at a cost of \$1600K/year.

Beyond the software R\&D scope envisioned for the Institute when
the \s2i2 conceptualization process started, we have considered the
possibility that a single institute might serve as an umbrella
organization with some OSG-like operational responsibilities related to
the LHC experiments, as well.  As indicated in Table \ref{tab:osg},
this would require supporting up to 13.3 FTE operations personnel at an
estimated cost of $\sim$\$2100K/year.

\begin{table}[ht]
\centering
\begin{tabular}{l| c c c c c}
\hline
         & core and & participant  &             &             &       \\
scenario & backbone & costs & focus areas & operations  & total \\ \hline
low  R\&D    & 1200     & 200               & 1600        &  & 3000  \\
medium R\&D  & 1400     & 300               & 2400        &  & 4100  \\
high R\&D    & 1800     & 400               & 3200        &  & 5400  \\ \hline
OSG-HEP  &          &                   &             & 2100  & 2100  \\
\hline
\end{tabular}
\caption{Three possible budget scenarios for the R\&D efforts, plus the 
OSG-HEP operations effort. All entries are k\$/year.
\label{tab:budgets}}
\end{table}

Three software R\&D scenarios (no OSG-like operations responsibilities)
are illustrated in Table \ref{tab:budgets}.  The numbers are rough
estimates.  Funding for OSG-like operations adds another \$2100K
to any of these.  A proposal responding to a solicitation will need
to provide better estimates of the funding required to cover the
proposed activities.  For the purposes of this Strategic Plan, we
tentatively identify the {\it Reconstruction and Trigger Algorithms}
and {\it Data Organization, Management and Access} focus areas to be
the very highest priority for \s2i2 funding.  The former is closest
to the core physics program, and it is where U.S.\ university groups
have the most expertise and interest.  The latter covers core
technologies tying together processing all the way from data
acquisition to final physics analysis.  It is inherently
cross-disciplinary, and will engage U.S.\ university  HEP, Computer
Science, and  Software Engineering researchers. Data Analysis Systems
R\&D is essential to the success of the HL-LHC. If
insufficient funding is available through this funding mechanism,
efforts in this area might be funded through other mechanisms or
might be deferred. However, continuity of effort from the existing 
NSF-funded DIANA-HEP project~\cite{diana} and the ability to test run
analysis system solutions during LHC Run 3 will be at
risk. Applications of Machine Learning garnered the highest level of
interest during the CWP and \s2i2 conceptualization processes, and it
is especially well suited to cross-disciplinary research.  Deciding
not to include this as  one of the two highest priority focus areas at
this stage was a close call. Depending on the details of a
solicitation and the anticipated funding level, it might displace one
of the focus areas identified as higher priority here.


%

\tempnewpage

\appendix
\section{Appendix - \s2i2 Strategic Plan Elements}

The original \s2i2-HEP proposal was written in response to solicitation NSF 15-553~\cite{NSF15553}. This solicitation specified that: ``The product of a conceptualization award will be a strategic plan for enabling science and education through a sustained software infrastructure that will be freely available to the community, and will address the following elements: \ldots''. The specified elements and the corresponding sections in this document which address them are:

\begin{itemize}
\item the science community and the specific grand challenge research questions that the \s2i2 
will support (see Section~\ref{science_drivers});
\item specific software elements and frameworks that are relevant to the community, 
the sustainability challenges that need to be addressed, and why addressing these challenges 
will be transformative (see Section~\ref{sec:focusareas});
\item appropriate software architectures and lifecycle processes, development, testing 
and deployment methodologies, validation and verification processes, end usability and 
interface considerations, and required infrastructure and technologies
(see Sections~\ref{sec:hepsw}, \ref{sec:s2i2role-sw}, and~\ref{sec:backbone});
\item the required organizational, personnel and management structures and operational processes (see Sections~\ref{sec:orggov} and~\ref{sec:funding});
\item the requirements and necessary mechanisms for human resource development, including 
integration of education and training, mentoring of students, postdoctoral fellows 
as well as software professionals, and proactively addressing diversity and broadening 
participation (see Sections~\ref{sec:training} and~\ref{sec:broadeningparticipation});
\item potential approaches for long-term sustainability of the software institute 
as well as the software (see Section~\ref{sec:sustainability}); and
\item potential risks including risks associated with establishment and execution, necessary 
infrastructure and associated technologies, community engagement, and 
long-term sustainability (see Section~\ref{sec:RisksAndMitigation}).
\end{itemize}

\noindent Moreover, the solicitation states that ``The strategic plan resulting from the 
conceptualization phase is expected to serve as the conceptual design upon which a 
subsequent \s2i2 Implementation proposal could be based''. 
This document attempts to provide such a conceptual design,
as indicated by the section pointers in the list above.
\vskip 0.10in

The same solicitation (NSF 15-553~\cite{NSF15553}) invited implementation 
proposals for ``Chemical and Materials Research'' and ``Science Gateways''. 
For these, the solicitation requested that the following elements 
be addressed in the (20 page) proposals.
This document attempts to address all of these issues as well, other than 
a few which relate to specific implementation details: 

\begin{itemize}
\item The overall rationale for the envisioned institute, its mission, and its goals
(see Section~\ref{sec:intro});
\item A set of software issues and needs and software sustainability challenges faced by a 
particular, well-defined yet broad community (that is clearly identified in the proposal) 
that can best be addressed by an institute of the type proposed, a compelling case these 
are the most important issues faced by the community, and that these issues are truly important
(see Section~\ref{sec:focusareas});
\item A clear and compelling plan of activities that shows how the proposed institute 
will address these issues and needs by involving (and leveraging) the community, 
including its software developers, in a way that will benefit the entire community
(again, see Section~\ref{sec:focusareas});
\item If there are other NSF-funded activities that might appear to overlap the 
institute's activities, a discussion clarifying how the funding of each activity will be 
independent and non-overlapping (see Sections~\ref{sec:fabric} and~\ref{sec:partnerships});
\item Metrics of how success will be measured, that include at least impact on the developer 
and user communities (see Section~\ref{sec:metrics});
\item Evidence that the people involved in planning and setting up the institute have 
the organizational, scientific, technical, and sociocultural skills to undertake such a task, 
and that they are trusted and respected by the community as a whole 
(see Section~\ref{sec:Conceptualization} and the list of Endorsers following the Executive Summary);
\item Evidence of a high degree of community buy in that a) these are the 
urgent/critical needs and b) this institute is the way to address them
(again, see Section~\ref{sec:Conceptualization} and the list of Endorsers following the Executive Summary);
\item A plan for management of the institute, including 1) the specific roles of the PI, 
co-PIs, other senior personnel and paid consultants at all institutions involved, 
2) how the project will be managed across institutions and disciplines, 
3) identification of the specific coordination mechanisms that will 
enable cross-institution and/or cross-discipline scientific integration, and 
4) pointers to the budget line items that support these management and coordination mechanisms
(this would be premature in a Strategic Plan,  
and will need to be addressed carefully in a proposal responding to a specific solicitation).
\item A steering committee composed of leading members of the targeted community that will 
assume key roles in the leadership and/or management of the institute. 
A brief biography of the members of the steering committee and their role in the 
conceptualization process should be included
(again, this would be premature in a Strategic Plan,
and will need to be addressed carefully in a proposal responding to a specific solicitation);
\item A plan for how the institute activities will continue and/or the value of the institute's 
products will be preserved after the award, particularly if it does not receive 
additional funds from NSF  
(see Section~\ref{sec:sustainability})
\end{itemize}

\vskip 0.1in
In addition, a National Academy of Science report, {\it Future
Directions for NSF Advanced Computing Infrastructure to Support
U.S. Science and Engineering in 2017-2020}~\cite{NAP21886}, appeared
shortly before the \s2i2-HEP project began. One of its general 
recommendations is that NSF ``collect community requirements and construct
and publish roadmaps to allow it to better set priorities and make
more strategic decisions about advanced computing'' and that these
roadmaps should ``would reflect the visions of the science communities
supported by NSF, including both large users and those (in the
``long- tail'') with more modest needs. The goal is to develop brief
documents that set forth the overall strategy and approach rather
than high-resolution details. They would look roughly 5 years
ahead and provide a vision that extends about 10 years ahead.'' The
\s2i2-HEP and CWP community processes should be seen as input regarding
the vision of the HEP community for the HL-LHC era.

\tempnewpage

\section{Appendix - Workshop List}
\label{workshoplist}

\vskip 0.10in
During the process we have organized a number of workshops and sessions at
preexisting meetings. These included (in chronological order):

\vskip 0.10in
\noindent{\bf \s2i2 HEP/CS Workshop} \\
\noindent{\it Date:} 7--9 Dec, 2016 \\
\noindent{\it Location:} University of Illinois at Urbana-Champaign \\
\noindent{\it URL:} \url{https://indico.cern.ch/event/575443/}\\
\noindent{\it Summary report:} \url{http://s2i2-hep.org/downloads/s2i2-hep-cs-workshop-summary.pdf}\\
\noindent{\it Description:} This workshop brought together attendees from both the particle physics and computer science (CS) communities to understand how the two communities could work together in the context of a future NSF Software Institute aimed at supporting particle physics research over the long term. While CS experience and expertise has been brought into the HEP community over the years, this was a fresh look at planned HEP and computer science research and brainstorm about engaging specific areas of effort, perspectives, synergies and expertise of mutual benefit to HEP and CS communities, especially as it relates to a future NSF Software Institute for HEP. \\
\vskip 0.01in

\vskip 0.1in
\noindent{\bf HEP Software Foundation Workshop} \\
\noindent{\it Date:} 23--26 Jan, 2017 \\
\noindent{\it Location:} UCSD/SDSC (La Jolla, CA) \\
\noindent{\it URL:} \url{http://indico.cern.ch/event/570249/}\\
\noindent{\it Description:} This HSF workshop at SDSC/UCSD was the first workshop supporting the CWP process. There were plenary sessions covering topics of general interest as well as parallel sessions for the many topical working groups in progress for the CWP.\\
\vskip 0.01in

\vskip 0.1in
\noindent{\bf \s2i2-HEP/OSG/US-CMS/US-ATLAS Panel} \\ 
\noindent{\it Date:} 8 Mar, 2017 \\
\noindent{\it Location:} UCSD/SDSC (La Jolla, CA)\\
\noindent{\it URL:} \url{https://indico.fnal.gov/conferenceTimeTable.py?confId=12973#20170308}\\
\noindent{\it Description:} This panel took place at Open Science Grid All Hands Meeting (OSG-AHM). Participants included Kaushik De (US-ATLAS), Peter Elmer (\s2i2-HEP, US-CMS), Oli Gutsche (US-CMS) and Mark Neubauer (\s2i2-HEP, US-ATLAS), with Frank Wuerthwein (OSG, US-CMS) as moderator. The goal was to inform the OSG community about the CWP and \s2i2-HEP processes and learn from the OSG experience.\\
\vskip 0.01in

\vskip 0.1in
\noindent{\bf Software Triggers and Event Reconstruction WG meeting}\\
\noindent{\it Date:} 9 Mar, 2017\\
\noindent{\it Location:} LAL-Orsay (Orsay, France)\\
\noindent{\it URL:} \url{https://indico.cern.ch/event/614111/}\\
\noindent{\it Description:} This was a meeting of the Software Triggers and Event Reconstruction CWP working group. It was held as a parallel session at the ``Connecting the Dots'' workshop, which focuses on forward-looking pattern recognition and machine learning algorithms for use in HEP.\\
\vskip 0.01in

\vskip 0.1in
\noindent{\bf IML Topical Machine Learning Workshop} \\
\noindent{\it Date:} 20--22 Mar, 2017\\
\noindent{\it Location:} CERN (Geneva, Switzerland)\\
\noindent{\it URL:} \url{https://indico.cern.ch/event/595059}\\
\noindent{\it Description:} This was a meeting of the Machine Learning CWP working group. It was held as a parallel session at the ``Inter-experimental Machine Learning (IML)'' workshop, an organization formed in 2016 to facilitate communication regarding R\&D on ML applications in the LHC experiments.\\
\vskip 0.01in

\vskip 0.1in
\noindent{\bf Community White Paper Follow-up at FNAL} \\
\noindent{\it Date:} 23 Mar, 2017\\
\noindent{\it Location:} FNAL (Batavia, IL)\\
\noindent{\it URL:} \url{https://indico.fnal.gov/conferenceDisplay.py?confId=14032}\\
\noindent{\it Description:} This one-day workshop was organized to engage with the experimental HEP community involved in computing and software for Intensity Frontier experiments at FNAL. Plans for the CWP and the \s2i2-HEP project were described, with discussion about commonalities between the HL-LHC challenges and the challenges of the FNAL neutrino and muon experiments.\\
\vskip 0.01in

\vskip 0.1in
\noindent{\bf CWP Visualization Workshop} \\
\noindent{\it Date:} 28--30 Mar, 2017\\
\noindent{\it Location:} CERN (Geneva, Switzerland)\\
\noindent{\it URL:} \url{https://indico.cern.ch/event/617054/}\\
\noindent{\it Description:} This workshop was organized by the Visualization CWP working group. It explored the current landscape of HEP visualization tools as well as visions for how these could evolve. There was participation both from HEP developers and industry.\\
\vskip 0.01in

\vskip 0.1in
\noindent{\bf 2nd \s2i2 HEP/CS Workshop} \\
\noindent{\it Date:} 1--3 May, 2017\\
\noindent{\it Location:} Princeton University (Princeton, NJ)\\
\noindent{\it URL:} \url{https://indico.cern.ch/event/622920/}\\
\noindent{\it Description:} This 2nd HEP/CS workshop built on the discussions which took place at the the first \s2i2 HEP/CS workshop to take a fresh look at planned HEP and computer science research and brainstorm about engaging specific areas of effort, perspectives, synergies and expertise of mutual benefit to HEP and CS communities, especially as it relates to a future NSF Software Institute for HEP.\\
\vskip 0.01in

\vskip 0.1in
\noindent{\bf DS@HEP 2017 (Data Science in High Energy Physics)} \\
\noindent{\it Date:} 8--12 May, 2017\\
\noindent{\it Location:} FNAL (Batava, IL)\\
\noindent{\it URL:} \url{https://indico.fnal.gov/conferenceDisplay.py?confId=13497}\\
\noindent{\it Description:} This was a meeting of the Machine Learning CWP working group. It was held as a parallel session at the ``Data Science in High Energy Physics (DS@HEP)'' workshop, a workshop series begun in 2015 to facilitate communication regarding R\&D on ML applications in HEP.\\
\vskip 0.01in

\vskip 0.1in
\noindent{\bf HEP Analysis Ecosystem Retreat} \\
\noindent{\it Date:} 22--24 May, 2017\\
\noindent{\it Location:} Amsterdam, the Netherlands\\
\noindent{\it URL:} \url{http://indico.cern.ch/event/613842/}\\
\noindent{\it Summary report:} \url{http://hepsoftwarefoundation.org/assets/AnalysisEcosystemReport20170804.pdf}\\
\noindent{\it Description:} This was a general workshop, organized about the HSF, about the ecosystem of analysis tools used in HEP and the ROOT software framework. The workshop focused both on the current status and the 5-10 year time scale covered by the CWP. \\
\vskip 0.01in

\vskip 0.1in
\noindent{\bf CWP Event Processing Frameworks Workshop}\\
\noindent{\it Date:} 5-6 Jun, 2017\\
\noindent{\it Location:} FNAL (Batavia, IL)\\
\noindent{\it URL:} \url{https://indico.fnal.gov/conferenceDisplay.py?confId=14186}\\
\noindent{\it Description:} This was a workshop held by the Event Processing Frameworks CWP working group.\\
\vskip 0.01in

\vskip 0.1in
\noindent{\bf HEP Software Foundation Workshop}\\
\noindent{\it Date:} 26--30 Jun, 2017 \\
\noindent{\it Location:} LAPP (Annecy, France)\\
\noindent{\it URL:} \url{https://indico.cern.ch/event/613093/}\\
\noindent{\it Description:} This was the final general workshop for the CWP process. The CWP working groups came together to present their status and plans, and develop consensus on the organization and context for the community roadmap. Plans were also made for the CWP writing phase that followed in the few months following this last workshop.\\
\vskip 0.01in

\vskip 0.1in
\noindent{\bf \s2i2-HEP Workshop}\\ 
\noindent{\it Date:} 23--26 Aug, 2017 \\
\noindent{\it Location:} University of Washington, Seattle (Seattle, WA)\\
\noindent{\it URL:} \url{https://indico.cern.ch/event/640290/}\\
\noindent{\it Description:} This final \s2i2-HEP workshop was held as a satellite workshop of the ACAT 2017 Conference. The workshop built on the emerging consensus from the CWP process and focused on the role an NSF-supported Software Institute could play. Specific discussions focused on establishing which areas would be both high impact and appropriate for leadership role in the U.S.\ universities. In addition the relative roles of an Institute, the US LHC Ops programs and the international LHC program were discussed, along with possible management structures for an Institute.\\
\vskip 0.01in

\vskip 0.1in
\noindent{\bf Data Organisation, Management and Access (DOMA) in Astronomy, Genomics and High Energy Physics}\\
\noindent{\it Date:} 16--17 Nov, 2017 \\
\noindent{\it Location:} Flatiron Institute (New York City, NY)\\
\noindent{\it URL:} \url{https://indico.cern.ch/event/669506/}\\
\noindent{\it Description:} This workshop was co-sponsored by the Simons Foundation and the \s2i2-HEP project. The workshop focused on the current research practices and future needs for Data Organization, Management and Access (DOMA) across the fields of Astronomy, Genomics and High Energy Physics. Discussions centered on identifying possibilities for integration across the different fields, as well as opportunities for common research and development activities.\\
\vskip 0.01in

\vskip 0.1in
\noindent{\bf \s2i2/DOE mini-workshop on HL-LHC Software and Computing R\&D}\\
\noindent{\it Date:} 28--29 Nov, 2017 \\
\noindent{\it Location:} Catholic University of America (Washington DC)\\
\noindent{\it URL:} \url{https://indico.cern.ch/event/678121/}\\
\noindent{\it Description:} The goals of this workshop was to {\em(1)} review 
the vision for the ensemble of possible R\&D efforts for the HL-LHC as 
articulated via the international CWP effort, and {\em(2)} articulate how R\&D 
efforts such as an NSF \s2i2 would interact with the US-LHC Operations 
programs and DOE efforts (in the context of the full, international efforts).
Discuss the broad scope of relevant capabilities and current DOE and NSF 
funded efforts. \\
\vskip 0.01in

\noindent This full list of workshops and meetings (with links) is also available on the \url{http://s2i2-hep.org} website. In addition there were ``internal'' sessions regarding the CWP in the LHC experiment collaboration meetings, which are not listed above.



\vskip 0.1in
More than 260 people participated in one or more of the workshops
which had an explicit registration and participant list. This does not 
include those who participated in the many
``outreach'' or panel sessions at pre-existing workshops/meetings
such as DS@HEP, the
OSG AHM, the IML Workshop or the sessions at LHC experiment collaboration meetings which not
listed above, for which no explicit participant list was tracked. The
combined list of known registered participants is:
\vskip 0.1in
\noindent Aaron Elliott (Aegis Research Labs), 
Aaron Dominguez (Catholic University of America),
Aaron Sauers (Fermilab),
Aashrita Mangu (California Institute of Technology),
Abid Patwa (DOE),
Adam Aurisano (University of Cincinnati),
Adam Lyon (FNAL),
Ajit Majumder (Wayne State), 
Alexei Klimentov (Brookhaven National Lab),
Alexey Svyatkovskiy (Princeton University),
Alja Mrak Tadel (Univerity California San Diego),
Amber Boehnlein (Jefferson Lab),
Amir Farbin (University of Texas at Arlington),
Amit Kumar (Southern Methodist), 
Andrea Dotti (SLAC National Accelerator Laboratory),
Andrea Rizzi (INFN-Pisa),
Andrea Valassi (CERN),
Andrei Gheata (CERN),
Andrew Gilbert (KIT),
Andrew Hanushevsky (SLAC National Accelerator Laboratiry),
Anton Burtsev (University of California, Irvine),
Anton Poluektov (University of Warwick),
Antonio Augusto Alves Junior (University of Cincinnati),
Antonio Limosani (CERN / University of Sydney),
Anyes Taffard (UC Irvine),
Ariel Schwartzman (SLAC),
Attila Krasznahorkay (CERN),
Avi Yagil (UCSD),
Axel Naumann (CERN),
Ben Hooberman (Illinois), 
Benedikt Hegner (CERN),
Benedikt Riedel (University of Chicago), 
Benjamin Couturier (CERN),
Bill Nitzberg (Altair),
Bo Jayatilaka (FNAL),
Bogdan Mihaila (NSF),
Brian Bockelman (University of Nebraska - Lincoln),
Brian O'Connor (University of California at Santa Cruz),
Burt Holzman (Fermilab),
Carlos Maltzahn (University of California - Santa Cruz),
Catherine Biscarat (CNRS),
Cecile Barbier (LAPP),
Charles Leggett (LBNL),
Charlotte Lee (University of Washington),
Chris Green (FNAL),
Chris Tunnell (University of Chicago, KICP),
Christopher Jones (FNAL),
Claudio Grandi (INFN),
Conor Fitzpatrick (EPFL),
Daniel S. Katz (University of Illinois at Urbana-Champaign/NCSA),
Dan Riley (Cornell University),
Daniel Whiteson (UC Irvine),
Daniele Bonacorsi (University of Bologna),
Danko Adrovic (DePaul), 
Dario Berzano (CERN),
Dario Menasce (INFN Milano-Bicocca),
David Abdurachmanov (University of Nebraska-Lincoln),
David Lange (Princeton University),
David Lesny (Illinois), 
David Malon (Argonne National Laboratory),
David Rousseau (LAL-Orsay),
David Smith (CERN),
Dick Greenwood (Louisiana Tech University),
Dirk Duellmann (CERN),
Dirk Hufnagel (Fermilab),
Don Petravick (Illinois/NCSA), 
Dorian Kcira (California Institute of Technology),
Doug Benjamin (Duke University),
Doug Thain (Notre Dame), 
Douglas Thain (University of Notre Dame),
Dustin Anderson (California Institute of Technology),
Dustin Tran (Columbia University),
Eduardo Rodrigues (University of Cincinnati),
Elizabeth Sexton-Kennedy (FNAL),
Enric Tejedor Saavedra (CERN),
Eric Lancon (BNL),
Eric Vaandering (FNAL),
Farah Hariri (CERN),
Federico Carminati (CERN),
Fernanda Psihas (Indiana University),
Fons Rademakers (CERN),
Frank Gaede (DESY),
Frank Wuerthwein (University of California at San Diego/SDSC),
Frederique Chollet (LAPP),
Gabriel Perdue (Fermilab),
Gerardo Ganis (CERN),
Gerhard Raven (Nikhef),
Giacomo Govi (FNAL),
Giacomo Tenaglia (CERN),
Gianluca Cerminara (CERN),
Giulio Eulisse (CERN),
Gloria Corti (CERN),
Gordon Watts (University of Washington),
Graeme Stewart (University of Glasgow),
Graham Mackintosh (IBM),
H. Birali Runesha (University of Chicago),
Hadrien Grasland (Universite de Paris-Sud),
Harvey Newman (Caltech),
Helge Meinhard (CERN),
Henry Schreiner III (University of Cincinnati),
Horst Severini (University of Oklahoma),
Ian Bird (CERN),
Ian Collier (RAL),
Ian Cosden (Princeton University),
Ian Fisk (Simons Foundation),
Ian Stockdale (Altair Engineering),
Ilija Vukotic (University of Chicago),
Isobel Ojalvo (Princeton University),
Ivo Jimenez UC (University of California - Santa Cruz),
Jakob Blomer (CERN),
Jamie Bedard (Siena College),
Jean Jacquemier (LAPP),
Jean-Roch Vlimant (California Institute of Technology),
Jeff Carver (University of Alabama),
Jeff Hammond (Intel),
Jeff Lefevre (University of California at Santa Cruz),
Jeff Porter (LBNL), 
Jeff Templon (Nikhef),
Jeffrey Carver (University of Alabama),
Jerome Lauret (BNL),
Jim Kowalkowski (FNAL),
Jim Pivarski (Princeton University),
Johannes Albrecht (TU Dortmund),
John Apostolakis (CERN),
John Harvey (CERN),
John Towns (Illinois/NCSA), 
Joon Kim (Princeton University),
Joseph Boudreau (University of Pittsburgh),
Justas Balcas (Caltech),
Justin Wozniak (University of Chicago/ANL),
Karan Bhatia (Google Cloud),
Karen Tomko (Ohio Supercomputer Center),
Kathryn Huff (Illinois), 
Kaushik De (University of Texas at Arlington),
Ken Bloom (University of Nebraska-Lincoln),
Kevin Jorissen (Amazon Web Services),
Kevin Lannon (University of Notre Dame),
Konstantin Toms (University of New Mexico),
Kurt Rinnert (U.Liverpool),
Kyle Chard (University of Chicago), 
Kyle Cranmer (New York University),
Kyle Knoepfel (FNAL),
Lauren Anderson (Flatiron Institute),
Lawrence R Frank (UCSD),
Lindsey Gray (Fermilab),
Liz Sexton-Kennedy (FNAL),
Lorenzo Moneta (CERN),
Lothar Bauerdick (FNAL),
Louis Capps (NVIDIA),
Lukas Heinrich (New York University),
Lukasz Kreczko (Bristol),
Madeline Hagen (Siena College),
Makoto Asai (SLAC),
Manish Parashar (Rutgers University),
Marc Paterno (FNAL),
Marc Verderi (Ecole Polytechnique),
Marcin Nowak (CERN),
Maria Girone (CERN),
Maria Spiropulu (Caltech),
Mario Lassnig (CERN),
Mark Neubauer (University of Illinois at Urbana-Champaign),
Markus Klute (MIT),
Markus Schulz (CERN),
Martin Ritter (LMU Munich),
Matevz Tadel (UCSD),
Matthew Bellis (Siena College),
Matt Zhang (Illinois),
Matthew Feickert (Southern Methodist University),
Matthew Turk (University of Illinois), 
Matthieu Lefebvre (Princeton University),
Max Baak (KPMG),
Meghan Frate (University of California, Irvine),
Meghan Kane (SoundCloud, MIT),
Michael Andrews (Carnegie Mellon University/CERN),
Michael Kirby (FNAL),
Michael Sevilla (University of California, Santa Cruz),
Michael Sokoloff (University of Cincinnati),
Michel Jouvin (LAL/Universite de Paris-Sud),
Michela Paganini (Yale University),
Michela Taufer (University of Delaware),
Mike Hildreth (University of Notre Dame),
Mike Williams (MIT),
Miron Livny (University of Wisconsin-Madison),
Mohammad Al-Turany (GSI),
Nadine Neyroud (LAPP),
Nan Niu (University of Cincinnati),
Nancy Wilkins-Diehr (University of California San Diego),
Natalia Volfovsky (Simons Foundation),
Nathalie Rauschmayr (CERN),
Neil Ernst (Software Engineering Institute), 
Noah Watkins (University of California, Santa Cruz),
Oliver Gutsche (FNAL),
Oliver Keeble (CERN),
Panagiotis Spentzouris (FNAL),
Paolo Calafiura (LBNL),
Parag Mhashilkar (Fermilab),
Patricia Mendez Lorenzo (CERN),
Patrick Bos (Netherlands eScience Center),
Patrick Skubic (University of Oklahoma),
Patrick de Perio (Columbia University),
Paul Laycock (CERN),
Paul Mattione (Jefferson Lab),
Paul Rossman (Google Inc.),
Pere Mato (CERN),
Peter Elmer (Princeton University),
Peter Hristov (CERN),
Peter Onyisi (University of Texas at Austin),
Philippe Canal (FNAL),
Pierre Aubert (LAPP),
Rajesh Ranganath (Princeton University),
Riccardo Maria Bianchi (University of Pittsburgh),
Richard Hay Jr (Princeton University),
Richard Mount (SLAC),
Rick Wagner (Globus),
Rob Gardner (University of Chicago),
Rob Quick (Indiana University),
Robert Illingworth (Fermilab),
Robert Kalescky (Southern Methodist), 
Robert Knight (Princeton University),
Robert Kutschke (Fermilab),
Roger Jones (Lancaster),
Ruslan Mashinistov (University of Texas at Arlington),
Sabine Elles (LAPP),
Sally Seidel (New Mexico), 
Sandra Gesing (University of Notre Dame), 
Sandro Wenzel (CERN),
Sascha Caron (Nikhef),
Sebastien Binet (IN2P3/LPC),
Sergei Gleyzer (University of Florida),
Shantenu Jha (Rutgers University),
Shaun Astarabadi (Western Digital),
Shawfeng Dong (University of California at Santa Cruz),
Shawn McKee (University of Michigan),
Shy Genel (Flatiron Institute),
Simone Campana (CERN),
Slava Krutelyov (University of California at San Diego),
Spencer Smith (McMaster University),
Stefan Roiser (CERN),
Steven Schramm (Universite de Geneve),
Sudhir Malik (University of Puerto Rico Mayaguez),
Sumanth Mannam (DePaul), 
Sumit Saluja (Princeton University),
Sunita Chandrasekaran (University of Delaware),
Tanu Malik (Depaul University),
Taylor Childers (Argonne Nat. Lab),
Thomas Hacker (Purdue University),
Thomas Kuhr (LMU),
Thomas McCauley (University of Notre Dame),
Thomas Vuillaume (LAPP),
Thorsten Kollegger (GSI),
Tom Gibbs (NVIDIA),
Tom Lecompte (DOE/ANL),
Tommaso Boccali (INFN Pisa),
Torre Wenaus (BNL),
V. Daniel Elvira (Fermilab),
Vakho Tsulaia (LBNL),
Valentin Kuznetsov (Cornell University),
Vassil Vassilev (Princeton University),
Vincent Croft (Nikhef),
Vinod Gupta (Princeton University),
Vladimir Gligorov (CNRS),
Wahid Bhimji (NERSC/LBNL),
Wenjing Wu (Institute of High Energy Physics, Beijing),
Wouter Verkerke (Nikhef)

\tempnewpage

\section{Appendix - Glossary of Acronyms}
\label{glossary}

\begin{description}[leftmargin=0pt]

\item[ABC] Approximate Bayesian Computation

\item[ACAT] A workshop series on Advanced Computing and Analysis Techniques in HEP.

\item[ALICE] A Large Ion Collider Experiment, an experiment at the LHC at CERN.

\item[ALPGEN] An event generator designed for the generation of Standard Model
processes in hadronic collisions, with emphasis on final states with
large jet multiplicities. It is based on the exact LO evaluation of
partonic matrix elements, as well as top quark and gauge boson decays
with helicity correlations.

\item[AOD] Analysis Object Data is a summary of the reconstructed event and
contains sufficient information for common physics analyses.

\item[ATLAS] A Toroidal LHC ApparatuS, an experiment at the LHC at CERN.

\item[BaBar] A large HEP experiment which ran at SLAC from 1999 through 2008.

\item[BSM] Physics beyond the Standard Model (BSM) refers to the theoretical
developments needed to explain the deficiencies of the Standard Model
(SM), such as the
\href{https://en.wikipedia.org/wiki/Origin_of_mass}{origin of mass}, the
\href{https://en.wikipedia.org/wiki/Strong_CP_problem}{strong CP
problem},
\href{https://en.wikipedia.org/wiki/Neutrino_oscillation}{neutrino
oscillations},
\href{https://en.wikipedia.org/wiki/Baryon_asymmetry}{matter--antimatter
asymmetry}, and the nature of
\href{https://en.wikipedia.org/wiki/Dark_matter}{dark matter} and
\href{https://en.wikipedia.org/wiki/Dark_energy}{dark energy}.

\item[CDN] Content Delivery Network

\item[CERN] The European Laboratory for Particle Physics, the host laboratory
for the LHC (and eventually HL-LHC) accelerators and the ALICE, ATLAS, CMS
and LHCb experiments.

\item[CHEP] An international conference series on Computing in High Energy and Nuclear 
Physics.

\item[CMS] Compact Muon Solenoid, an experiment at the LHC at CERN.

\item[CMSSW] Application software for the CMS experiment including the processing framework itself and components relevant for
event reconstruction, high-level trigger, analysis, hardware trigger emulation, simulation, 
and visualization workflows.

\item[CMSDAS] The CMS Data Analysis School

\item[CoDaS-HEP] The COmputational and DAta Science in HEP school.




\item[CP] Charge and Parity conjugation symmetry

\item[CPV] CP violation

\item[CS] Computer Science

\item[CRSG] Computing Resources Scrutiny Group, a WLCG committee
in charge of scrutinizing and assessing LHC experiment yearly
resource requests to prepare funding agency decisions.


\item[CTDR] Computing Technical Design Report, a document written by one of
the experiments to describe the experiment's technical blueprint for building 
the software and computing system

\item[CVMFS] The CERN Virtual Machine File System is a network file system
based on HTTP and optimised to deliver experiment software in a fast,
scalable, and reliable way through sophisticated caching strategies.

\item[CVS] Concurrent Versions System, a source code version control system

\item[CWP] The Community White Paper is the result of an organised effort to 
describe the community strategy and a roadmap for software and computing R\&D 
in HEP for the 2020s. This activity is organised under the umbrella of the HSF.

\item[DASPOS] the Data And Software Preservation for Open Science project

\item[Deep Learning] one class of Machine Learning algorithms, based on a
high number of neural network layers.

\item[DES] The Dark Energy Survey

\item[DIANA-HEP] the Data Intensive Analysis for High Energy Physics project,
funded by NSF as part of the SI2 program


\item[DOE] The Department of Energy

\item[DHTC] Distributed High Throughput Computing

\item[DOMA] Data Organization, Management and Access, a term for an integrated
view of all aspects of how a project interacts with and uses data.


\item[EFT] the Effective Field Theory, an extension of the Standard Model


\item[EYETS] Extended Year End Technical Stop, used to denote a period 
(typically several months) in the winter when small upgrades and maintenance
are performed on the CERN accelerator complex and detectors





\item[FNAL] Fermi National Accelerator Laboratory, also known as Fermilab,
the primary US High Energy Physics Laboratory, funded by the US Department
of Energy

\item[FPGA] Field Programmable Gate Array

\item[FTE] Full Time Equivalent

\item[FTS] File Transfer Service

\item[GAN] Generative Adversarial Networks are a class of
\href{https://en.wikipedia.org/wiki/Artificial_intelligence}{artificial
intelligence} algorithms used in
\href{https://en.wikipedia.org/wiki/Unsupervised_machine_learning}{unsupervised
machine learning}, implemented by a system of two
\href{https://en.wikipedia.org/wiki/Neural_network}{neural networks}
contesting with each other in a
\href{https://en.wikipedia.org/wiki/Zero-sum_game}{zero-sum game}
framework.

\item[GAUDI] An event processing application framework developed by CERN

\item[Geant4] A toolkit for the simulation of the passage of
particles through matter.

\item[GeantV] An R\&D project that aims to fully exploit the
parallelism, which is increasingly offered by the new generations
of CPUs, in the field of detector simulation.

\item[GPGPU] General-Purpose computing on Graphics Processing Units
is the use of a Graphics Processing Unit (GPU), which typically
handles computation only for computer graphics, to perform computation
in applications traditionally handled by the Central Processing
Unit (CPU). Programming for GPUs is typically more challenging, but
can offer significant gains in arithmetic throughput.

\item[HEP] High Energy Physics 

\item[HEP-CCE] the HEP Center for Computational Excellence, a DOE-funded
cross-cutting initiative to promote excellence in high performance
computing (HPC) including data-intensive applications, scientific
simulations, and data movement and storage

\item[HEPData] The Durham High Energy Physics Database is an open access
repository for scattering data from experimental particle physics.

\item[HEPiX] A series of twice-annual workshops which bring together IT
staff and HEP personnel involved in HEP computing


\item[HL-LHC] The High Luminosity Large Hadron Collider is a proposed
upgrade to the Large Hadron Collider to be made in 2026. The upgrade
aims at increasing the luminosity of the machine by a factor of 10,
up to $10^{35}\mathrm{cm}^{-2}\mathrm{s}^{-1}$, providing a better
chance to see rare processes and improving statistically marginal
measurements.

\item[HLT] High Level Trigger. Software trigger system generally using a large computing cluster located close 
to the detector. Events are processed in real-time (or within the latency defined by small buffers) 
and select those who must be stored for further processing offline.

\item[HPC] High Performance Computing.

\item[HS06] HEP-wide benchmark for measuring CPU performance based on the 
SPEC2006 benchmark (\href{https://www.spec.org}{{https://www.spec.org}}).

\item[HSF] The HEP Software Foundation facilitates coordination and common
efforts in high energy physics (HEP) software and computing
internationally.

\item[IgProf] The Ignominius Profiler, a tool for exploring the CPU and 
memory use performance of very large C++ applications like those used in HEP

\item[IML] The Inter-experimental LHC Machine Learning (IML) Working Group is
focused on the development of modern state-of-the art machine learning
methods, techniques and practices for high-energy physics problems.

\item[INFN] The Istituto Nazionale di Fisica Nucleare, the main funding agency 
and series of laboratories involved in High Energy Physics research in Italy


\item[JavaScript] A high-level, dynamic, weakly typed,
prototype-based, multi-paradigm, and interpreted programming language.
Alongside HTML and CSS, JavaScript is one of the three core technologies
of World Wide Web content production.

\item[Jupyter Notebook] This is a server-client application that allows
editing and running notebook documents via a web browser. Notebooks are
documents produced by the Jupyter Notebook App, which contain both
computer code (e.g., python) and rich text elements (paragraph,
equations, figures, links, etc...). Notebook documents are both
human-readable documents containing the analysis description and the
results (figures, tables, etc..) as well as executable documents which
can be run to perform data analysis.

\item[LEP] The Large Electron-Positron Collider, the original accelerator
which occupied the 27km circular tunnel at CERN now occupied by the Large
Hadron Collider

\item[LHC] Large Hadron Collider, the main particle accelerator at CERN.

\item[LHCb] Large Hadron Collider beauty, an experiment at the LHC at CERN



\item[LIGO] The Laser Interferometer Gravitational-Wave Observatory

\item[LS] Long Shutdown, used to denote a period (typically 1 or more years) 
in which the LHC is not producing data and the CERN accelerator complex and
detectors are being upgraded.

\item[LSST] The Large Synoptic Survey Telescope



\item[ML] Machine learning is a field of computer science that gives computers
the ability to learn without being explicitly programmed. It focuses on
prediction making through the use of computers and encompasses a lot of
algorithm classes (boosted decision trees, neural networks\ldots{}).


\item[MREFC] Major Research Equipment and Facilities Construction, an NSF
mechanism for large construction projects

\item[NAS] The National Academy of Sciences

\item[NCSA] National Center of Supercomputing Applications, at the University 
of Illinois at Urbana-Champaign

\item[NDN] Named Data Networking

\item[NSF] The National Science Foundation

\item[ONNX] Open Neural Network Exchange, an evolving open-source standard for exchanging AI models


\item[Openlab] CERN openlab is a public-private partnership that accelerates
the development of cutting-edge solutions for the worldwide LHC
community and wider scientific research.

\item[OSG] The Open Science Grid

\item[P5] The Particle Physics Project Prioritization Panel is a scientific
advisory panel tasked with recommending plans for U.S.\ investment in
particle physics research over the next ten years.

\item[PI] Principal Investigator




\item[QA] Quality Assurance

\item[QC] Quality Control

\item[QCD] Quantum Chromodynamics, the theory describing the strong
interaction between quarks and gluons.

\item[REANA] REusable ANAlyses, a system to preserve and instantiate analysis workflows 

\item[REU] Research Experience for Undergraduates, an NSF program to fund undergraduate participation in research projects

\item[RRB] Resources Review Board, a CERN committee made up of representatives of funding agencies participating in the LHC collaborations, the CERN management and the experiment's management.


\item[ROOT] A scientific software framework widely used in HEP data
processing applications.


\item[SciDAC] Scientific Discovery through Advanced Computing, a DOE program
to fund advanced R\&D on computing topics relevant to the DOE Office of
Science


\item[SDSC] San Diego Supercomputer Center, at the University of California
at San Diego

\item[SHERPA] Sherpa is a Monte Carlo event generator for the Simulation of
High-Energy Reactions of PArticles in lepton-lepton, lepton-photon,
photon-photon, lepton-hadron and hadron-hadron collisions.

\item[SIMD] Single instruction, multiple data (\textbf{SIMD}), describes
computers with multiple processing elements that perform the same
operation on multiple data points simultaneously.

\item[SI2] The Software Infrastructure for Sustained Innovation program at NSF

\item[SKA] The Square Kilometer Array

\item[SLAC] The Stanford Linear Accelerator Center, a laboratory funded by the
US Department of Energy

\item[SM] The Standard Model is the name given in the 1970s to a theory of
fundamental particles and how they interact. It is the currently
dominant theory explaining the elementary particles and their dynamics.

\item[SOW] Statement of Work, a mechanism used to define the expected activities and deliverables of individuals funded from a subaward with a multi-institutional project. The SOW is typically revised annually, along with the corresponding budgets.

\item[SSI] The Software Sustainability Institute, an organization in the UK dedicated to fostering better, and more sustainable, software for research.

\item[SWAN] Service for Web based ANalysis is a platform for interactive data
mining in the CERN cloud using the Jupyter notebook interface.


\item[TMVA] The Toolkit for Multivariate Data Analysis with ROOT is a
standalone project that provides a ROOT-integrated machine learning
environment for the processing and parallel evaluation of sophisticated
multivariate classification techniques.

\item[TPU] Tensor Processing Unit, an application-specific integrated circuit 
by Google designed for use with Machine Learning applications

\item[URSSI] the US Software Sustainability Institute, an \s2i2 conceptualization activity recommended for funding by NSF



\item[WAN] Wide Area Network


\item[WLCG] The Worldwide LHC Computing Grid project is a global collaboration
of more than 170 computing centres in 42 countries, linking up national
and international grid infrastructures. The mission of the WLCG project
is to provide global computing resources to store, distribute and
analyse data generated by the Large Hadron Collider (LHC) at CERN.


\item[x86\_64] 64-bit version of the x86 instruction set, which originated wiht the Intel 8086, but has
now been implemented on processors from a range of companies, including the Intel and AMD processors that 
make up the vast majority of computing resources used by HEP today.

\item[XRootD] Software framework that is a fully generic suite for
fast, low latency and scalable data access.

\end{description}

\newpage 
\bibliographystyle{unsrt}
\bibliography{s2i2-hep-strategic-plan}

\end{document}